\documentclass[onecolumn,journal,twoside]{IEEEtran}
\usepackage{cite}

\ifCLASSINFOpdf
\else
   \usepackage[dvips]{graphicx}
\fi
\usepackage[cmex10]{amsmath}
\usepackage{amssymb,amsthm}
\usepackage{multirow,bigdelim}
\newtheorem{theorem}{Theorem} 
 
\newtheorem{lemma}{Lemma} 
\newtheorem{example}{Example} 
\newtheorem{remark}{Remark} 
\newtheorem{definition}{Definition} 
\newtheorem{proposition}{Proposition} 
\newtheorem{corollary}{Corollary}

\begin{document}
%
\title{Performance Improvement of Iterative Multiuser Detection for Large 
Sparsely-Spread CDMA Systems by Spatial Coupling}
%
%
%

\author{Keigo~Takeuchi,~\IEEEmembership{Member,~IEEE,}
        Toshiyuki~Tanaka,~\IEEEmembership{Member,~IEEE,}
        and~Tsutomu~Kawabata,~\IEEEmembership{Member,~IEEE,}
\thanks{Manuscript received June, 2012. 
The work of K.~Takeuchi was in part supported by the Grant-in-Aid for 
Young Scientists (B) (No.\ 23760329) from MEXT, Japan. 
The material in this paper was presented in part at 2011 IEEE International 
Symposium on Information Theory, Saint-Petersburg, Russia, Aug. 2011.}
\thanks{K.~Takeuchi and T.~Kawabata are with the Department of Communication 
Engineering and Informatics, the University of Electro-Communications, 
Tokyo 182-8585, Japan (e-mail: ktakeuchi@uec.ac.jp, kawabata@uec.ac.jp).}
\thanks{T.~Tanaka is with the Department of Systems Science, 
Graduate School of Informatics, Kyoto University, Kyoto 606-8501, 
Japan (e-mail: tt@i.kyoto-u.ac.jp).}
}

%
%

\markboth{IEEE transactions on information theory,~Vol.~, No.~, 2012}%
{Takeuchi \MakeLowercase{\textit{et al.}}:Performance Improvement of Iterative 
Multiuser Detection for Large Sparse CDMA by Spatial Coupling}
%

\IEEEpubid{0000--0000/00\$00.00~\copyright~2012 IEEE}


\maketitle

\begin{abstract}
Kudekar et al.\ proved that the belief-propagation (BP) performance for 
low-density parity check (LDPC) codes can be boosted up to the 
maximum-a-posteriori (MAP) performance by spatial coupling. In this paper, 
spatial coupling is applied to sparsely-spread code-division multiple-access 
(CDMA) systems to improve the performance of iterative multiuser detection 
based on BP. Two iterative receivers based on BP are considered: One receiver 
is based on exact BP and the other on an approximate BP with Gaussian 
approximation. The performance of the two BP receivers is evaluated via 
density evolution (DE) in the {\em dense} limit after taking the large-system 
limit, in which the number of users and the spreading factor tend to infinity 
while their ratio is kept constant. 
The two BP receivers are shown to achieve the same performance as 
each other in these limits. Furthermore, taking a continuum limit  
for the obtained DE equations implies that the performance of the two BP 
receivers can be improved up to the performance achieved by the 
symbol-wise MAP detection, called individually-optimal detection, via 
spatial coupling. Numerical simulations show that spatial coupling can 
provide a significant improvement in bit error rate for finite-sized systems 
especially in the region of high system loads. 
\end{abstract}

\begin{IEEEkeywords}
Code-division multiple-access (CDMA) systems, sparse spreading, 
spatial coupling, threshold saturation, iterative multiuser detection, 
belief propagation, individually-optimal (IO) detection, 
large-system analysis, density evolution, continuum limit. 
\end{IEEEkeywords}

%
\IEEEpeerreviewmaketitle

\section{Introduction}
\IEEEPARstart{C}{ode}-division multiple-access (CDMA) systems have been 
used in the air interface of third-generation (3G) mobile communication 
systems~\cite{Adachi98,Dahlman98,Ojanperae98}. In CDMA uplink, multiple 
users simultaneously communicate with one base station in the same frequency 
band. The base station is required to mitigate multiple-access interference 
(MAI) to detect the desired signal for each user. Multiuser detection 
(MUD) is a sophisticated method for mitigating MAI by utilizing the 
statistical properties of the MAI~\cite{Verdu98}. Two optimal receivers 
were proposed~\cite{Verdu98}: One optimal receiver, called 
individually-optimal (IO) receiver, performs the symbol-wise 
maximum-a-posteriori (MAP) detection. The other optimal receiver, called 
jointly-optimal (JO) receiver, is based on block-wise MAP detection. 
Since the two receivers are infeasible in terms of the computational 
complexity for practical modulation schemes, the main issue in MUD is to 
construct a suboptimal scheme that can achieve a good tradeoff between 
performance and complexity. 

As suboptimal MUD, linear receivers with low complexity, such as the 
decorrelator (DEC)~\cite{Lupas89} and the linear minimum mean-squared error 
(LMMSE) receiver~\cite{Xie90,Madhow94}, were proposed. An idealized 
assumption for analyzing the performance of MUD is random 
spreading: All spreading sequences have independent and 
identically-distributed (i.i.d.) elements, whereas pseudo-random sequences 
are used in practice. The LMMSE receiver for randomly-spread CDMA systems 
can achieve nearly optimal performance for low system load in the 
large-system limit~\cite{Tse99,Verdu99,Tanaka02,Mueller04,Guo05}, 
where the number of users $K$ and the spreading factor $N$ tend to infinity 
while the system load $\beta=K/N$ is kept constant. However, the performance 
of the LMMSE receiver degrades significantly for moderate-to-high system load, 
compared to the IO receiver. Thus, it is an important issue to construct a 
low-complexity scheme that can achieve nearly optimal performance for 
moderate-to-high system load. The precise meaning of low or high 
will be noted shortly. 

\IEEEpubidadjcol

A breakthrough is the use of iterative receivers based on belief propagation 
(BP)~\cite{Pearl88,Richardson08}. Iterative receivers are classified into 
two groups. In one group an iterative receiver 
performs iterative joint MUD and decoding~\cite{Wang99,Boutros02,Caire04}, in 
which the detector subtracts the MAI by using decisions fed back from the 
decoders. In these works, conventional non-iterative detectors were used in 
iterative MUD and decoding. In the other group an iterative receiver performs  
MUD with inner iterations~\cite{Kabashima03}, in which decisions in the 
detector are directly utilized to mitigate the MAI. 
In this paper, we focus on the latter group of iterative receivers, and  
only consider the uncoded case since coding makes no essential change in 
analysis of the latter iterative receivers. 
See \cite{Takeuchi13} for an application of methodology in this paper  
to the former group of iterative receivers. 

Kabashima~\cite{Kabashima03} has proposed a BP-based iterative receiver 
with Gaussian approximation (GA) for uncoded CDMA systems. It was shown 
that the proposed receiver achieves nearly optimal performance for moderate 
system load in spite of the low complexity. 
As a method for guaranteeing the convergence of the BP receiver, 
sparsely-spread CDMA systems, or sparse CDMA (SCDMA) systems, have been 
considered~\cite{Montanari06,Yoshida06,Raymond07,Guo08}. 
In SCDMA systems the spreading sequence used by each user is 
sparse: Only $c$ chips out of $N$ chips are non-zero, whereas the conventional 
CDMA system uses dense spreading sequences whose chips are all non-zero. 
The main advantage of SCDMA systems is that the convergence of BP-based 
receivers is guaranteed in the large-system limit with $c$ fixed. 
Montanari and Tse~\cite{Montanari06} proved that the performance of an 
iterative receiver based on exact BP is equal to the performance of the 
(soft) IO receiver for the conventional (dense) CDMA system below a critical 
system load $\beta_{\mathrm{BP}}$, which will be explained shortly, in the 
large-sparse-system limit, where the dense limit $c\to\infty$ is 
taken {\em after} the large-system limit. 
Note that the system is sparse even in the dense limit since the large-system 
limit is taken first. Furthermore, a regular ensemble for sparse spreading 
sequences has been considered in \cite{Raymond07,Guo08}. The BP receiver for 
the regular ensemble can achieve nearly optimal 
performance for moderate system load. 

\begin{figure}[t]
\begin{center}
\includegraphics[width=0.5\hsize]{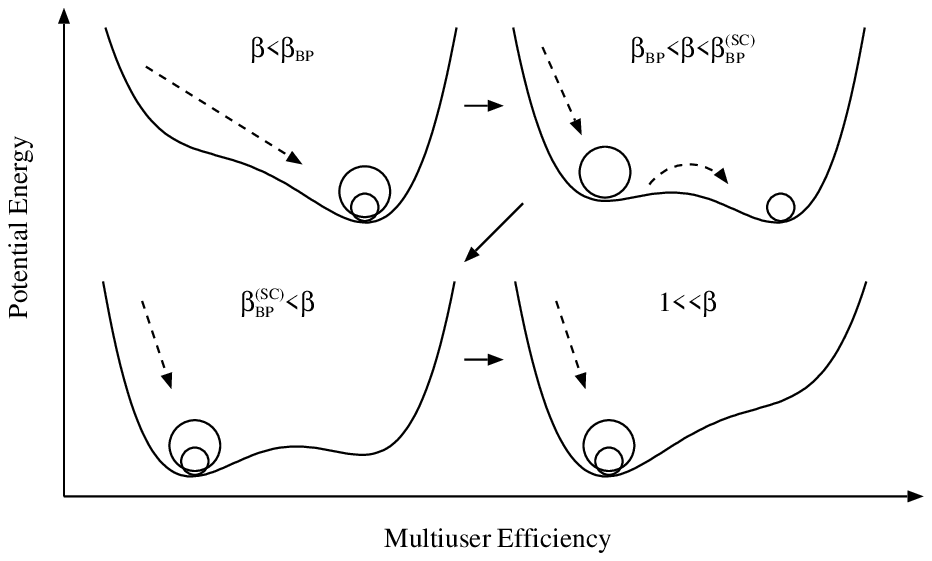}
\end{center}
\caption{
Landscape of potential energy as a function of multiuser efficiency. 
The multiuser efficiency for the BP receiver after infinite iterations 
is represented by the large (respectively (resp.) small) balls for uncoupled 
(resp.\ spatially-coupled) SCDMA systems in the large-sparse-system-limit. 
}
\label{fig1} 
\end{figure}

In order to present the aim of this paper, we shall explain 
the definition of the critical system load $\beta_{\mathrm{BP}}$, called 
the BP threshold in this paper. Let us consider the multiuser efficiency (ME), 
which is a normalized signal-to-interference ratio (SIR), as a performance 
measure of MUD. The density-evolution (DE) equations for the BP 
receiver in the large-sparse-system limit, derived in 
\cite{Montanari06,Guo08}, characterize the dynamics of the ME. The DE 
equations can be regarded as a discrete-time gradient system with a potential 
energy function. Time-evolution of the system represents the dynamics of 
the ME as the iteration of the BP receiver proceeds. As shown by the potential 
at the upper left side in Fig.~\ref{fig1}, the potential energy 
as a function of the ME has the unique stable solution for small system 
load~$\beta$. As $\beta$ increases across 
a critical system load, a metastable solution emerges at an ME value smaller 
than the original stable solution for high signal-to-noise ratio (SNR), 
that is, to the left side of the original stable solution, 
as shown in the potential landscape at the 
upper right side in Fig.~\ref{fig1}. A metastable solution means a stable solution 
at which the potential energy is minimized locally, whereas a global stable 
solution means a global minimizer of the potential energy. 
The BP threshold $\beta_{\mathrm{BP}}$ is 
defined as the supremum of $\beta_{\mathrm{th}}$ such that the BP receiver can 
achieve the ME corresponding to the rightmost\footnote{
If the potential energy is monostable, the rightmost stable solution 
corresponds to the unique stable solution. 
} stable solution after infinite iterations 
for all $\beta\in(0,\beta_{\mathrm{th}})$. 
Note that the infinite-iteration limit is considered {\em after} 
the large-sparse-system limit. The BP threshold 
$\beta_{\mathrm{BP}}$ is characterized as the bifurcation point at which the 
stability of the potential energy changes: The potential energy has one stable 
solution for $\beta<\beta_{\mathrm{BP}}$, whereas it has two stable solutions 
and one unstable solution for $\beta>\beta_{\mathrm{BP}}$. In fact, the ME 
for the BP receiver converges to the unique stable solution after infinite 
iterations for $\beta<\beta_{\mathrm{BP}}$. 
For $\beta>\beta_{\mathrm{BP}}$, on the other hand, 
the ME converges to the left stable solution after infinite iterations, since 
the initial ME is commonly a smaller value than the unstable solution 
(See the large balls in Fig.~\ref{fig1}). Thus, the BP threshold 
$\beta_{\mathrm{BP}}$ is equal to the bifurcation point between the 
monostability and the bistability. 

The aim of this paper is to construct a novel SCDMA system that improves the 
BP threshold $\beta_{\mathrm{BP}}$. For that purpose, we utilize a recent 
excellent achievement on coding theory: Kudekar et al.~\cite{Kudekar111} have 
proved that the BP threshold of a low-density parity-check (LDPC) convolutional 
code~\cite{Felstrom99} over the binary erasure channel (BEC) is equal to the 
MAP threshold of the corresponding LDPC block code (See also 
\cite{Lentmaier10}). Since an LDPC convolutional code can be regarded as a 
spatially-coupled (SC) chain of LDPC block codes, this result was  
referred to as ``threshold saturation via spatial 
coupling''~\cite{Kudekar111}. An improvement of the BP threshold via spatial 
coupling is believed to be a universal 
phenomenon~\cite{Hassani10,Hassani12,Takeuchi111,Yedla12,Kudekar12,Schlegel132}. 
The same phenomenon has been observed 
in many other problems~\cite{Kudekar102,Krzakala12,Donoho13,Hagiwara11,Kasai11,Kudekar112,Kudekar113,Rathi11,Schlegel11,Schlegel131,Takeuchi112,Uchikawa11,Yedla11,Aref11,Truhachev12}. 
In this paper, we propose spatially-coupled SCDMA (SC-SCDMA) systems to 
improve the BP threshold.  

We shall develop a simple and generic method for characterizing 
the position of the BP threshold for SC-SCDMA systems by using a potential for 
the corresponding {\em un}coupled system. In this paper, potential for the uncoupled system is 
simply referred to as potential, since potential for the coupled system is not 
considered. Recently, we have presented a phenomenological study on 
threshold improvement via spatial coupling~\cite{Takeuchi111}. The study 
allows us to specify a (probably tight) lower bound on the BP threshold 
$\beta_{\mathrm{BP}}^{(\mathrm{SC})}$ for SC-SCDMA systems via 
the shape of a potential energy function: The BP 
threshold is larger than or equal to a system load 
$\tilde{\beta}_{\mathrm{BP}}^{(\mathrm{SC})}(>\beta_{\mathrm{BP}})$, called potential 
threshold, at which the heights 
of the potential energy at the two stable solutions coincide with each other. 
The ME for $\beta<\tilde{\beta}_{\mathrm{BP}}^{(\mathrm{SC})}$ converges to the right  
stable solution after infinite iterations, whereas the ME for 
$\beta>\tilde{\beta}_{\mathrm{BP}}^{(\mathrm{SC})}$ may be trapped in the other stable 
solution (See the small balls in Fig.~\ref{fig1}). Thus, the BP receiver for 
SC-SCDMA systems can achieve better performance than that for 
the corresponding uncoupled SCDMA system in the large-sparse-system limit, 
when the system load is between 
$\beta_{\mathrm{BP}}$ and $\tilde{\beta}_{\mathrm{BP}}^{(\mathrm{SC})}$. 
The main contribution of this paper is to prove that the lower bound 
$\tilde{\beta}_{\mathrm{BP}}^{(\mathrm{SC})}$ is equal to a critical threshold 
$\beta_{\mathrm{IO}}$ for the uncoupled CDMA system, 
called the IO threshold in this paper. 

The IO threshold $\beta_{\mathrm{IO}}$ ($>\beta_{\mathrm{BP}}$) has been 
specified via the large-system analysis of the IO receiver based on the 
non-rigorous replica method~\cite{Tanaka02}. The ME of the IO receiver for 
the uncoupled CDMA system is characterized via essentially the same potential 
energy as that for determining the conventional BP threshold 
$\beta_{\mathrm{BP}}$. The IO threshold $\beta_{\mathrm{IO}}$ is defined as the 
system load at which the heights of the potential energy at the two stable 
solutions coincide with each other. 
The ME achieved by the IO receiver corresponds to the rightmost 
stable solution of the potential energy for $\beta<\beta_{\mathrm{IO}}$, 
whereas it corresponds to the left stable 
solution for $\beta>\beta_{\mathrm{IO}}$. 

In summary, the BP receiver for the 
uncoupled SCDMA system is inferior to the IO receiver for system loads between 
$\beta_{\mathrm{BP}}$ and $\beta_{\mathrm{IO}}$, since the definition 
of the BP threshold $\beta_{\mathrm{BP}}$ implies that the ME achieved by the 
BP receiver corresponds to the left stable solution of the potential 
energy for $\beta>\beta_{\mathrm{BP}}$. On the other hand, we will show that 
the potential threshold $\tilde{\beta}_{\mathrm{BP}}^{(\mathrm{SC})}$ is equal to 
the IO threshold $\beta_{\mathrm{IO}}$, by proving that the potential for 
characterizing $\tilde{\beta}_{\mathrm{BP}}^{(\mathrm{SC})}$ is essentially the same 
as for determining $\beta_{\mathrm{IO}}$. Thus, 
the BP receiver for SC-SCDMA systems can achieve the same performance as the IO 
receiver for the uncoupled CDMA system when $\beta$ is smaller than 
$\beta_{\mathrm{IO}}$. In this paper, 
small system load means that $\beta$ is smaller than one. 
We refer to SCDMA systems as moderately loaded systems if $\beta$ is between 
one and the conventional BP threshold $\beta_{\mathrm{BP}}$. High system load 
means that $\beta$ is between the conventional BP threshold 
$\beta_{\mathrm{BP}}$ and the IO threshold $\beta_{\mathrm{IO}}$.  

We have so far focused on the thresholds in the large-sparse-system limit. 
It is worth investigating what they indicate for the performance of 
finite-sized systems. The definition of the BP threshold $\beta_{\mathrm{BP}}$ 
implies that the asymptotic ME changes discontinuously at 
$\beta=\beta_{\mathrm{BP}}$ as $\beta$ grows. What does this phenomenological 
picture indicate for finite-sized systems? 
The ME for the BP receiver never changes discontinuously for finite-sized 
systems. Rather, numerical simulations in \cite{Kabashima03,Montanari06} 
implied that the ME decreases rapidly like a waterfall 
when the system load moves from below to above the BP threshold 
$\beta_{\mathrm{BP}}$. The slope of the ME as a function of $\beta$ becomes 
steep around the critical point $\beta=\beta_{\mathrm{BP}}$ as the system size 
grows. Thus, the system size required for achieving an ME close to the 
asymptotic one increases as the system load gets closer to the BP threshold 
$\beta_{\mathrm{BP}}$ from below. In other words, the performance for a fixed 
finite-sized system gets away from the asymptotic one as the system load gets 
closer to the BP threshold. These arguments may indicate that increasing the 
BP threshold results in improving the performance for a fixed finite-sized 
system. Numerical simulations will show that spatial coupling can improve 
the performance of the BP receiver for a finite-sized system especially 
in the region of high system loads. 

We would like to refer to an independent work~\cite{Schlegel11,Schlegel131}: 
Schlegel and Truhachev proposed another SC-SCDMA system based on graph lifting, 
while we consider sparse spreading~\cite{Takeuchi112}. Interestingly, 
the obtained DE equations are the same as those derived in this paper.    
They analyzed the BP threshold for the coupled case in the high SNR limit, 
whereas we investigate its position for {\em any} SNR. 
 
The rest of this paper is organized as follows: After summarizing the notation 
used in this paper, we first consider the conventional SCDMA system 
in Section~\ref{sec2}. After introducing its factor-graph representation, 
SC-SCDMA systems are defined on the basis of two operations with respect to 
the factor graph. In Section~\ref{sec3} two BP-based iterative receivers are  
derived. One receiver is based on exact BP~\cite{Montanari06,Guo08}, and 
the other receiver is a BP receiver with GA~\cite{Kabashima03}. 
Section~\ref{sec4} presents the main results of this paper. 
The main theorem on spatial coupling is proved as a general framework 
in Section~\ref{sec5}. The section is organized as an independent section, 
so that it should be possible to skip Sections~\ref{sec2}--\ref{sec4} and to 
read Section~\ref{sec5}. In Section~\ref{sec6} the performance of the SC-SCDMA 
systems is investigated numerically. Section~\ref{sec7} concludes this paper.

\subsection{Notation}
For a matrix $\boldsymbol{A}$, $\boldsymbol{A}^{\mathrm{T}}$ denotes the 
transpose of $\boldsymbol{A}$. $\boldsymbol{I}_{N}$ stands for the 
$N\times N$ identity matrix. The Kronecker delta is denoted by $\delta_{i,j}$. 
For a natural number $L$ and an integer $l$, the remainder $(l)_{L}=l\mod L$ 
for the division of $l$ by $L$ is equal to $l+kL$ for 
an integer $k$ such that $0\leq l+kL\leq L-1$. 

For a random variable $x$, $\mathbb{E}[x]$ and $\mathbb{V}[x]$ denote the 
mean and variance of $x$, respectively. The notation 
$p(x)$ stands for the probability density function (pdf) of a continuous 
random variable $x$. We use the same notation $p(x)$ for the probability mass 
function (pmf) of a discrete random variable $x$. The notation  
$x\sim p(x)$ indicates that the pdf or pmf of a random variable~$x$ is equal  
to $p(x)$. The real Gaussian pdf with a mean vector~$\boldsymbol{m}$ and a 
covariance matrix~$\boldsymbol{\Sigma}$ is denoted by 
$\mathcal{N}(\boldsymbol{m},\boldsymbol{\Sigma})$. In particular, the pdf 
for a zero-mean Gaussian random variable $x$ with variance $\sigma^{2}$ is 
written as  
\begin{equation} \label{Gauss_function} 
g(x;\sigma^{2}) = \frac{1}{\sqrt{2\pi\sigma^{2}}}\exp\left(
 -\frac{x^{2}}{2\sigma^{2}} 
\right). 
\end{equation}
For variables $\{a_{k}\in\mathcal{M}\}$ on a finite alphabet $\mathcal{M}$, 
the sum $\sum_{\{a_{k}\}}f(\{a_{k}\})$ denotes the marginalization with respect 
to $\{a_{k}\}$---the summation of a function $f(\{a_{k}\})$ over all possible 
configurations of $\{a_{k}\}$. 
Furthermore, the sum $\sum_{\backslash a_{k}}$ stands for the summation 
over all possible configurations of $\{a_{k'}\}$ except for $a_{k}$.  
For a conditional pdf or pmf $p(x|y)$, $p(x|y)\propto f(x)$ means that 
$p(x|y)$ is proportional to $f(x)$, i.e.\ there is an $x$-independent 
constant $C(y)$ such that $p(x|y)=C(y)f(x)$. 

Graphs with nodes specified by two indices $i$ and $j$ are considered for 
SC systems. Thus, the pair $(i, j)$ represents not an edge but a node for the 
SC systems. Furthermore, $\partial(i,j)$ stands for the neighborhood of the 
node~$(i,j)$, i.e.\ the set of nodes that are directly connected to the 
node~$(i,j)$.  

\section{System Model} \label{sec2} 
\subsection{Sparsely-Spread CDMA Systems} 
We introduce conventional synchronous SCDMA systems before presenting 
SC-SCDMA systems. 
In this paper, the receive power is assumed to be identical for all users. 
Let $K$ and $N$ denote the number of users and the spreading factor, 
respectively. Without loss of generality, we focus 
on one symbol period. User~$k$ sends the product of the unbiased 
binary phase shift keying (BPSK) data 
symbol $b_{k}\in\{-1,1\}$ and a sparse spreading sequence 
$\boldsymbol{s}_{k}=(s_{1,k},\ldots,s_{N,k})^{\mathrm{T}}$ with 
$\bar{c}_{k}=\mathbb{E}[\|\boldsymbol{s}_{k}\|^{2}]$, in which the 
statistics of $\{\boldsymbol{s}_{k}\}$ will be defined shortly. 
Under the assumption of unfaded channels, the received 
vector~$\boldsymbol{y}=(y_{1},\ldots,y_{N})^{\mathrm{T}}\in\mathbb{R}^{N}$ is 
given by 
\begin{equation} \label{SCDMA} 
\boldsymbol{y} = \sum_{k\in\mathcal{K}}\frac{1}{\sqrt{\bar{c}_{k}}}
\boldsymbol{s}_{k}b_{k} + \boldsymbol{w},  
\end{equation}
with $\mathcal{K}=\{1,\ldots,K\}$. 
In (\ref{SCDMA}), the $N$-dimensional vector 
$\boldsymbol{w}\sim\mathcal{N}(0,\sigma_{\mathrm{n}}^{2}\boldsymbol{I}_{N})$ denotes additive 
white Gaussian noise (AWGN) with variance~$\sigma_{\mathrm{n}}^{2}$. 
The expression~(\ref{SCDMA}) can be re-written as 
\begin{equation}
\boldsymbol{y} = \boldsymbol{S}\boldsymbol{b} + \boldsymbol{w}, 
\end{equation}
with $\boldsymbol{S}=(\bar{c}_{1}^{-1/2}\boldsymbol{s}_{1},\ldots,
\bar{c}_{K}^{-1/2}\boldsymbol{s}_{K})$ and 
$\boldsymbol{b}=(b_{1},\ldots,b_{K})^{\mathrm{T}}$. 

The conventional CDMA systems use dense spreading sequences whose elements 
are all non-zero. In the SCDMA system~(\ref{SCDMA}), on the other hand,  
user~$k$ utilizes the sparse spreading sequence $\boldsymbol{s}_{k}$ with 
$c_{k}$ ($\ll N$) non-zero elements. The number $c_{k}$ is equal to the weight  
(number of non-zero elements) of the $k$th column of the spreading 
matrix $\boldsymbol{S}$. For simplicity, we assume sparse spreading with 
binary antipodal chips as non-zero chips: 
Non-zero elements of $\boldsymbol{s}_{k}$ take $\pm 1$ 
with equal probability. Then, the normalization constant 
$\bar{c}_{k}=\mathbb{E}[\|\boldsymbol{s}_{k}\|^{2}]$ is equal to the 
average of the $k$th column weight of $\boldsymbol{S}$. 
Let $r_{n}$ denote the $n$th row 
weight for the spreading matrix $\boldsymbol{S}$. 
A constraint with respect to the number of non-zero elements imposes 
$\sum_{k=1}^{K}c_{k}=\sum_{n=1}^{N}r_{n}$. In this paper, we only 
consider regular and quasi-regular ensembles of the spreading matrix. 

\begin{example}[Regular Ensemble]
In the $(c, r)$-regular ensemble of $\boldsymbol{S}$, 
all column weights $\{c_{k}\}$ and all row weights $\{r_{n}\}$ are equal to 
$c$ and $r$, respectively. 
The constraint $\sum_{k=1}^{K}c_{k}=\sum_{n=1}^{N}r_{n}$ implies that 
$K$ and $N$ must satisfy the constraint on the system load $\beta=K/N=r/c$. 
One regular spreading matrix is obtained as follows: 
\begin{enumerate}
\item Pick up a matrix from all possible binary ($0$ or $1$) matrices with 
row weight~$r$ and column weight~$c$ uniformly and randomly. 
\item Replace each non-zero element of the obtained binary matrix by 
$\pm c^{-1/2}$ independently and with equal probability.  
\end{enumerate}
The $(c, r)$-regular ensemble of the spreading matrix is composed of 
all possible spreading matrices obtained in the above-mentioned manner.  
\end{example}

The $(c, r)$-regular ensemble is well-defined when $K$ and $N$ satisfy 
the constraint on the system load $\beta=r/c$. The following $r$-quasi-regular 
ensemble is well-defined for any $K$ and $N$. 

\begin{example}[Quasi-Regular Ensemble] \label{example2} 
In the $r$-quasi-regular ensemble, all row weights $\{r_{k}\}$ are fixed to 
$r$. Let $N_{\mathrm{w}}=rN$ denote the number of non-zero elements 
in $\boldsymbol{S}$. 
The column vectors of $\boldsymbol{S}$ are classified into two groups with 
column weights~$c=\lfloor N_{\mathrm{w}}/K\rfloor$ and $c+1$: 
One group consists of $(N_{\mathrm{w}}-cK)$ column vectors with weight~$c+1$. 
The other group is composed of the remaining column vectors with weight~$c$. 
The average column weight $\bar{c}$ is given by $\bar{c}=r/\beta$.  
It is straightforward to confirm that the constraint 
$N_{\mathrm{w}}=\sum_{k=1}^{K}c_{k}$ is satisfied. 
One spreading matrix is obtained as follows: 
A binary matrix is uniformly and randomly picked up from all possible binary 
matrices satisfying the conditions above. Subsequently, 
the gain $\pm\bar{c}^{-1/2}$ is associated with each non-zero element of the 
obtained binary matrix independently and with equal probability. 
The $r$-quasi-regular ensemble of the spreading matrix consists of all 
possible spreading matrices obtained in this manner.  
\end{example}
Obviously, the $r$-quasi-regular ensemble reduces 
to the $(rN/K, r)$-regular ensemble when $N_{\mathrm{w}}=r N$ is a multiple of 
$K$. Thus, we only consider the quasi-regular ensemble in this paper. 

\subsection{Factor Graph Representation} 
\begin{figure}[t]
\begin{center}
\includegraphics[width=0.5\hsize]{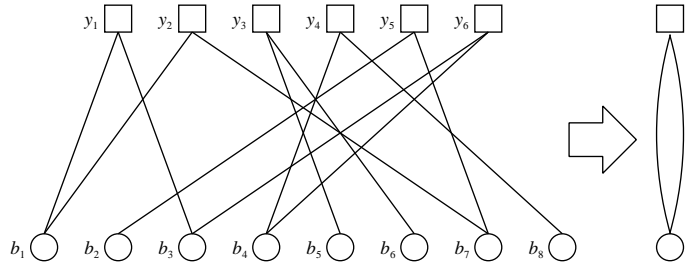}
\end{center}
\caption{
An example of the factor graph picked up from the $2$-quasi-regular ensemble 
for $K=8$ and $N=6$ (left). The graph on the right represents a simplified 
graph representation for the same ensemble. 
}
\label{fig2} 
\end{figure}

We next introduce the factor-graph representation~\cite{Richardson08} for the 
SCDMA system~(\ref{SCDMA}), shown in Fig.~\ref{fig2}. 
Each data symbol $b_{k}$ corresponds to a variable node represented by a 
circle, whereas each received signal $y_{n}$ is associated with a 
function node shown by a square. If the $n$th chip $s_{n,k}$ for user~$k$ 
is non-zero, there exists an edge connecting function node~$n$ and variable 
node~$k$ in the factor graph. Furthermore, the edge is associated with the 
corresponding gain $\bar{c}_{k}^{-1/2}s_{n,k}$.   
Let $\partial k\subset\mathcal{N}=\{1,\ldots,N\}$ denote the neighborhood of 
variable node~$k$, i.e.\ the set of function nodes that are directly connected 
to variable node~$k$. The degree $|\partial k|$ of variable 
node~$k$ corresponds to the $k$th column weight $c_{k}$ of the spreading 
matrix $\boldsymbol{S}$. Similarly, let $\partial n\subset
\mathcal{K}$ denote the neighborhood of function node~$n$. 
The degree $|\partial n|$ of function node~$n$ is equal to the $n$th row 
weight~$r_{n}$ of the spreading matrix. 

There is a one-to-one correspondence between a spreading matrix and 
a factor graph. Thus, we can consider an ensemble of factor graphs 
corresponding to the $r$-quasi-regular ensemble of spreading matrices. This 
ensemble is referred to as the $r$-quasi-regular ensemble of factor graphs 
in this paper, or simply as the $r$-quasi-regular ensemble if it is obvious 
that the ensemble is an ensemble of factor graphs. 

The crucial property of factor graphs picked up from the $r$-quasi-regular 
ensemble is the asymptotic cycle-free (ACF) property in the large-system 
limit, where $K$ and $N$ tend to infinity while the system load $\beta=K/N$ 
is kept constant. The length of a cycle is defined as the number of edges 
included in the cycle. A factor graph picked up from the $r$-quasi-regular 
ensemble has no cycles with {\em finite} length with probability one in the 
large-system limit~\cite{Richardson08,Montanari06,Guo08}. This ACF property 
guarantees the convergence of iterative detection based on BP in the 
large-system limit. 

\subsection{Spatial Coupling} \label{sec2_3}
In this section, we present an intuitive explanation for 
spatial coupling as a brief introduction, instead of presenting a precise 
definition of SC-SCDMA systems. The precise definition will be presented in 
the next section. We use a simplified 
graph representation for the $r$-quasi-regular ensemble 
(See the graph on the right in Fig.~\ref{fig2}). 
The graph consists of one function node represented by a square and of one 
variable node shown by a circle. The number of edges is equal to the 
degree~$r$ of the function nodes in the original factor graph. In other words, 
the simplified graph only represents the fact that the degree of the function 
nodes is equal to $r$. 
In introducing SC-SCDMA systems, multiple simplified graphs are used. Thus, 
we refer to each simplified graph as a subgraph. 
An SC-SCDMA system with coupling width~$W$ and the number of 
subgraphs~$L$ is constructed from two operations with respect to the 
simplified graph representation. 

\begin{figure}[t]
\begin{center}
\includegraphics[width=0.5\hsize]{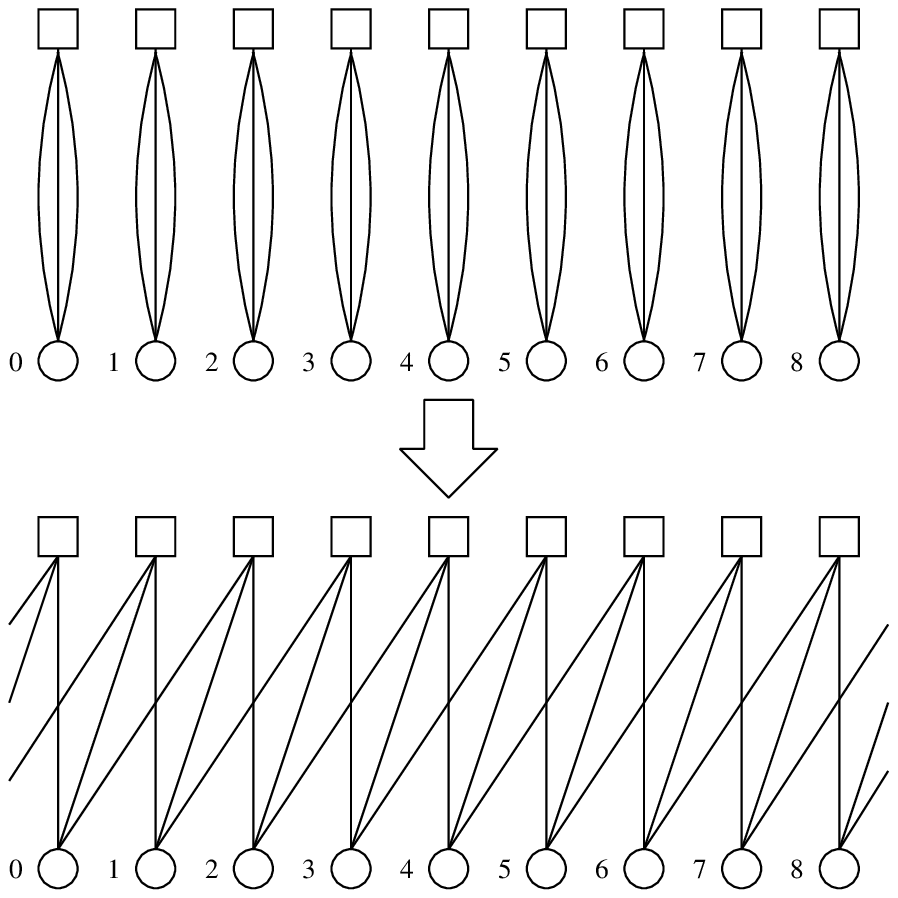}
\end{center}
\caption{
Spatial coupling for $r=3$, $L=9$, and $W=2$. 
}
\label{fig3} 
\end{figure}

\begin{enumerate}
\item Generate $L$ uncoupled subgraphs. Different subgraphs may 
have different spreading factors. 
\item \label{step2}
Disconnect $Wr/(W+1)$ edges out of $r$ edges from the variable 
node at position~$l$ for $l=0,\ldots,L-1$. Subsequently, reconnect $r/(W+1)$ 
edges out of the disconnected edges to the variable nodes  
at position~$(\tilde{l}')_{L}$ for $\tilde{l}'=l-W,\ldots,l-1$ 
(See Fig.~\ref{fig3}). 
The subgraphs are connected circularly. 
\end{enumerate}

The degree of the function nodes~$r$ is restricted to a multiple of $W+1$. 
In Step~\ref{step2}, $L$ subgraphs have been coupled circularly, whereas 
Kudekar et al.~\cite{Kudekar111} considered termination at both ends. 
The point of spatial coupling is that the data symbols at both ends are known 
to the receiver or can be detected well. Reliable information about the 
data symbols at both ends is expected to spread over the whole system by 
spatial coupling. 

In order to allow the receiver to detect the data symbols at both ends, 
we reduce the system loads for positions~$l=0,\ldots,W-1$. It would be 
possible to send known symbols in these positions instead. This scheme is 
equivalent to assuming noiseless channels with zero system load 
for the positions. We use 
small system load to reduce the influence of rate loss in the case of 
finite $L$. In the next section, we present the detailed definition of 
SC-SCDMA systems. 

\subsection{Spatially-Coupled SCDMA Systems} 
MUD for SC-SCDMA systems is performed for every $L$ symbol 
periods\footnote{{\em Temporal} coupling may be an appropriate naming, rather 
than spatial coupling. Nonetheless, we follow \cite{Kudekar111} to use 
the term ``spatial coupling'' in this paper.}, 
whereas it is done for every symbol period in the conventional SCDMA 
system~(\ref{SCDMA}). This implies that the detection delay increases 
linearly in $L$ for SC-SCDMA systems. In practice, the detection delay 
does not necessarily result in the overall delay for coded systems: If $L$ is 
smaller than the code length, the overall delay is dominated by the delay 
due to decoding. 

Let $N_{l}$ denote the spreading factor in symbol period~$l$. The received 
vector $\boldsymbol{y}_{l}=(y_{1,l},\ldots,y_{N_{l},l})^{\mathrm{T}}
\in\mathbb{R}^{N_{l}}$ in symbol period~$l$ is given by 
\begin{equation} \label{SC_SCDMA}
\boldsymbol{y}_{l} = \sum_{k\in\mathcal{K}}\sum_{l'\in\mathcal{L}}
\frac{1}{\sqrt{\bar{c}_{l,k,l'}}}h_{(l-l')_{L}}\boldsymbol{s}_{l,k,l'}
b_{k,l'} + \boldsymbol{w}_{l}, 
\end{equation}
for $l\in\mathcal{L}=\{0,\ldots,L-1\}$, with 
\begin{equation}
h_{l}=\left\{
 \begin{array}{ll}
 1/\sqrt{W+1} & \hbox{for $l=0,\ldots,W$} \\ 
 0 & \hbox{for $l=W+1,\ldots,L-1$.} 
 \end{array}
\right.
\end{equation}
In (\ref{SC_SCDMA}), the vector $\boldsymbol{w}_{l}\sim
\mathcal{N}(\boldsymbol{0},\sigma_{\mathrm{n}}^{2}\boldsymbol{I}_{N_{l}})$ 
denotes the AWGN 
vector with variance~$\sigma_{\mathrm{n}}^{2}$. The $N_{l}$-dimensional vector 
$\boldsymbol{s}_{l,k,l'}=(s_{1,l,k,l'},\ldots,s_{N_{l},l,k,l'})^{\mathrm{T}}$ 
represents the $l'$th sparse spreading sequence of user~$k$ for symbol 
period~$l$, which will be defined shortly. The normalization constant 
$\bar{c}_{l,k,l'}$ is given by $\bar{c}_{l,k,l'}=
\mathbb{E}[\|\boldsymbol{s}_{l,k,l'}\|^{2}]$. 
Finally, $b_{k,l}\in\{-1,1\}$ denotes the $l$th BPSK 
data symbol for user~$k$. Let 
$\boldsymbol{b}_{l}=(b_{1,l},\ldots,b_{K,l})^{\mathrm{T}}$ and 
$\boldsymbol{S}_{l,l'}=(\bar{c}_{l,1,l'}^{-1/2}\boldsymbol{s}_{l,1,l'},\ldots,
\bar{c}_{l,K,l'}^{-1/2}\boldsymbol{s}_{l,K,l'})$. 
The system~(\ref{SC_SCDMA}) can be represented as 
\begin{equation} \label{large_SC_SCDMA}
\begin{bmatrix}
\boldsymbol{y}_{0} \\
\vdots \\
\boldsymbol{y}_{L-1} 
\end{bmatrix}
= \frac{1}{\sqrt{W+1}}\boldsymbol{G}\begin{bmatrix}
\boldsymbol{b}_{0} \\
\vdots \\
\boldsymbol{b}_{L-1} 
\end{bmatrix}
+ \begin{bmatrix}
\boldsymbol{w}_{0} \\
\vdots \\
\boldsymbol{w}_{L-1} 
\end{bmatrix}, 
\end{equation}
with 
\begin{equation} \label{large_spreading_matrix} 
\boldsymbol{G}=
\begin{bmatrix}
\boldsymbol{S}_{0,0} & & & \boldsymbol{S}_{0,L-W} & \cdots & 
\boldsymbol{S}_{0,L-1} \\ 
\vdots & \ddots & &  & \ddots &\vdots \\
\vdots & & \ddots & &  & \boldsymbol{S}_{W-1,L-1} \\ 
\boldsymbol{S}_{W,0} &  & & \ddots & &  \\ 
 & \ddots & & & \ddots & \\ 
& & \boldsymbol{S}_{L-1,L-W-1} & \cdots & \cdots & \boldsymbol{S}_{L-1,L-1} 
\end{bmatrix}. 
\end{equation}
It is straightforward to confirm that the simplified graph representation in 
Fig.~\ref{fig3} corresponds to the one for the SC-SCDMA system with $r=3$, 
$L=9$, and $W=2$. 

We reduce the system loads for positions~$l=0,\ldots,W-1$.  
This corresponds to the situation under which 
the SC-SCDMA system~(\ref{SC_SCDMA}) has two phases: 
initialization and communication phases. The spreading factors $N_{l}$ for the 
initialization phase $l=0,\ldots,W-1$ are fixed to a large 
value~$N_{\mathrm{init}}$ to allow the receiver to detect the data symbols 
transmitted in the phase. On the other hand, the spreading factors $N_{l}$ for 
the communication phase $l=W,\ldots,L-1$ are set to a small value $N$ to 
increase the sum rate. 

The average system load $\bar{\beta}$ of the SC-SCDMA system is given by 
\begin{IEEEeqnarray}{rl}
\bar{\beta} &= \frac{KL}{N_{\mathrm{init}}W + N(L-W)} \nonumber \\ 
&= \frac{1}{\beta_{\mathrm{init}}^{-1}(W/L) + \beta^{-1}\{1-(W/L)\}}, 
\label{sum_rate} 
\end{IEEEeqnarray}
where $\beta_{\mathrm{init}}=K/N_{\mathrm{init}}$ and $\beta=K/N$ denote the 
system loads for the initialization and communication phases, respectively. 
The average system load~(\ref{sum_rate}) converges to the system load~$\beta$ 
for the conventional SCDMA system~(\ref{SCDMA}) when $\gamma=W/L$
tends to zero. 

We only consider the quasi-regular ensemble with spatial coupling. 
Throughout this paper, the matrix~(\ref{large_spreading_matrix}) is assumed 
to be drawn from the $(r, L, W)$-quasi-regular ensemble below. 

\begin{example}[Quasi-Regular Ensemble with Spatial Coupling] \label{example3} 
In the $(r, L, W)$-quasi-regular ensemble with spatial coupling, all row 
weights of the matrix~(\ref{large_spreading_matrix}) are fixed to $r$. 
Each non-zero submatrix $\boldsymbol{S}_{l,l'}$ has the row weight~$r/(W+1)$ 
for all rows. Thus, the row weight $r$ of (\ref{large_spreading_matrix}) must 
be a multiple of $W+1$. Each submatrix $\boldsymbol{S}_{l,l'}$ is a member 
of the $r/(W+1)$-quasi-regular ensemble with the number of users~$K$ 
and the spreading factor~$N_{l}$, presented in Example~\ref{example2}. 
The average column weights $\{\bar{c}_{l,k,l'}\}$ are equal to 
$\bar{c}_{l,k,l'}=r/\{(W+1)\beta_{\mathrm{init}}\}$ for the initialization 
phase $l=0,\ldots,W-1$ and 
$\bar{c}_{l,k,l'}=r/\{(W+1)\beta\}$ for the communication phase 
$l=W,\ldots,L-1$. The $(r,L,W)$-ensemble consists of all possible 
matrices~(\ref{large_spreading_matrix}) that satisfy the conditions above. 
It is guaranteed that the $(r,L,W)$-ensemble of factor graphs corresponding 
to the $(r,L,W)$-ensemble of (\ref{large_spreading_matrix}) has the ACF 
property in the large-system limit, although its proof is omitted. 
\end{example}

\section{Iterative Receivers} \label{sec3} 
\subsection{Belief Propagation} 
The goal of the receiver is to compute the marginal posterior probability 
for each data symbol~$b_{k,l'}$, given by 
\begin{equation} \label{marginal_posterior} 
p(b_{k,l'}|\mathcal{Y},\boldsymbol{G}) = 
\sum_{\backslash b_{k,l'}}
p(\mathcal{B} | \mathcal{Y},\boldsymbol{G}), 
\end{equation}
where the joint posterior probability 
$p(\mathcal{B} | \mathcal{Y},\boldsymbol{G})$ of 
$\mathcal{B}=\{\boldsymbol{b}_{l'}:l'\in\mathcal{L}\}$ given  
$\mathcal{Y}=\{\boldsymbol{y}_{l}:l\in\mathcal{L}\}$ and $\boldsymbol{G}$ 
is defined as 
\begin{equation} \label{joint_posterior} 
p(\mathcal{B} | \mathcal{Y},\boldsymbol{G}) 
= \frac{
 p(\mathcal{Y}|\boldsymbol{G},\mathcal{B})
 p(\mathcal{B})
}
{
 p(\mathcal{Y}|\boldsymbol{G}) 
}, 
\end{equation}
with 
\begin{equation}
p(\mathcal{Y}|\boldsymbol{G}) 
= \sum_{\mathcal{B}}p(\mathcal{Y}|\boldsymbol{G},\mathcal{B})p(\mathcal{B}). 
\end{equation}
In (\ref{joint_posterior}), the conditional pdf 
$p(\mathcal{Y}|\boldsymbol{G},\mathcal{B})$ 
represents the SC-SCDMA system~(\ref{large_SC_SCDMA}). 
It is well-known that the IO decision $\hat{b}_{k,l'}^{(\mathrm{IO})}=
\mathrm{argmax}_{b_{k,l'}\in\{-1,1\}}
p(b_{k,l'}|\mathcal{Y},\boldsymbol{G})$ 
and the soft IO decision $\hat{b}_{k,l'}^{(\mathrm{SIO})}=
\sum_{b_{k,l'}\in\{-1,1\}}b_{k,l'}p(b_{k,l'}|\mathcal{Y},\boldsymbol{G})$ 
minimize the bit error rate (BER) and the mean-squared error (MSE), 
respectively~\cite{Verdu98}. 

BP is an iterative algorithm for computing the marginal posterior 
probability~(\ref{marginal_posterior}) efficiently. Messages are exchanged 
between the variable and function nodes on the factor graph for the 
SC-SCDMA system~(\ref{SC_SCDMA}). Let $q_{n,l}^{(i)}(b_{k,l'})$ 
(resp.\ $m_{n,l}^{(i)}(b_{k,l'})$) denote the message passed 
from the function node $y_{n,l}$ (resp.\ variable node $b_{k,l'}$) to the 
variable node $b_{k,l'}$ (resp.\ function node $y_{n,l}$) in iteration~$i$. 
A tentative marginal posterior probability of $b_{k,l'}$ in 
iteration~$i$ is given by the product of all incoming messages to 
the variable node~$b_{k,l'}$,  
\begin{equation} \label{BP} 
p^{(i)}(b_{k,l'}|\mathcal{Y},\boldsymbol{G}) 
\propto\prod_{(\tilde{n},\tilde{l})\in\partial(k,l')}
q_{\tilde{n},\tilde{l}}^{(i)}(b_{k,l'}), 
\end{equation} 
where the set of index pairs 
$\partial(k,l')\subset\mathcal{N}_{\mathrm{max}}\times\mathcal{L}$, 
with $\mathcal{N}_{\mathrm{max}}=\{1,\ldots,\max_{l}N_{l}\}$, denotes the 
neighborhood of the variable node $b_{k,l'}$. 
The BP decision is defined as the hard (or soft) decision based on the 
marginal posterior probability~(\ref{BP}) in each iteration.  

The messages $q_{n,l}^{(i)}(b_{k,l'})$ and $m_{n,l}^{(i)}(b_{k,l'})$ 
are updated as follows: 
\begin{IEEEeqnarray}{rl} 
q_{n,l}^{(i)}(b_{k,l'})=& 
\sum_{\backslash b_{k,l'}}p\left(
 y_{n,l}\left|
  \sum_{(\tilde{k},\tilde{l}')\in\partial(n,l)}\frac{s_{n,l,\tilde{k},\tilde{l}'}
  b_{\tilde{k},\tilde{l}'}}{\sqrt{(W+1)\bar{c}_{l,\tilde{k},\tilde{l}'}}}
 \right.
\right) \nonumber \\ 
&\cdot\prod_{(\tilde{k},\tilde{l}')\in\partial(n,l)\backslash(k,l')}
m_{n,l}^{(i-1)}(b_{\tilde{k},\tilde{l}'}), \label{sum_step} 
\end{IEEEeqnarray}  
\begin{equation} \label{product_step} 
m_{n,l}^{(i)}(b_{k,l'})= \alpha_{n,l,k,l'}^{(i)}
\prod_{(\tilde{n},\tilde{l})\in\partial(k,l')
\backslash(n,l)}q_{\tilde{n},\tilde{l}}^{(i)}(b_{k,l'}), 
\end{equation}
with the initial values $m_{n,l}^{(0)}(b_{k,l'})=1/2$. 
In (\ref{sum_step}), the set of index pairs 
$\partial(n,l)\subset\mathcal{K}\times\mathcal{L}$ denotes the neighborhood 
of the function node $y_{n,l}$. Furthermore, the conditional pdf 
in (\ref{sum_step}) represents the SC-SCDMA system~(\ref{SC_SCDMA}) for the 
$n$th received signal in symbol period~$l$. In (\ref{product_step}), 
$\alpha_{n,l,k,l'}^{(i)}$ denotes the normalization constant. 
We hereafter refer to the update rules (\ref{sum_step}) and 
(\ref{product_step}) as the sum and product steps, respectively. 

It is known that BP computes the exact marginal posterior probabilities if 
there are no cycles in the factor graph~\cite{Pearl88}. When the factor 
graphs have cycles, the convergence of BP is not guaranteed in general. 
Even if BP has converged, the computed marginal posterior 
probabilities~(\ref{BP}) are approximate. Fortunately, the $(r,L,W)$-ensemble 
of factor graphs has the ACF property in the large-system limit: There are no 
cycles with finite length in the large-system limit. Thus, the BP  
receiver~(\ref{BP}) is guaranteed to converge in the infinite-iteration 
limit $i\to\infty$ {\em after} taking the large-system limit. 
Note that the two limits do not commute with each other. 

\subsection{Gaussian Approximation} \label{sec_GA} 
The computational complexity of the BP receiver is exponential 
in the row weight $r$, whereas the complexity is linear in $K$ and $L$.  
Consequently, a large row weight~$r$ cannot be used in terms of the 
complexity. 
We derive an approximate BP-based iterative receiver that works efficiently 
for large $r$, following \cite{Kabashima03}. The update rule~(\ref{sum_step}) 
can be regarded as a marginalization with respect to the independent 
variables $b_{\tilde{k},\tilde{l}'}\sim m_{n,l}^{(i)}(b_{\tilde{k},\tilde{l}'})$ 
for all $(\tilde{k},\tilde{l}')\in\partial(n,l)\backslash(k,l')$. 
Let us define the postulated interference to the data symbol $b_{k,l'}$ 
in the function node~$y_{n,l}$ as 
\begin{equation} \label{postulated_interference} 
\tilde{I}_{n,l,k,l'}^{(i)}=
\sum_{(\tilde{k},\tilde{l}')\in\partial(n,l)\backslash(k,l')}
\frac{s_{n,l,\tilde{k},\tilde{l}'}\tilde{b}_{\tilde{k},\tilde{l}'}^{(i)}}
{\sqrt{(W+1)\bar{c}_{l,\tilde{k},\tilde{l}'}}},  
\end{equation} 
where $\tilde{b}_{\tilde{k},\tilde{l}'}^{(i)}
\sim m_{n,l}^{(i)}(b_{\tilde{k},\tilde{l}'})$. 
The central limit theorem implies that the 
interference~(\ref{postulated_interference}) converges in law to a Gaussian 
random variable in the limit $|\partial(n,l)|=r\to\infty$. The 
mean $\tilde{\mu}_{n,l,k,l'}^{(i)}$ and variance $\tilde{v}_{n,l,k,l'}^{(i)}$ 
of (\ref{postulated_interference}) are given by 
\begin{equation} \label{mean}
\tilde{\mu}_{n,l,k,l'}^{(i)} 
= \sum_{(\tilde{k},\tilde{l}')\in\partial(n,l)\backslash(k,l')}
\frac{s_{n,l,\tilde{k},\tilde{l}'}
\mathbb{E}[\tilde{b}_{\tilde{k},\tilde{l}'}^{(i)}]}
{\sqrt{(W+1)\bar{c}_{l,\tilde{k},\tilde{l}'}}}, 
\end{equation}
\begin{equation} \label{variance} 
\tilde{v}_{n,l,k,l'}^{(i)} 
= \sum_{(\tilde{k},\tilde{l}')\in\partial(n,l)\backslash(k,l')}
\frac{s_{n,l,\tilde{k},\tilde{l}'}^{2}\mathbb{E}[
 (\tilde{b}_{\tilde{k},\tilde{l}'}^{(i)}
 -\mathbb{E}[\tilde{b}_{\tilde{k},\tilde{l}'}^{(i)}])^{2}]}
{(W+1)\bar{c}_{l,\tilde{k},\tilde{l}'}}, 
\end{equation}
respectively. 
We use the GA of (\ref{postulated_interference}) to 
approximate the update rule~(\ref{sum_step}) by  
\begin{IEEEeqnarray}{rl} 
& q_{n,l}^{(i)}(b_{k,l'}) \nonumber \\ 
=& \mathbb{E}_{\tilde{I}_{n,l,k,l'}^{(i-1)}}\left[
 p\left(
  y_{n,l}\left|
   \frac{s_{n,l,k,l'}b_{k,l'}}{\sqrt{(W+1)\bar{c}_{l,k,l'}}} 
   + \tilde{I}_{n,l,k,l'}^{(i-1)} 
  \right. 
 \right) 
\right] \nonumber \\ 
\approx& 
g\left(
 y_{n,l} - \frac{s_{n,l,k,l'}b_{k,l'}}{\sqrt{(W+1)\bar{c}_{l,k,l'}}} 
 - \tilde{\mu}_{n,l,k,l'}^{(i-1)};\tilde{v}_{n,l,k,l'}^{(i-1)}+\sigma_{\mathrm{n}}^{2}
\right), \label{GA} 
\end{IEEEeqnarray}
where $g(x;\sigma^{2})$ denotes the pdf~(\ref{Gauss_function}) for a 
zero-mean Gaussian random variable with variance~$\sigma^{2}$. 
The complexity of the BP receiver with GA~(\ref{GA}) is 
linear in the row weight~$r$, as well as in $K$ and $L$. 

\section{Main Results} \label{sec4} 
\subsection{Density Evolution Analysis} 
The asymptotic properties of the BP receiver~(\ref{BP}) are analyzed in this 
section. Let us define the equivalent channel between $b_{k,l'}$ and the 
corresponding output in iteration~$i$, denoted by $b_{k,l'}^{(i)}$, as 
\begin{equation} \label{equivalent_channel}  
p(b_{k,l'}^{(i)}|b_{k,l'}) 
= \overline{
 \int p^{(i)}(b_{k,l'}=b_{k,l'}^{(i)}|\mathcal{Y},\boldsymbol{G})
 p(\mathcal{Y}|\boldsymbol{G},b_{k,l'})d\mathcal{Y}
}, 
\end{equation}
where the overline represents the expectation with respect to 
$\boldsymbol{G}$. The average BER and SIR can be calculated from the equivalent 
channel~(\ref{equivalent_channel}). 

We consider five limits: the large-system limit, the dense limit 
$r\to\infty$, the continuum limit~$L, W\to\infty$ with $\gamma=W/L$ kept 
constant, the infinite-iteration limit $i\to\infty$, and $\gamma\to0$. 
We first present the main 
result in the first two limits, i.e.\ in the large-sparse-system limit where 
the dense limit~$r\to\infty$ is taken {\em after} the large-system 
limit. The main result is that the equivalent 
channel~(\ref{equivalent_channel}) converges to the one for a scalar AWGN 
channel in the large-sparse-system limit. 
The remaining three limits will be investigated in the next subsection.  

We first introduce the equivalent AWGN channel for iteration~$i$, 
\begin{equation} \label{AWGN} 
z_{k,l'}^{(i)} = b_{k,l'} + w_{k,l'}^{(i)}, 
\end{equation}
with $w_{k,l'}\sim\mathcal{N}(0,(\mathrm{sir}_{l'}^{(i)})^{-1})$, in which 
$\mathrm{sir}_{l'}^{(i)}$ will be defined shortly.  
Let $\langle b_{k,l'} \rangle_{i}$ denote the posterior mean estimator of 
$b_{k,l'}$ in iteration~$i$, 
\begin{equation} \label{posterior_mean_estimator} 
\langle b_{k,l'} \rangle_{i} = 
\sum_{b_{k,l'}=\pm1}b_{k,l'}p(b_{k,l'}|z_{k,l'}^{(i)}), 
\end{equation}
where the posterior probability $p(b_{k,l'}|z_{k,l'}^{(i)})$ in 
iteration~$i$ is defined as  
\begin{equation} \label{posterior} 
p(b_{k,l'}|z_{k,l'}^{(i)})
= \frac{
 p(z_{k,l'}^{(i)}|b_{k,l'})p(b_{k,l'})
}
{
 p(z_{k,l'}^{(i)}) 
},  
\end{equation} 
with 
\begin{equation}
p(z_{k,l'}^{(i)}) 
= \sum_{b_{k,l'}=\pm1}p(z_{k,l'}^{(i)}|b_{k,l'})p(b_{k,l'}). 
\end{equation} 
The MSE $\xi(\mathrm{sir}_{l'}^{(i)})$ for the posterior mean 
estimator~(\ref{posterior_mean_estimator}) in iteration~$i$ is given by 
\begin{equation} \label{MSE} 
\xi(\mathrm{sir}_{l'}^{(i)})
= \mathbb{E}\left[
 (b_{k,l'} - \langle b_{k,l'} \rangle_{i})^{2}
\right]. 
\end{equation} 

\begin{theorem} \label{theorem1} 
Suppose that (\ref{large_spreading_matrix}) is picked up from the 
$(r,L,W)$-ensemble, presented in Example~\ref{example3}. Then, the equivalent 
channel~(\ref{equivalent_channel}) for the BP receiver 
in iteration~$i$ converges to the equivalent channel for the scalar AWGN 
channel~(\ref{AWGN}) in the large-sparse-system limit: 
\begin{equation} \label{asymptotic_equivalent_channel} 
p(b_{k,l'}^{(i)}|b_{k,l'})\to 
\int p(b_{k,l'}=b_{k,l'}^{(i)}|z_{k,l'}^{(i)})
p(z_{k,l'}^{(i)}|b_{k,l'})dz_{k,l'}^{(i)}. 
\end{equation} 
In (\ref{asymptotic_equivalent_channel}), the posterior probability 
$p(b_{k,l'}|z_{k,l'}^{(i)})$ is given by (\ref{posterior}). 
The conditional pdf $p(z_{k,l'}^{(i)}|b_{k,l'})$ represents the scalar 
AWGN channel~(\ref{AWGN}) in iteration~$i$. In evaluating these expressions, 
the asymptotic SIR $\mathrm{sir}_{l'}^{(i)}$ is given via the coupled 
equations 
\begin{equation} \label{SIR} 
\mathrm{sir}_{l'}^{(i)} = \frac{1}{W+1}\sum_{w=0}^{W}
\frac{1}{\sigma_{(l'+w)_{L}}^{2}(i)}, 
\end{equation}
\begin{equation} \label{sigma_evolution} 
\sigma_{l}^{2}(i) = \sigma_{\mathrm{n}}^{2} + \frac{\beta_{l}}{W+1}\sum_{w=0}^{W}
\xi\left(
 \mathrm{sir}_{(l-w)_{L}}^{(i-1)}
\right), 
\end{equation}
with $\mathrm{sir}_{l'}^{(0)}=0$ for all $l'\in\mathcal{L}$. 
In (\ref{sigma_evolution}), $\beta_{l}=K/N_{l}$ is equal to 
$\beta_{\mathrm{init}}$ for $l=0,\ldots,W-1$ and to $\beta$  
for $l=W,\ldots,L-1$, respectively. 
\end{theorem}
\begin{IEEEproof}
See Appendix~\ref{proof_theorem1}. 
\end{IEEEproof}
The expressions~(\ref{SIR}) and (\ref{sigma_evolution}) determine the 
evolution of the asymptotic equivalent 
channel~(\ref{asymptotic_equivalent_channel}) with respect to $i$. 
Thus, they are referred to as DE equations. 

As is obvious from the proof of Theorem~\ref{theorem1}, the GA for the 
postulated interference~(\ref{postulated_interference}) is exact in the 
large-sparse-system limit. Thus, we have the following result. 

\begin{theorem} \label{theorem2} 
Suppose that (\ref{large_spreading_matrix}) is picked up from the 
$(r,L,W)$-ensemble, presented in Example~\ref{example3}. Then, 
the equivalent channel for the BP receiver with GA converges to the 
asymptotic equivalent channel for the true BP receiver, i.e.\  
the right-hand side (RHS) of (\ref{asymptotic_equivalent_channel})  
in the large-sparse-system limit. 
\end{theorem}
\begin{IEEEproof}
Repeat the proof of Theorem~\ref{theorem1}. 
\end{IEEEproof}
From Theorem~\ref{theorem2}, the performance of the BP receiver with GA is 
indistinguishable from that of the true BP receiver in the large-sparse-system 
limit. Thus, we hereafter focus on the true BP receiver. 

We here investigate the performance of the BP receivers for finite $L$ and 
$W$. It is proved that the asymptotic SIRs~(\ref{SIR}) 
monotonically converge toward a fixed-point of the DE 
equations~(\ref{SIR}) and (\ref{sigma_evolution}) as $i\to\infty$. 

\begin{proposition} \label{proposition_conv} 
For all positions~$l'$, 
\begin{equation}
\mathrm{sir}_{l'}^{(0)}\leq \mathrm{sir}_{l'}^{(1)} \leq \cdots 
\leq \mathrm{sir}_{l'}^{(\infty)}, 
\end{equation}
where $\{\mathrm{sir}_{l'}^{(\infty)}:l'\in\mathcal{L}\}$ denotes a fixed-point 
of the DE equations~(\ref{SIR}) and (\ref{sigma_evolution}). 
\end{proposition}
\begin{IEEEproof}
See the proof of Lemma~\ref{lemma_convergence} in Section~\ref{sec5}. 
\end{IEEEproof}

It is shown that the BP receiver can achieve the same performance as that 
for the (soft) IO receiver if the fixed-point is unique. 

\begin{theorem} \label{theorem3} 
Suppose that (\ref{large_spreading_matrix}) is picked up from the 
$(r,L,W)$-ensemble, presented in Example~\ref{example3}. 
If the fixed-point of the DE equations~(\ref{SIR}) and (\ref{sigma_evolution}) 
is unique, the asymptotic SIR for the BP receiver converges to that for 
the (soft) IO receiver in the infinite-iteration limit $i\to\infty$ 
after taking the large-sparse-system limit. 
\end{theorem}
\begin{IEEEproof}
We follow an argument based on a genie-aided BP receiver in \cite{Guo08} to 
prove Theorem~\ref{theorem3}. For a fixed number of iterations~$i$,   
let $\mathcal{Y}_{k,l'}^{(i)}$ denote the set of the received signals 
$\{y_{n,l}\}$ utilized in the BP detection of $b_{k,l'}$. The data symbols 
are classified into two groups: the data symbols $\mathcal{X}_{k,l'}^{(i)}$ 
that connect only to the function nodes in $\mathcal{Y}_{k,l'}^{(i)}$, and 
the remaining data symbols. The ACF property of the $(r,L,W)$ ensemble 
implies that $\mathcal{X}_{k,l'}^{(i)}$ consists of the data symbols whose 
depth (distance from the root $b_{k,l'}$) is less than or equal to $2(i-1)$. 
Furthermore, $\mathcal{Y}_{k,l'}^{(i)}$ is composed of the function nodes that 
have depth less than $2i$. The BP detection of $b_{k,l'}$ with the number of 
iterations~$i$ is equivalent to the soft IO detection of $b_{k,l'}$ based on 
$\mathcal{Y}_{k,l'}^{(i)}$, whereas the soft IO detection of $b_{k,l'}$ is 
based on the entire received signals $\mathcal{Y}$. We use a genie-aided 
BP receiver to obtain an upper bound on the asymptotic SIR of the soft IO 
receiver. Let a genie inform the BP receiver about the data symbols not in 
$\mathcal{X}_{k,l'}^{(i)}$. Since information about the received signals not 
in $\mathcal{Y}_{k,l'}^{(i)}$ is passed only through the {\em known} data 
symbols with depth $2i$, using the information $\mathcal{Y}\backslash
\mathcal{Y}_{k,l'}^{(i)}$ does not improve the performance of the genie-aided 
BP receiver. Thus, the soft IO receiver cannot outperform the genie-aided BP 
receiver. In other words, the asymptotic SIR for the genie-aided BP receiver 
provides an upper bound on that for the soft IO receiver. 

The performance of the genie-aided BP receiver can be evaluated in the 
large-sparse-system limit by repeating the proof of Theorem~\ref{theorem1}. 
The asymptotic SIR converges to (\ref{SIR}), which are determined by the DE 
equations~(\ref{SIR}) and (\ref{sigma_evolution}). The only difference is 
that the initial condition is not $\mathrm{sir}_{l'}^{(0)}=0$ but 
$\mathrm{sir}_{l'}^{(0)}=\infty$, 
because the data symbols with depth $2i$ are known to the receiver. 
Let us take the infinite-iteration limit $i\to\infty$. Since we 
have assumed that the DE equations have the unique fixed-point, the solution 
to the DE equations converges to the unique fixed-point as  
$i\to\infty$, regardless of the initial condition.  
This observation implies that the performance of the genie-aided BP 
receiver coincides with that of the BP receiver as $i\to\infty$ 
after taking the large-sparse-system limit. Since the performance of the 
soft IO receiver is sandwiched between the performance of the two BP 
receivers, the BP receiver can achieve the same performance as that for 
the soft IO receiver.  
\end{IEEEproof}

\subsection{Threshold Analysis} 
We have so far considered the two limits: the large-system limit and 
the dense limit. In this section, the remaining limits are investigated. 
We start with the definition of the BP threshold. 

\begin{definition}[BP Threshold] \label{def_BP_threshold}
Let $\{\mathrm{sir}_{l'}^{(\mathrm{opt})}\}$ denote the fixed-point of the DE 
equations~(\ref{SIR}) and (\ref{sigma_evolution}) that has the largest SIR 
at the middle point $l'/L=1/2$ among all possible fixed-points.  
The BP threshold is defined as the supremum of 
$\beta_{\mathrm{th}}$ such that 
$L^{-1}\sum_{l'\in\mathcal{L}}|\mathrm{sir}_{l'}^{(i)}-\mathrm{sir}_{l'}^{(\mathrm{opt})}|$ 
converges to zero for all $\beta\in(0,\beta_{\mathrm{th}})$. 
\end{definition}

Let us postulate that the DE equations~(\ref{SIR}) and (\ref{sigma_evolution}) 
for the uncoupled case $W=0$ have multiple fixed-points. For the SC-SCDMA 
system with $W>0$, reliable information about the data symbols should 
propagate toward the middle point $l'/L=1/2$. Thus, the asymptotic SIR at 
the middle point should be worst among those for all positions. 
Furthermore, the SIR $\mathrm{sir}_{L/2}^{(\mathrm{opt})}$ at the middle point 
is close to $1/\sigma_{\mathrm{n}}^{2}$ for high SNR when $\beta$ is below the BP 
threshold, as shown in Section~\ref{sec6}, so that the SIRs 
$\{\mathrm{sir}_{l'}^{(\mathrm{opt})}\}$ are close to $1/\sigma_{\mathrm{n}}^{2}$ 
at all positions. 
This implies that the BP threshold corresponds to a boundary between the 
interference-limited region and the non-limited region for the BP receivers. 
The BP receivers can mitigate the MAI well when $\beta$ is below the BP 
threshold. 
 
If the DE equations~(\ref{SIR}) and (\ref{sigma_evolution}) for $W>0$ have 
the unique fixed-point $\{\mathrm{sir}_{l'}^{(\mathrm{opt})}\}$, the asymptotic SIRs 
converge to $\{\mathrm{sir}_{l'}^{(\mathrm{opt})}\}$ as $i\to\infty$. Otherwise, 
the asymptotic SIRs are expected to be trapped in the other fixed-point, 
as noted in Section~\ref{sec5}. Thus, the BP threshold should be equal to 
the supremum of $\beta_{\mathrm{th}}$ such that the DE equations~(\ref{SIR}) and 
(\ref{sigma_evolution}) have the unique fixed-point for all system loads 
$\beta\in(0,\beta_{\mathrm{th}})$. 

We focus on the BP threshold for the SC system in the limit 
$L,W\to\infty$ with $\gamma=W/L\to0$, since analytical evaluation of the 
BP threshold for finite $L$ and $W>0$ is intractable.  
In order to distinguish the BP threshold for the SC system from that for the 
uncoupled system, the one for the uncoupled system $W=0$ is denoted by 
$\beta_{\mathrm{BP}}$. On the other hand, the BP threshold for the SC system 
in the limit $L,W\to\infty$ with $\gamma=W/L\to0$ is written as 
$\beta_{\mathrm{BP}}^{(\mathrm{SC})}$. 

Before evaluating the BP threshold $\beta_{\mathrm{BP}}^{(\mathrm{SC})}$ 
for the SC-SCDMA system, we shall review the performance assessment of 
the soft IO receiver for the uncoupled dense CDMA system based on the 
non-rigorous replica method~\cite{Tanaka02,Guo05}, and define the IO threshold. 

\begin{proposition}[Tanaka 2002] \label{proposition1} 
The asymptotic SIR of the soft IO receiver for the uncoupled 
dense BPSK-input CDMA system with system load~$\beta$ converges to $s$ in 
the large-system limit, in which $s$ is a solution to the following 
fixed-point equation,  
\begin{equation} \label{fixed_point} 
\frac{1}{s} = \sigma_{\mathrm{n}}^{2} + \beta\xi(s), 
\end{equation} 
where $\xi(s)$ denotes the MSE for the posterior mean estimator of the 
BPSK data symbol transmitted through the scalar AWGN channel with SNR $s$. 
If the fixed-point equation has multiple solutions, the solution $s$ 
is selected to minimize the free energy  
\begin{equation} \label{free_energy} 
F(s) = \beta C(s) + \frac{1}{2}\left[
 \sigma_{\mathrm{n}}^{2}s - \ln(\sigma_{\mathrm{n}}^{2}s) - 1 
\right], 
\end{equation}
where $C(s)$ denotes the input-output mutual information in nats 
for the BPSK-input scalar AWGN channel with SNR~$s$. 
\end{proposition} 

The fixed-points of the DE equations~(\ref{SIR}) and (\ref{sigma_evolution}) 
for the uncoupled SCDMA system~$W=0$ coincide with the solutions to the 
fixed-point equation~(\ref{fixed_point}). When the fixed-point 
equation~(\ref{fixed_point}) has multiple solutions, there is a difference in 
performance between the BP and IO receivers for the uncoupled case. 
The IO receiver can achieve the largest solution to (\ref{fixed_point}) 
if it is the global minimum of the free energy~(\ref{free_energy}), 
whereas the BP receivers cannot. 

It is straightforward to find that the solution $s$ to the fixed-point 
equation~(\ref{fixed_point}) corresponds to a stationary point of the free 
energy~(\ref{free_energy}), with a general relationship proved 
in \cite{Guo052} between the mutual information $C(s)$ and the MSE $\xi(s)$ 
for the scalar AWGN channel  
\begin{equation} \label{relationship} 
\frac{dC}{ds}(s) = \frac{1}{2}\xi(s).  
\end{equation}
Korada and Montanari~\cite{Korada11} proved that the minimum of the free 
energy~(\ref{free_energy}) over $s$ is equal to the sum capacity 
in nats for the uncoupled dense CDMA system with BPSK inputs in the 
large-system limit. Unfortunately, it is still open whether or not the 
asymptotic SIR for the soft IO receiver coincides with the solution $s$ to 
minimize the free energy~(\ref{free_energy}) when the fixed-point 
equation~(\ref{fixed_point}) has multiple solutions, although the 
non-rigorous replica analysis~\cite{Tanaka02,Guo05} suggests so. 

The fixed-point equation~(\ref{fixed_point}) has the unique solution for 
all system loads in the low-to-moderate SNR regime. On the other hand, 
it has multiple solutions for high system loads\footnote{
This is the definition of the term ``high system load'' in this 
paper.} in the high SNR regime. In other words, the free 
energy~(\ref{free_energy}) is bistable for high system loads, 
as shown in Fig.~\ref{fig1}. The latter situation is the target of spatial 
coupling. 

Only the free energy~(\ref{free_energy}) at the solutions to 
(\ref{fixed_point}) is used in Proposition~\ref{proposition1}. 
Consequently, one can apply any change of variables as long as 
it maps the global stable solution of the original free energy to 
that of the obtained one.  
We use this ambiguity to derive another expression of the free energy that is 
suitable for understanding the BP threshold for the SC-SCDMA system. 
Let us consider the free energy $\tilde{F}(s)$ obtained by 
substituting (\ref{fixed_point}) into $s$ in the second term of 
(\ref{free_energy}), 
\begin{IEEEeqnarray}{rl} 
\tilde{F}(s) =& \beta C(s) + \frac{1}{2}\left[
 \frac{\sigma_{\mathrm{n}}^{2}}{\sigma_{\mathrm{n}}^{2}+\beta\xi(s)} 
 - \ln\frac{\sigma_{\mathrm{n}}^{2}}{\sigma_{\mathrm{n}}^{2}+\beta\xi(s)} - 1
\right] \nonumber \\ 
=& \beta C(s) + \frac{1}{2}\left[
 \ln\frac{\sigma_{\mathrm{n}}^{2}+\beta\xi(s)}{\sigma_{\mathrm{n}}^{2}} 
 - \beta s\xi(s) 
\right]. \label{another_free_energy} 
\end{IEEEeqnarray}
In the derivation of the last expression, we have used (\ref{fixed_point}) 
again. The statement of Proposition~\ref{proposition1} would be 
unchanged even if the free energy~(\ref{another_free_energy}) were used 
instead of (\ref{free_energy}). 

\begin{remark} 
The shape of the free energy~(\ref{another_free_energy}) as a function of $s$ 
is qualitatively the same as that of the original one~(\ref{free_energy}). 
In fact, calculating the stationarity condition for 
(\ref{another_free_energy}) yields 
\begin{equation} \label{stationarity} 
\tilde{F}'(s) = \frac{\beta\xi'(s)}{2}\left[
 \frac{1}{\sigma_{\mathrm{n}}^{2}+\beta\xi(s)} - s
\right] = 0, 
\end{equation}
where we have used (\ref{relationship}). 
Since the MSE $\xi(s)$ is a monotonically decreasing function of SNR~$s$, 
the stationarity condition~(\ref{stationarity}) reduces to the original 
one~(\ref{fixed_point}). The values of the free 
energy~(\ref{another_free_energy}) at the stationary points coincide with 
those of the original one~(\ref{free_energy}) at the same stationary points. 
Furthermore, any two adjacent stationary points in the free energy are 
connected by a monotonic curve. These observations imply that the metastable, 
unstable, and global stable solutions to the free 
energy~(\ref{another_free_energy}) are equal to the corresponding 
solutions of the original one~(\ref{free_energy}), respectively.  
\end{remark}

We shall present the definition of the IO threshold for the uncoupled CDMA 
system. 

\begin{definition}[IO Threshold] \label{IO_threshold}
The IO threshold $\beta_{\mathrm{IO}}$ for the uncoupled CDMA system is defined 
as the supremum of $\beta_{\mathrm{th}}$ such that the asymptotic SIR for the 
IO receiver is equal to the largest solution of the free 
energy~(\ref{free_energy}) or (\ref{another_free_energy}) 
for all $\beta\in(0,\beta_{\mathrm{th}})$.  
\end{definition}

Proposition~\ref{proposition1} implies that the IO threshold 
$\beta_{\mathrm{IO}}$ is equal to the system load $\beta$ such that the free 
energy~(\ref{free_energy}) or (\ref{another_free_energy}) has two global 
minima. It is obvious that the IO threshold $\beta_{\mathrm{IO}}$ is larger 
than the conventional BP threshold $\beta_{\mathrm{BP}}$. 
The IO threshold corresponds to a boundary between the 
interference-limited region and the non-limited region for the IO receiver. 
The IO receiver can mitigate the MAI well when $\beta$ is below the IO  
threshold. 
 
We move on to the evaluation of the BP threshold for the SC-SCDMA system. 
The following result implies that the BP threshold can be improved up to 
the IO threshold by spatial coupling. 

\begin{theorem} \label{theorem4} 
Let $\beta_{\mathrm{init}}=0$ and take the continuum limit $L,W\to\infty$ 
with $\gamma=W/L$ kept constant, $i\to\infty$, and finally $\gamma\to0$. 
Then, 
\begin{equation}
\beta_{\mathrm{IO}}\leq\beta_{\mathrm{BP}}^{(\mathrm{SC})}. 
\end{equation}
\end{theorem}
\begin{IEEEproof}  
We use the two functions $\psi(v)=-\xi(v)$ and 
$\varphi(u)=1/(\sigma_{\mathrm{n}}^{2}-u)$ to define a potential energy 
function as 
\begin{equation} \label{potential_CDMA} 
V(u) = vu - \int\beta\psi(v)dv - \int\varphi(u)du, 
\end{equation}
with $v=\psi^{-1}(u/\beta)$. 
In (\ref{potential_CDMA}), the integrals denote indefinite integrals. 
Let $\tilde{\beta}_{\mathrm{BP}}^{(\mathrm{SC})}$ denote 
the potential threshold that is defined as $\beta$ such that 
the potential~(\ref{potential_CDMA}) has two global minima.  
In Section~\ref{sec5} we will prove 
$\tilde{\beta}_{\mathrm{BP}}^{(\mathrm{SC})}\leq\beta_{\mathrm{BP}}^{(\mathrm{SC})}$. 
Thus, it is sufficient to show $\tilde{\beta}_{\mathrm{BP}}^{(\mathrm{SC})}
=\beta_{\mathrm{IO}}$.  

Calculating the RHS of (\ref{potential_CDMA}) with (\ref{relationship}) 
and $u=\beta\psi(v)$, we obtain
\begin{equation}
V(u) = -\beta v\xi(v) + 2\beta C(v) 
+ \ln(\sigma_{\mathrm{n}}^{2}+\beta\xi(v)) + A, 
\end{equation}
with any constant $A$. Setting $A=-\ln\sigma_{\mathrm{n}}^{2}$ yields 
$V(u)=2\tilde{F}(\psi^{-1}(u/\beta))$, given by (\ref{another_free_energy}). 
Since the transformation of variables $v=\psi^{-1}(u/\beta)$ does not change 
the qualitative shape of the free energy~ (\ref{another_free_energy}), 
from the definition of the IO threshold 
we find $\tilde{\beta}_{\mathrm{BP}}^{(\mathrm{SC})}=\beta_{\mathrm{IO}}$.
\end{IEEEproof}

As shown in Section~\ref{sec5}, the DE 
equations~(\ref{SIR}) and (\ref{sigma_evolution}) have the unique fixed-point 
in the limit $W,L\to\infty$ with $\gamma\to0$ if $\beta$ is smaller than the 
IO threshold $\beta_{\mathrm{IO}}$.  
From Theorem~\ref{theorem3}, the BP receiver is optimal in the limit 
$W,L\to\infty$ with $\gamma\to0$ for $\beta<\beta_{\mathrm{IO}}$ if 
the limit $W,L\to\infty$ with $\gamma=W/L$ fixed commutes with 
the infinite-iteration limit $i\to\infty$ in Theorem~\ref{theorem3}, 
whereas Theorem~\ref{theorem3} was proved in the limit $i\to\infty$ 
for finite $L$ and $W$. 

\begin{remark} \label{remark2} 
We shall conjecture the position of the BP threshold 
$\beta_{\mathrm{BP}}^{(\mathrm{SC})}$ for the SC-SCDMA system. 
The non-rigorous replica analysis presented in \cite{Takeuchi111} implies 
that the IO threshold $\beta_{\mathrm{IO}}^{(\mathrm{SC})}$ for the 
SC-CDMA system converges to the conventional IO threshold 
$\beta_{\mathrm{IO}}$ from above in the limit $L,W\to\infty$ with 
$\gamma\to0$. Since the BP threshold $\beta_{\mathrm{BP}}^{(\mathrm{SC})}$ 
is bounded from above by $\beta_{\mathrm{IO}}^{(\mathrm{SC})}$, 
we obtain $\beta_{\mathrm{BP}}^{(\mathrm{SC})}\leq\beta_{\mathrm{IO}}$ 
in the limit $L,W\to\infty$ with $\gamma\to0$. 
Combining this upper bound and Theorem~\ref{theorem4}, 
we can conclude 
\begin{equation}
\beta_{\mathrm{BP}}^{(\mathrm{SC})}=\beta_{\mathrm{IO}}, 
\end{equation} 
if the replica analysis provides a correct result. Thus, we hereafter refer to 
the potential threshold $\tilde{\beta}_{\mathrm{BP}}^{(\mathrm{SC})}$ 
as the BP threshold for the SC-SCDMA system. 

The convergence of $\beta_{\mathrm{IO}}^{(\mathrm{SC})}$ to $\beta_{\mathrm{IO}}$ 
implies that spatial coupling never improves the performance of IO detection. 
In other words, spatial coupling should be regarded as a method for improving 
the performance of iterative detection. Unfortunately, we can provide no 
rigorous proof for the convergence, which may be intuitively understood as 
follows: The reason why $\beta_{\mathrm{IO}}^{(\mathrm{SC})}$ is above 
$\beta_{\mathrm{IO}}$ is due to the rate loss, which vanishes 
in the limit $L,W\to\infty$ with $\gamma\to0$. 
\end{remark}


\section{Phenomenological Study on Spatial Coupling} 
\label{sec5} 
\subsection{Continuum Limit} \label{sec5_0}
\subsubsection{Density-Evolution Equations}
We shall present the proof of Theorem~\ref{theorem4} in a general setting. 
We assume that two functions $\varphi(u)$ and $\psi(v)$ are 
bounded, strictly increasing, twice continuously differentiable. 
Let $\mathcal{D}\subset\mathbb{R}$ 
and $\tilde{\mathcal{D}}\subset\mathbb{R}$ denote the images of $\varphi$ and 
$\psi$, respectively. We assume that $\mathcal{D}$ and $\tilde{\mathcal{D}}$ 
are bounded, and that the supremum $u_{\mathrm{max}}$ of $\tilde{\mathcal{D}}$ is 
equal to $u_{\mathrm{max}}=0$, without loss of generality. 
Let $(v_{l}(i),u_{l}(i))\in\mathcal{D}\times\tilde{\mathcal{D}}$ denote the 
state in iteration~$i\geq0$ at position~$l\in\mathcal{L}=\{0,\ldots,L-1\}$ of 
an SC system with the number of subsystems~$L$ and coupling width~$W$, 
governed by the following DE equations 
\begin{equation} \label{DE_v_dis_tmp} 
v_{l}(i) = \frac{1}{W+1}\sum_{w=0}^{W}\varphi(u_{(l+w)_{L}}(i)), 
\end{equation}
\begin{equation} \label{DE_u_dis_tmp} 
u_{l}(i) = \frac{\beta_{l}}{W+1}\sum_{w=0}^{W}\psi(v_{(l-w)_{L}}(i-1)), 
\end{equation}
with the initial condition $v_{l}(0)=v_{\mathrm{min}}\equiv
\inf\mathcal{D}$. 
In (\ref{DE_u_dis_tmp}), the parameter $\beta_{l}\geq0$ is given by 
\begin{equation} \label{beta} 
\beta_{l} = \left\{
\begin{array}{cc}
 0 & l\in\{0,\ldots,W-1\} \\ 
 \beta & l\in\{W,\ldots,L-1\}. \\ 
\end{array}
\right. 
\end{equation}

The DE equations~(\ref{DE_v_dis_tmp}) and (\ref{DE_u_dis_tmp}) include 
(\ref{SIR}) and (\ref{sigma_evolution}) for the SC-SCDMA system as a special 
case, which can be confirmed by letting $v_{l}(i) = \mathrm{sir}_{l}^{(i)}$, 
$u_{l}(i)=\sigma_{\mathrm{n}}^{2}-\sigma_{l}^{2}(i)\leq0$, 
$\varphi(u)=1/(\sigma_{\mathrm{n}}^{2} - u)$, and $\psi(v)=-\xi(v)$.   

From (\ref{beta}) the DE equations~(\ref{DE_v_dis_tmp}) and 
(\ref{DE_u_dis_tmp}) can be represented as 
\begin{equation} \label{DE_v_dis} 
v_{l}(i) = \frac{1}{W+1}\sum_{w=0}^{W}\varphi(u_{l+w}(i)) 
\quad l\in\{0,\ldots,L-1\},  
\end{equation}
\begin{equation} \label{DE_u_dis} 
u_{l}(i) = \frac{\beta}{W+1}\sum_{w=0}^{W}\psi(v_{l-w}(i-1)) 
\quad l\in\{W,\ldots,L-1\},  
\end{equation}
with $u_{l}(i)=0$ for $l\notin\{W,\ldots,L-1\}$. Note that the boundaries are 
fixed to the supremum $u_{\mathrm{max}}=0$ of the set $\tilde{\mathcal{D}}$. 
The monotonicity $\varphi'(u)>0$ implies that 
$v_{\mathrm{max}}=\varphi(u_{\mathrm{max}})$ is also the supremum of $\mathcal{D}$. 

Let $v_{\mathrm{r}}$ denote the largest solution to the fixed-point 
equation $v=\varphi(\beta\psi(v))$ for the uncoupled case. 
The solution $(v_{\mathrm{r}}, u_{\mathrm{r}})$ 
satisfies the following fixed-point equations:  
\begin{equation} \label{uncoupled_FP} 
v_{\mathrm{r}} = \varphi(u_{\mathrm{r}}), 
\quad u_{\mathrm{r}} = \beta\psi(v_{\mathrm{r}}). 
\end{equation}
We assume that $(v_{\mathrm{r}}, u_{\mathrm{r}})$ is a stable fixed-point for 
the DE equations~(\ref{DE_v_dis}) and (\ref{DE_u_dis}) 
in the uncoupled case $W=0$. 

We first prove that the DE equations~(\ref{DE_v_dis}) and (\ref{DE_u_dis}) 
are convergent as $i\to\infty$. 
\begin{lemma} \label{lemma_convergence} 
For any $l\in\mathcal{L}$ and $i$, 
\begin{equation}
v_{l}(i)\leq v_{l}(i+1), \quad 
u_{l}(i)\leq u_{l}(i+1). 
\end{equation}
\end{lemma}
\begin{IEEEproof}
We follow \cite{Donoho13} to prove the statement by induction. 
For $i=0$ the statement $v_{l}(0)\leq v_{l}(1)$ holds for any $l$ 
because of $v_{l}(0)=v_{\mathrm{min}}$. 
Assume $v_{l}(i-1)\leq v_{l}(i)$ for all $x$. 
From (\ref{DE_u_dis}), we obtain 
\begin{IEEEeqnarray}{rl}
&u_{l}(i+1) - u_{l}(i) 
\nonumber \\ 
=& \frac{\beta}{W+1}\sum_{w=0}^{W}\left\{
 \psi(v_{l-w}(i)) - \psi(v_{l-w}(i-1))
\right\}
\geq0,
\end{IEEEeqnarray}
for all $l\in\{W,\ldots,L-1\}$. 
In the derivation of the inequality, we have used the assumption 
$v_{l}(i-1)\leq v_{l}(i)$ and $\psi'(v)>0$. Combining this observation and 
the boundary condition $u_{l}(i)=u_{\mathrm{max}}$ for any $i$ and 
$l\notin\{W,\ldots,L-1\}$, 
we obtain $u_{l}(i)\leq u_{l}(i+1)$ for any $l\in\mathcal{L}$. 
Repeating the same argument for (\ref{DE_v_dis}), we find 
$v_{l}(i)\leq v_{l}(i+1)$ for any $l\in\mathcal{L}$. 
By induction, the statement holds for any $i$. 
\end{IEEEproof}

We take three limits to analyze the DE equations~(\ref{DE_v_dis}) and 
(\ref{DE_u_dis}): In a first limit called {\em continuum limit}, $L$ and $W$ 
tend to infinity while the ratio $\gamma=W/L$ is kept constant. 
A second limit is the infinite-iteration limit $i\to\infty$. 
In the last limit, $\gamma$ tends to zero. 
The goal of Section~\ref{sec5_0} is to prove that the state governed by the DE 
equations converges to a stationary solution of a temporally-continuous 
and spatially-continuous partial differential equation in the limits above. 
The proof strategy is as follows: We first take the continuum limit 
to reduce the DE equations~(\ref{DE_v_dis}) and (\ref{DE_u_dis}) 
to discrete-time and spatially-continuous integral systems. 
Subsequently, we approximate the integral systems by a continuous-time 
partial differential equation as $\gamma\to0$ after taking $i\to\infty$, 
whereas Donoho et al.~\cite{Donoho13} analyzed the integral systems directly. 
The main theorem in Section~\ref{sec5_0} can be rigorously proved by deriving 
the partial differential equation via the integral systems. Furthermore, 
the partial differential equation provides an intuitive understanding of 
spatial coupling, as shown in Section~\ref{sec5_1}. 

\subsubsection{Integral Systems} 
We define two spatially-continuous state functions $v_{\gamma}(x,i)$ and 
$u_{\gamma}(x,i)$ as  
\begin{equation} \label{DE_v}
v_{\gamma}(x,i)
= \mathfrak{F}[u_{\gamma}(\cdot,i);\varphi](x),  
\quad |x|\leq1,  
\end{equation}
\begin{equation} \label{DE_u} 
u_{\gamma}(x,i) 
= \left\{
 \begin{array}{ll} 
 \beta\mathfrak{F}[v_{\gamma}(\cdot,i-1);\psi](x) & |x|<1-\gamma \\ 
 u_{\mathrm{max}} & |x|\geq1-\gamma,  
 \end{array} 
\right.  
\end{equation}
with 
\begin{equation} \label{integral_operator} 
\mathfrak{F}[u;\varphi](x) 
= \frac{1}{2\gamma}\int_{-\gamma}^{\gamma}\varphi(u(x+\omega))d\omega. 
\end{equation}
We impose the initial condition $v_{\gamma}(x,0)=v_{\mathrm{min}}$ for $|x|\leq1$. 

Let $\mathcal{C}_{1-2\gamma}^{2}$ denote the space of continuous even 
functions on $[-1,1]$ that are twice continuously differentiable on 
$(-1,1)-\{\pm(1-2\gamma)\}$. 
The function $v_{\gamma}(x,i)$ is shown to be 
contained in the space $\mathcal{C}_{1-2\gamma}^{2}$ for any $i$.   
\begin{lemma} \label{lemma_cont} 
\begin{enumerate}
\item For any~$x$, $i$, and any $\gamma>0$,  
\begin{equation}
v_{\gamma}(x,i)\leq v_{\gamma}(x,i+1), 
\quad u_{\gamma}(x,i)\leq u_{\gamma}(x,i+1). 
\end{equation}
\item For any $i$, $u_{\gamma}(x,i)$ is an even function and continuous 
on $\mathbb{R}-\{\pm(1-\gamma)\}$. 
Furthermore, $v_{\gamma}(x,i)\in\mathcal{C}_{1-2\gamma}^{2}$ for any $i$. 
\item For any $i$ and $\gamma>0$, 
\begin{equation} \label{v_dif}
\lim_{W=\gamma L\to\infty}\frac{1}{L}\sum_{l\in\mathcal{L}}\left|
 v_{l}(i) - v_{\gamma}\left(
  \frac{2l}{L}-1,i
 \right)
\right|=0, 
\end{equation}
\begin{equation} \label{u_dif} 
\lim_{W=\gamma L\to\infty}\frac{1}{L}\sum_{l\in\mathcal{L}}\left|
 u_{l}(i) - u_{\gamma}\left(
  \frac{2l}{L}-1-\gamma,i
 \right) 
\right|=0. 
\end{equation}
\end{enumerate} 
\end{lemma}

\begin{IEEEproof}
The first property is proved by repeating the proof of 
Lemma~\ref{lemma_convergence}. 
Thus, we shall prove the second property. 
The symmetry of $v_{\gamma}(x,i)$ and $u_{\gamma}(x,i)$ follows from the 
symmetries of the initial condition $v_{\gamma}(x,0)=v_{\mathrm{min}}$ and of 
the integral systems~(\ref{DE_v}) and (\ref{DE_u}). 
It is straightforward to observe that, for any even function $u(x)$, 
the function $\mathfrak{F}[u;\varphi](x)$ is also an even function.  
Indeed, one has 
\begin{IEEEeqnarray}{rl}
\mathfrak{F}[u;\varphi](-x)
&=\frac{1}{2\gamma}\int_{-\gamma}^\gamma
\varphi(u(-x+\omega))d\omega
\nonumber\\
&=\frac{1}{2\gamma}\int_{-\gamma}^\gamma
\varphi(u(-x-\omega))d\omega
\nonumber\\
&=\frac{1}{2\gamma}\int_{-\gamma}^\gamma
\varphi(u(x+\omega))d\omega
=\mathfrak{F}[u;\varphi](x).
\end{IEEEeqnarray}
Thus, the integral systems~(\ref{DE_v}) and (\ref{DE_u}) with the even 
initial function $v_{\gamma}(x,0)=v_{\mathrm{min}}$ define even functions  
$v_{\gamma}(x,i)$ and $u_{\gamma}(x,i)$ for any $i$.  

Let us show the statement on continuity. 
Expression~(\ref{DE_v}) can be represented as 
\begin{equation} \label{DE_v_tmp} 
v_{\gamma}(x,i) 
= \frac{1}{2\gamma}\int_{x-\gamma}^{x+\gamma}\varphi\left(
 u_{\gamma}(\omega,i)
\right)d\omega, \quad |x|\leq1. 
\end{equation}
Since the initial function $v_{\gamma}(x,0)=v_{\mathrm{min}}$ is measurable, 
the integral systems~(\ref{DE_v}) and (\ref{DE_u}) with the bounded functions 
$\varphi$ and $\psi$ define measurable functions 
$v_{\gamma}(x,i)$ and $u_{\gamma}(x,i)$ for any $i$. This observation and the 
boundedness of $\varphi$ and $\psi$ imply  
that the integrand in (\ref{DE_v_tmp}) is Lebesgue-integrable. 
Thus, $v_{\gamma}(x,i)$ is (absolutely) continuous~\cite{Ash99} on $[-1,1]$ 
for any~$i$. Repeating the same argument for (\ref{DE_u}), we find 
that $u_{\gamma}(x,i)$ is continuous for $|x|<1-\gamma$. Combining this 
observation and the boundary condition $u_{\gamma}(x,i)=u_{\mathrm{max}}$ for 
$|x|\geq1-\gamma$, 
the function $u_{\gamma}(x,i)$ is continuous on $\mathbb{R}-\{\pm(1-\gamma)\}$.  

The statement on differentiability follows from (\ref{DE_v_tmp}). 
When the integrand in (\ref{DE_v_tmp}) is continuous at $\omega=x\pm\gamma$, 
from the fundamental theorem of calculus~\cite{Strang10}, the derivative of 
(\ref{DE_v_tmp}) exists and is given by 
\begin{equation}
\frac{dv_{\gamma}}{dx}(x,i) 
= \frac{\varphi(u_{\gamma}(x+\gamma,i)) - \varphi(u_{\gamma}(x-\gamma,i))}
{2\gamma}. 
\end{equation}
Since $u_{\gamma}(x,i)$ is continuous with the exception of $x=\pm(1-\gamma)$, 
$v_{\gamma}(x,i)$ is continuously differentiable on $(-1,1)-\{\pm(1-2\gamma)\}$. 
Repeating the same argument for $u_{\gamma}(x,i)$, we find that $u_{\gamma}(x,i)$ 
is continuously differentiable with the exception of the discontinuous points 
$x=\pm(1-\gamma)$. The function $v_{\gamma}(x,i)$ is twice continuously 
differentiable when the RHS of (\ref{DE_v_tmp}) is continuously 
differentiable. Thus, $v_{\gamma}(x,i)$ is twice continuously differentiable on 
$(-1,1)-\{\pm(1-2\gamma)\}$. 

The last property holds from the definition of the Riemann integral. 
See \cite{Donoho13} for a formal proof of the last property by induction. 
We here present a sketch of the proof. 
Assume that (\ref{v_dif}) holds for some $i$. From (\ref{DE_u_dis}) and 
(\ref{DE_u}), calcuating the difference 
$|u_{l}(i+1) - u_{\gamma}(x_{l}-\gamma,i+1)|$ for $x_{l}=(2l/L)-1$ yields 
\begin{IEEEeqnarray}{rl}
&|u_{l}(i+1) - u_{\gamma}(x_{l}-\gamma,i+1)| 
\nonumber \\
<& \frac{1}{W+1}\sum_{w=0}^{W}\left|
 \psi(v_{l-w}(i)) - \psi(v_{\gamma}(x_{l-w},i))
\right| 
\nonumber \\ 
+& \left|
 \frac{1}{W+1}\sum_{w=0}^{W}\psi(v_{\gamma}(x_{l-w},i))
\right. \nonumber \\ 
&-\left. 
  \frac{1}{2\gamma}\int_{-\gamma}^{\gamma}
 \psi(v_{\gamma}(x_{l}-\gamma+\omega,i))d\omega
\right|. 
\end{IEEEeqnarray}
Taking the sum $L^{-1}\sum_{l\in\mathcal{L}}$, letting $\omega=-(2w/L)+\gamma$, and 
considering the continuum limit, it is possible to show  
that the first term tends to zero from the assumption~(\ref{v_dif}). 
Furthermore, the second term is also proved to converge to zero from the 
definition of the Riemann integral. 
Repeating the same argument for (\ref{DE_v_dis}) and (\ref{DE_v}) results in  
the last property of Lemma~\ref{lemma_cont}.  
\end{IEEEproof}

From the first property in Lemma~\ref{lemma_cont}, it is guaranteed that the 
state $(v_{\gamma}(x,i),u_{\gamma}(x,i))$ of the integral systems~(\ref{DE_v}) 
and (\ref{DE_u}) converges a stationary solution $(v_{\gamma}(x),u_{\gamma}(x))$ 
as $i\to\infty$. Since the integrands in (\ref{DE_v}) and (\ref{DE_u}) 
are bounded, we can use the dominated convergence 
theorem~\cite{Ash99} to exchange the order of the limit $i\to\infty$ and the 
integrals. Thus, any stationary solution $(v_{\gamma}(x),u_{\gamma}(x))$ satisfies 
the fixed-point equations 
\begin{equation} \label{FP_v}
v_{\gamma}(x)
= \mathfrak{F}[u_{\gamma}(\cdot);\varphi](x),  
\quad |x|\leq1,  
\end{equation}
\begin{equation} \label{FP_u} 
u_{\gamma}(x) 
= \left\{
 \begin{array}{ll} 
 \beta\mathfrak{F}[v_{\gamma}(\cdot);\psi](x) & |x|<1-\gamma \\ 
 u_{\mathrm{max}} & |x|\geq1-\gamma,  
 \end{array} 
\right.  
\end{equation}
with $\mathfrak{F}$ defined in (\ref{integral_operator}). 
Although differentiability for 
stationary solutions is non-trivial in general, we can prove that any 
stationary solution $(v_{\gamma}(x),u_{\gamma}(x))$ has the same differentiability 
as $(v_{\gamma}(x,i),u_{\gamma}(x,i))$.  

\begin{lemma} \label{lemma_FP}
Suppose that $v_{\gamma}(x)$ is any stationary solution to the fixed-point 
equations~(\ref{FP_v}) and (\ref{FP_u}). Then, 
$v_{\gamma}(x)\in\mathcal{C}_{1-2\gamma}^{2}$. 
\end{lemma}  
\begin{IEEEproof}
Repeat the proof of the second property in Lemma~\ref{lemma_cont}, by using 
the fact that the sequence of measurable functions converges to a measurable 
function.  
\end{IEEEproof}

\subsubsection{Differential Systems} 
We study stationary solutions to the fixed-point 
equations~(\ref{FP_v}) and (\ref{FP_u}) in the limit $\gamma\to0$.  
It is done by introducing a continuous-time system with 
a state function $u(x,t)$ for $x\in[-1,1]$ at time~$t\geq0$, 
whose time evolution is governed by a partial differential equation.  
The continuous-time system is to be constructed so that 
any stable solution to the fixed-point equations~(\ref{FP_v}) 
and (\ref{FP_u}) is characterized in the limit $\gamma\to0$ 
by a stationary solution to the partial differential equation.  
Intuitively, one may regard the derivative $\partial u(x,t)/\partial t$ as 
an approximation of the difference $u_{\gamma}(x,t+1)-u_{\gamma}(x,t)$. 

Let us define a potential function $V(u)$ as 
\begin{equation} \label{potential}
V(u) = - D(\psi^{-1}(u/\beta)\|u), 
\end{equation}
where $D(v\|u)$ similar to the divergence\footnote{
The indefinite integrals of $\varphi$ and $\beta\psi$ are connected to each 
other via the Legendre transform in information geometry, whereas the two 
functions are independent functions in this paper.
} in information geometry~\cite{Amari00} is given by 
\begin{equation} \label{divergence} 
D(v\|u) = \int\beta\psi(v)dv + \int\varphi(u)du - vu.  
\end{equation}
The integrals in (\ref{divergence}) denote indefinite integrals, so that 
the first and second terms in (\ref{divergence}) are functions of 
$v$ and $u$, respectively. 
Furthermore, we define a differential operator $\mathfrak{L}$ as   
\begin{equation} \label{operator}  
\mathfrak{L}[u](x)  
= \frac{B'(u(x))}{2}\left(
 \frac{\partial u}{\partial x}(x)
\right)^{2} 
+ B(u(x))\frac{\partial^{2}u}{\partial x^{2}}(x), 
\end{equation}
for a twice continuously differential function $u(x)$ on $[-1,1]$, with 
\begin{equation}
B(u) = \frac{1}{3}\varphi'(u)>0. 
\end{equation}
Then, the partial differential equation that governs $u(x,t)$ is defined as 
\begin{equation} \label{single_system}
\frac{\partial u}{\partial t} 
= A(u(x,t))\left\{
 - V'(u(x,t)) + \gamma^{2}\mathfrak{L}[u(\cdot,t)](x)
\right\}, 
\end{equation}
with
\begin{equation} \label{function_A} 
A(u) 
= \beta\psi'\left(
 \psi^{-1}\left(
 \frac{u}{\beta}
 \right)
\right)>0. 
\end{equation}

We impose the boundary condition $u(\pm1,t)=u_{\mathrm{max}}$ and 
the initial condition $u(x,0)=\beta\psi(v_{\mathrm{init}}(x))$. 
In the initial condition, 
$v_{\mathrm{init}}(x)$ is a twice continuously differentiable function that 
satisfies $|v_{\mathrm{init}}(x)-v_{\gamma}(x,\infty)|<\epsilon_{\mathrm{init}}$ for 
any $\epsilon_{\mathrm{init}}>0$, with $v_{\gamma}(x,\infty)$ denoting the 
fixed-point of the integral systems~(\ref{DE_v}) and (\ref{DE_u}) 
as $i\to\infty$. We note that such a function $v_{\mathrm{init}}(x)$ exists 
from Lemma~\ref{lemma_FP}. 

\begin{lemma} \label{lemma_stability_bulk} 
For any $\epsilon>0$ and $x\in[-1,1]$, there exist some $t_{0}>0$ and 
stationary solution $u(x)$ such that 
\begin{equation}
|u(x,t) - u(x)| < \epsilon,
\end{equation}
for all $t\geq t_{0}$ and $\gamma>0$. 
\end{lemma}
\begin{IEEEproof}
See Appendix~\ref{deriv_gradient_system} for a sketch of the proof.  
\end{IEEEproof}

The goal of Section~\ref{sec5_0} is to prove the following theorem. 

\begin{theorem} \label{theorem5}
Let $\tilde{v}(x)=\lim_{t\to\infty}\psi^{-1}(u(x,t)/\beta)$. Then, 
\begin{equation}
\lim_{\gamma\to0}\lim_{i\to\infty}\lim_{W=\gamma L\to\infty}
\frac{1}{L}\sum_{l\in\mathcal{L}}\left|
 v_{l}(i) - \tilde{v}\left(
  \frac{2l}{L}-1
 \right)
\right| = 0. 
\end{equation}
\end{theorem}
The potential~(\ref{potential}) was originally defined in \cite{Yedla12}. 
The contribution of this paper is to provide a systematic derivation of 
the potential via the approximation of the DE equations~(\ref{DE_v_dis}) and 
(\ref{DE_u_dis}) by the partial differential 
equation~(\ref{single_system}). 

Theorem~\ref{theorem5} implies that the analysis of fixed-points to 
the DE equations~(\ref{DE_v_dis}) and (\ref{DE_u_dis}) reduces to that of the 
structure of stationary solutions to the partial differential 
equation~(\ref{single_system}). The analysis will be presented 
in the next section to prove Theorem~\ref{theorem4}. 

\subsubsection{Proof of Theorem~\ref{theorem5}}
The proof strategy of Theorem~\ref{theorem5} is as follows: 
The relationship between the DE equations and the integral systems has 
been established in Lemma~\ref{lemma_cont}. Thus, we need to assess the 
relationship between the stationary solution $u(x)$ of the differential 
system~(\ref{single_system}) as $t\to\infty$ and that of the integral 
systems~(\ref{DE_v}) and (\ref{DE_u}) as $i\to\infty$. 
In order to show that the difference between the two stationary solutions is 
negligibly small for sufficiently small $\gamma$, we use the two properties 
of the differential system: One property is the asymptotic stability of the 
differential system shown in Lemma~\ref{lemma_stability_bulk}. This implies 
that there exists some time $t_{0}>0$ such that the difference between the 
stationary solution $u(x)$ and the state $u(x,t_{0})$ at time $t=t_{0}$ is 
negligibly small. The other property is that the state of the differential 
system moves very slowly for sufficiently small $\gamma$, since the 
differential system is an approximation of the integral systems as 
$\gamma\to0$, and since the initial state of the differential system is very 
close to the fixed-point solution of the integral systems.  
This property implies a negligibly small change of the state for the 
differential system as long as finite time-evolution is considered. 
Combining the two properties yields Theorem~\ref{theorem5}.   

We first prove the latter property. 
Let $v_{\gamma}(x)=\mathfrak{G}[v_{\gamma}(\cdot)](x)$ denote a single 
fixed-point equation for $v_{\gamma}(x)$ obtained by eliminating $u_{\gamma}(x)$ 
from the fixed-point equations~(\ref{FP_v}) and (\ref{FP_u}). In order to 
evaluate the operator $\mathfrak{G}$, we define a continuous-time 
differential system as 
\begin{equation}  
\frac{\partial\tilde{v}}{\partial t} 
= - \tilde{v}(x,t) + \tilde{\mathfrak{G}}[\tilde{v}(\cdot,t)](x), 
\label{differential_system}  
\end{equation}
for $x\in[-1,1]$.  
In (\ref{differential_system}), the operator $\tilde{\mathfrak{G}}$ 
is given by  
\begin{equation} \label{differential_system_bulk} 
\tilde{\mathfrak{G}}[\tilde{v}(\cdot,t)](x)
= \varphi(\beta\psi(\tilde{v}(x,t)))
+ \gamma^{2}\mathfrak{L}[\beta\psi(\tilde{v}(\cdot,t))](x), 
\end{equation}
with $\mathfrak{L}$ defined in (\ref{operator}). 
We impose the boundary condition $\tilde{v}(\pm1,t)=v_{\mathrm{max}}$ 
for any $t$ and the initial condition 
$\tilde{v}(x,0)=v_{\mathrm{init}}(x)$, defined below (\ref{function_A}). 

We first confirm that the system~(\ref{differential_system}) 
is equivalent to the partial differential 
equation~(\ref{single_system}) under the change of variables 
$u=\beta\psi(\tilde{v})$. This property and 
Lemma~\ref{lemma_stability_bulk} imply that the differential 
system~(\ref{differential_system}) is convergent as $t\to\infty$ for any 
$\gamma>0$. The coefficient $A(u)$ in (\ref{single_system}) is due to 
the chain rule $\partial u/\partial t=A(u)\partial\tilde{v}/\partial t$. 
Let us show that $\tilde{v}-\varphi(\beta\psi(\tilde{v}))$ corresponds to 
the derivative of the potential~(\ref{potential}) 
under the change of variables. By definition, 
\begin{IEEEeqnarray}{rl}
\int\{\tilde{v}-\varphi(\beta\psi(\tilde{v}))\}du 
=& \int\{\tilde{v} 
 - \varphi(\beta\psi(\tilde{v}))\}\beta\psi'(\tilde{v})d\tilde{v} 
\nonumber \\ 
=& -D(\tilde{v}\|\beta\psi(\tilde{v})),  
\end{IEEEeqnarray}
which is equal to $V(u)$, because of $\tilde{v}=\psi^{-1}(u/\beta)$. 
Thus, the partial differential equation~(\ref{differential_system}) is 
equivalent to (\ref{single_system}). 

The differential system~(\ref{differential_system}) is obtained by 
Taylor-expanding the RHSs of (\ref{DE_v}) and (\ref{DE_u}) with respect to 
$\gamma$ around $\gamma=0$ up to the second order for the bulk region 
$\mathcal{X}=(-(1-2\gamma),1-2\gamma)$. 

\begin{proposition} \label{lemma_deriv}
Suppose that $v(x)$ is any twice continuously differentiable function on 
$[-1,1]$. For any $\epsilon>0$, there exists some $\gamma_{0}>0$ such that 
\begin{equation} \label{norm_v}
\int_{-1}^{1}\left|
 \mathfrak{G}[v](x) - \tilde{\mathfrak{G}}[v](x)
\right|dx < \epsilon, 
\end{equation}
for all $\gamma\in(0,\gamma_{0})$. 
\end{proposition}
\begin{IEEEproof}
Decomposing the integral~(\ref{norm_v}) into two parts, we obtain 
\begin{IEEEeqnarray}{rl}
&\int_{-1}^{1}\left|
 \mathfrak{G}[v](x) - \tilde{\mathfrak{G}}[v](x)
\right|dx
\nonumber \\ 
=& \int_{\mathcal{X}}\left|
 \mathfrak{G}[v](x) - \tilde{\mathfrak{G}}[v](x)
\right|dx
+ \int_{\bar{\mathcal{X}}}\left|
 \mathfrak{G}[v](x) - \tilde{\mathfrak{G}}[v](x)
\right|dx, 
\nonumber \\
\label{norm_v_tmp} 
\end{IEEEeqnarray}
where $\bar{\mathcal{X}}=[-1,-(1-2\gamma)]\cup[1-2\gamma,1]$ denotes the 
boundary region. We first upper-bound the second term. 
The second property of Lemma~\ref{lemma_cont} implies that 
$\mathfrak{G}[v](x)$ is bounded on $[-1,1]$ for given $v(x)$. 
Furthermore, from Lemma~\ref{lemma_stability_bulk} we find that 
$\tilde{\mathfrak{G}}[v](x)$ is also bounded. Thus, 
the second term on the RHS of (\ref{norm_v_tmp}) is bounded from above by 
\begin{IEEEeqnarray}{rl}
&\int_{\bar{\mathcal{X}}}\left|
 \mathfrak{G}[v](x) - \tilde{\mathfrak{G}}[v](x)
\right|dx 
\nonumber \\ 
<& 4\gamma\left(
 \sup_{x\in\bar{\mathcal{X}}}\left|
  \mathfrak{G}[v](x)
 \right| 
 + \sup_{x\in\bar{\mathcal{X}}}\left|
  \tilde{\mathfrak{G}}[v](x)
 \right| 
\right), 
\end{IEEEeqnarray}
which tends to zero as $\gamma\to0$. 

We next prove that the integrand 
$|\mathfrak{G}[v](x) - \tilde{\mathfrak{G}}[v](x)|$ in the first term of 
(\ref{norm_v_tmp}) tends to zero as $\gamma\to0$ for $x\in\mathcal{X}$ 
in the bulk region. 
This will complete the proof of Proposition~\ref{lemma_deriv} 
because of the following argument: 
Since $|\mathfrak{G}[v](x) - \tilde{\mathfrak{G}}[v](x)|$ is bounded on 
$\mathcal{X}$ for given $v(x)$, from the dominated convergence theorem 
we can exchange the order of the limit $\gamma\to0$ and the integral 
$\int_{\mathcal{X}}dx$. These observations imply that the first term on the 
RHS of (\ref{norm_v_tmp}) tends to zero as $\gamma\to0$. Thus, 
Proposition~\ref{lemma_deriv} holds. 

Let us prove that $|\mathfrak{G}[v](x) - \tilde{\mathfrak{G}}[v](x)|$ 
tends to zero as $\gamma\to0$ for $x\in\mathcal{X}$. 
Since $v(x)$ is twice continuously differentiable, we expand the integrand 
$\psi(v(x+\omega))$ in $\mathfrak{F}[v;\psi](x)$ given by 
(\ref{integral_operator}) with respect to $\omega$ up to the second order 
to obtain 
\begin{equation}
\mathfrak{F}[v;\psi](x)
= \left(
1 + \frac{\gamma^{2}}{6}\frac{d^{2}}{dx^{2}}
\right)\psi(v(x)) + o(\gamma^{2}),  
\label{expansion_u} 
\end{equation}
for $x\in\mathcal{X}$. 

We next expand (\ref{DE_v}) for the bulk region $x\in\mathcal{X}$ to derive 
\begin{equation} \label{expansion_v_bulk} 
\mathfrak{G}[v](x) 
= \left(
1 + \frac{\gamma^{2}}{6}\frac{d^{2}}{dx^{2}}
\right)\varphi(\beta\mathfrak{F}[v;\psi](x))
+ o(\gamma^{2}),  
\end{equation}
for $x\in\mathcal{X}$. Substituting (\ref{expansion_u}) into 
(\ref{expansion_v_bulk}) and expanding the obtained expression with respect to 
$\gamma$, we arrive at $\mathfrak{G}[v](x)=
\tilde{\mathfrak{G}}[v](x)+o(\gamma^{2})$ given by 
(\ref{differential_system_bulk}) for the bulk region $x\in\mathcal{X}$. 
Thus, the difference $|\mathfrak{G}[v](x)-\tilde{\mathfrak{G}}[v](x)|$ 
converges to zero as $\gamma\to0$ for $x\in\mathcal{X}$. 
\end{IEEEproof}

It is expected from Proposition~\ref{lemma_deriv} that the solution 
$\tilde{v}(x,t)$ to the partial differential 
equation~(\ref{differential_system}) is very close to the initial state 
for sufficiently small $\gamma$ as long as $t$ is finite, since the initial 
state corresponds to the fixed-point of the integral systems~(\ref{DE_v}) and 
(\ref{DE_u}) as $i\to\infty$. More precisely, we have the following lemma. 

\begin{lemma} \label{lemma_difference} 
For any $t_{0}\geq0$ and $\epsilon>0$, 
there exists some $\gamma_{0}>0$ such that 
\begin{equation} \label{difference} 
\int_{-1}^{1}|\tilde{v}(x,t_{0}) - \tilde{v}(x,0)|dx<\epsilon, 
\end{equation}
for all $\gamma\in(0,\gamma_{0})$. 
\end{lemma}
\begin{IEEEproof}
See Appendix~\ref{lemma_difference_proof} for a proof based on 
Proposition~\ref{lemma_deriv}.  
\end{IEEEproof}


We are ready to prove Theorem~\ref{theorem5}. 

\begin{IEEEproof}[Proof of Theorem~\ref{theorem5}]
Let $v_{\gamma}(x)=\lim_{i\to\infty}v_{\gamma}(x,i)$ and 
$\tilde{v}(x)=\lim_{t\to\infty}\tilde{v}(x,t)$. 
From Lemma~\ref{lemma_convergence} and the first property of 
Lemma~\ref{lemma_cont}, for any $x\in[-1,1]$, $l\in\mathcal{L}$,  
and $\epsilon>0$ there exists some $I\in\mathbb{N}$ such that 
\begin{equation} \label{discrete_system_convergence} 
|v_{l}(i) - v_{l}(I)| < \epsilon, 
\end{equation}
\begin{equation} \label{integral_system_convergence} 
|v_{\gamma}(x,i) - v_{\gamma}(x)| < \epsilon, 
\end{equation}
for all $i\geq I$. For this number $I$ of iterations and $x_{l}=(2l/L)-1$, 
we use the triangle inequality and 
(\ref{discrete_system_convergence}) to obtain 
\begin{IEEEeqnarray}{rl}
&\frac{1}{L}\sum_{l\in\mathcal{L}}|v_{l}(i) - \tilde{v}(x_{l})|
\nonumber \\ 
<& \frac{1}{L}\sum_{l\in\mathcal{L}}|v_{l}(I) - v_{\gamma}(x_{l},I)| 
+ \frac{1}{L}\sum_{l\in\mathcal{L}}|v_{\gamma}(x_{l},I) - \tilde{v}(x_{l})| 
+ \epsilon. 
\nonumber \\ 
\label{upper_bound}
\end{IEEEeqnarray}
The last property of Lemma~\ref{lemma_cont} implies that the first term on 
the upper bound~(\ref{upper_bound}) tends to zero in the continuum limit. 
From the definition of the Riemann integral, the sum 
$L^{-1}\sum_{l\in\mathcal{L}}$ in the second term can be replaced by the integral 
$2^{-1}\int_{-1}^{1}dx$. More precisely, we have 
\begin{equation}
\lim_{W=\gamma L\to\infty}\frac{2}{L}\sum_{l\in\mathcal{L}}
|v_{\gamma}(x_{l},I) - \tilde{v}(x_{l})| 
= \int_{-1}^{1}|v_{\gamma}(x,I) - \tilde{v}(x)|dx.  
\end{equation}
From (\ref{integral_system_convergence}) we have the following 
bound for the integrand 
\begin{equation}
|v_{\gamma}(x,I) - \tilde{v}(x)| 
< |v_{\gamma}(x) - \tilde{v}(x)| + \epsilon. 
\end{equation}
Applying these observations to (\ref{upper_bound}) yields 
\begin{IEEEeqnarray}{rl}
&\lim_{\gamma\to0}\lim_{i\to\infty}\lim_{W=\gamma L\to\infty}
\frac{1}{L}\sum_{l\in\mathcal{L}}|v_{l}(i) - \tilde{v}(x_{l})| 
\nonumber \\ 
<& \frac{1}{2}\lim_{\gamma\to0}\int_{-1}^{1}|v_{\gamma}(x) - \tilde{v}(x)|dx 
+ 2\epsilon. \label{target_upper_bound} 
\end{IEEEeqnarray} 
Thus, it is sufficient to prove that the first term on the upper 
bound~(\ref{target_upper_bound}) is equal to to zero. 

Lemma~\ref{lemma_stability_bulk} implies that 
$\tilde{v}(x,t)$ converges uniformly to $\tilde{v}(x)$ as $t\to\infty$ 
with respect to $\gamma>0$. Since $|\tilde{v}(x,t)-\tilde{v}(x)|$ is bounded, 
from the dominated convergence theorem we find that for any $\epsilon_{1}>0$ 
there exists some $t_{0}>0$ such that 
\begin{equation}
\frac{1}{2}\int_{-1}^{1}|\tilde{v}(x,t) - \tilde{v}(x)|dx 
<\epsilon_{1}, 
\end{equation}
for all $t\geq t_{0}$ and $\gamma>0$. 
From this observation, we use the triangle inequality to obtain 
\begin{equation}
\frac{1}{2}\int_{-1}^{1}|v_{\gamma}(x) - \tilde{v}(x)|dx  
< \frac{1}{2}\int_{-1}^{1}|v_{\gamma}(x) - \tilde{v}(x,t_{0})|dx  
+ \epsilon_{1}.  
\end{equation}
From the initial condition $|\tilde{v}(x,0)-v_{\gamma}(x)|
<\epsilon_{\mathrm{init}}$, Lemma~\ref{lemma_difference} implies that the first 
term on the upper bound converges to zero as $\gamma\to0$. Thus, 
the upper bound~(\ref{target_upper_bound}) tends to zero. 
\end{IEEEproof}

\subsection{Review of Phenomenological Study} \label{sec5_1} 
We shall review our phenomenological study on spatial 
coupling~\cite{Takeuchi111}. This section is organized as an independent 
section of Section~\ref{sec5_0}. 
The study characterizes the position of the BP 
threshold for SC systems. Furthermore, it helps us understand 
why spatial coupling improves the conventional BP threshold. 
We first explain the dynamics of the partial differential 
equation~(\ref{single_system}), although it is sufficient to 
investigate the properties of stationary solutions from Theorem~\ref{theorem5}. 

We start with the following partial differential equation: 
\begin{equation} \label{phenomenological_system} 
\frac{\partial u}{\partial t} 
= A(u(x,t))\left\{
 - V'(u(x,t)) + \gamma^{2}\mathfrak{L}[u(\cdot,t)](x)
\right\}, 
\end{equation}
with $\mathfrak{L}$ defined in (\ref{operator}). 
In (\ref{phenomenological_system}), 
$t\geq0$ and $x\in(-1,1)$ denote the temporal and spatial 
variables, respectively. 
The state $u(x,t)$ is associated with a performance measure, 
such as SIR, ME, and so on. Without loss of generality, 
we assume that larger $u$ implies better performance. 
The parameter $\gamma>0$ represents the {\em strength} of spatial coupling. 
The dynamics of the uncoupled system $\gamma=0$ is characterized by a 
potential energy function $V(u)$, which is assumed to be bounded below. 
The two functions $A(u)>0$ and $B(u)>0$ in (\ref{operator}) and 
(\ref{phenomenological_system}) are arbitrary smooth functions. 
These assumptions hold for the CDMA case, in which $\psi(v)=-\xi(v)$ and 
$\varphi(u)=1/(\sigma_{\mathrm{n}}^{2} - u)$ for $u<0$.    

Let us consider the uncoupled system $\gamma=0$. In this case, the partial 
differential equation~(\ref{phenomenological_system}) reduces to the ordinary 
differential equation
\begin{equation}
\frac{\partial u}{\partial t} = - A(u)V'(u). 
\end{equation}
Since the state $u(x,t)$ does not depend on the spatial variable~$x$ anymore, 
we re-write it as $u(t)$ for the uncoupled system. 
It is straightforward to find  
\begin{equation}
\frac{d}{dt}V(u(t)) = -A(u)\{V'(u(t))\}^{2}\leq0, 
\end{equation}
where the equality holds if and only if $V'(u(t))=0$. This implies that 
the energy $V(u(t))$ monotonically decreases with the time-evolution of the 
state $u(t)$. Since the potential is bounded below, the 
state $u(t)$ converges to a (local) minimum of the potential $V(u)$ 
as $t\to\infty$. 

Suppose that the potential $V(u)$ has a parameter $\beta$, and 
that the shape of the potential as a function of $u$ changes with the 
increase of $\beta$, as shown in Fig.~\ref{fig1}. 
The potential $V(u)$ is assumed to have the unique stable solution for small 
$\beta$. As $\beta$ increases across a critical value of $\beta$, denoted 
by $\beta_{\mathrm{BP}}$, a metastable solution emerges to the left side of 
the global stable solution. The BP threshold for the uncoupled system 
$\gamma=0$ is defined as the supremum of $\beta_{\mathrm{th}}$ 
such that the state converges to the rightmost stable solution as $t\to\infty$ 
for all $\beta\in(0,\beta_{\mathrm{th}})$. When the initial state is smaller 
than the infimum of the unstable solution of $V(u)$ over all 
$\beta>\beta_{\mathrm{BP}}$, the BP threshold is equal to the critical value 
$\beta_{\mathrm{BP}}$ such that the potential $V(u)$ is monostable 
(resp.\ bistable) for all $\beta<\beta_{\mathrm{BP}}$ 
(resp.\ $\beta>\beta_{\mathrm{BP}}$). In fact, 
the state for the uncoupled system $\gamma=0$ is trapped in the left stable  
solution for $\beta>\beta_{\mathrm{BP}}$, since the initial state is smaller 
than the unstable solution of $V(u)$, whereas it can arrive at the rightmost  
stable solution for $\beta<\beta_{\mathrm{BP}}$. Spatial coupling allows   
the state to escape from the left stable solution and to arrive at the 
right stable solution for $\beta\in(\beta_{\mathrm{BP}},
\tilde{\beta}_{\mathrm{BP}}^{(\mathrm{SC})})$, in which 
the potential threshold $\tilde{\beta}_{\mathrm{BP}}^{(\mathrm{SC})}$ 
will be specified shortly, whereas the state may be trapped in the left stable 
solution for $\beta>\tilde{\beta}_{\mathrm{BP}}^{(\mathrm{SC})}$. 
The goal of this section is to elucidate the 
mechanism of escaping from the left stable solution and to specify the 
position of $\tilde{\beta}_{\mathrm{BP}}^{(\mathrm{SC})}$. 
We hereafter focus on the case $\beta>\beta_{\mathrm{BP}}$, in which the 
potential $V(u)$ is bistable. 

Let $u_{\mathrm{l}}$ and $u_{\mathrm{r}}$ denote the left (smaller) and right  
(larger) stable solutions of $V(u)$, respectively. The boundaries of the 
state $u(x,t)$ are assumed to be fixed to the right stable solution, i.e.\ 
$u(\pm1,t)=u_{\mathrm{r}}$ for all $t\geq0$. The {\em correct} boundary 
condition $u(\pm1,t)=u_{\mathrm{max}}$ will be considered shortly. 
Furthermore, we impose an initial condition $u(x,0)=u_{\mathrm{init}}(x)$, with  
some function $u_{\mathrm{init}}(x)$. 
The state $u(x,t)$ governed by (\ref{phenomenological_system}) moves around 
in a space of functions on $[-1,1]$ as $t$ increases. 
If $u_{\mathrm{init}}(x)$ is smaller than the unstable solution of $V(u)$ for any 
$x\in(-1,1)$, the state for the uncoupled system $\gamma=0$ is 
trapped in $u(x)=u_{\mathrm{l}}$ for all $x\in(-1,1)$. Why can the state 
escape from the left stable solution for SC systems? When can the state arrive 
at the right stable solution $u_{\mathrm{r}}$ for all $x$ as $t\to\infty$?  
Our phenomenological study provides answers to these questions.

In order to answer the former question, we represent the 
system~(\ref{phenomenological_system}) as a gradient system 
\begin{equation} \label{gradient_system}  
\frac{\partial u}{\partial t} = 
-A(u(x,t))\frac{\delta H}{\delta u}[u(\cdot,t)](x), 
\end{equation}
where the energy functional $H[u]$ is given by  
\begin{equation} \label{energy_functional} 
H[u] = \int_{-1}^{1}\left[
 V(u(x)) + \frac{\gamma^{2}B(u(x))}{2}\left(
  \frac{\partial u}{\partial x} 
 \right)^{2}
\right]dx. 
\end{equation}
In (\ref{gradient_system}), $\delta/\delta u$ denotes the functional 
derivative with respect to $u$. See Appendix~\ref{deriv_gradient_system} 
for the derivation of (\ref{gradient_system}). As shown in the same 
appendix, it is straightforward to find 
\begin{equation} \label{stability}
\frac{dH}{dt}[u(\cdot,t)] = - \int_{-1}^{1}A(u(x,t))\left(
 \frac{\delta H}{\delta u}[u(\cdot,t)](x)
\right)^{2}dx \leq 0, 
\end{equation}
where the equality holds if and only if $\delta H/\delta u=0$. This implies 
that the energy functional~(\ref{energy_functional}) monotonically decreases 
with the time-evolution of the state. Since (\ref{energy_functional}) 
is bounded below, $H[u(\cdot,t)]$ converges to a finite value as $t\to\infty$. 
Thus, the state $u(x,t)$ is guaranteed to converge to a stationary 
state $u(x)$ as $t\to\infty$, which is a local minimum of the energy 
functional~(\ref{energy_functional}). 
It is obvious that the uniform solution $u(x)=u_{\mathrm{r}}$ is 
a stable stationary solution to (\ref{gradient_system}), since the 
boundaries are fixed to the right stable solution $u_{\mathrm{r}}$. 
The second term in the integrand of (\ref{energy_functional}) smooths 
the state $u(x,t)$ spatially. This smoothing effect helps the state escape 
from the left stable solution and move toward the uniform solution for all 
$x\in[-1,1]$. Surprisingly, the smoothing effect will be shown to work even 
in the limit $\gamma\to0$.  

We next elucidate the answer to the latter question. The following theorem 
presents a partial answer to the question. 

\begin{theorem} \label{theorem_review}  
Suppose that the boundary is fixed to the right 
stable solution $u_{\mathrm{r}}$ of the potential $V(u)$. If and only if 
$u_{\mathrm{r}}$ is the unique global stable solution of $V(u)$, the uniform 
solution $u(x)=u_{\mathrm{r}}$ is the unique stationary solution to the 
system~(\ref{phenomenological_system}) in the limit $\gamma\to0$. 
\end{theorem}
\begin{IEEEproof}
See Section~\ref{sec5_2}. 
\end{IEEEproof}

As shown from (\ref{stability}), the state $u(x,t)$ converges to a 
stable stationary solution as $t\to\infty$. 
Combining this observation and 
Theorem~\ref{theorem_review} implies that the state $u(x,t)$ converges to 
the uniform solution if $u_{\mathrm{r}}$ is the unique global stable 
solution of the potential $V(u)$. In other words, the state $u(x,t)$ can 
escape from the left stable solution and arrive at the right stable solution 
$u_{\mathrm{r}}$ for all $x$, when $u_{\mathrm{r}}$ is the unique global stable 
solution of the potential $V(u)$. 

Recall that the shape of the potential $V(u)$ with a positive parameter 
$\beta$ is assumed to change with the increase of $\beta$, as shown in 
Fig.~\ref{fig1}. The BP threshold for the SC system is defined as the parameter 
$\beta_{\mathrm{th}}$ such that the state $u(x,t)$ converges to the uniform 
solution for $\beta<\beta_{\mathrm{th}}$, whereas it is trapped in a 
non-uniform stationary solution for $\beta>\beta_{\mathrm{th}}$. 
Suppose that $V(u_{\mathrm{l}})=V(u_{\mathrm{r}})$ holds at 
$\beta=\tilde{\beta}_{\mathrm{BP}}^{(\mathrm{SC})}$. Theorem~\ref{theorem_review} 
implies that $\tilde{\beta}_{\mathrm{BP}}^{(\mathrm{SC})}$ is a lower bound on 
the BP threshold, since the state is guaranteed to converge to the uniform 
solution for $\beta<\tilde{\beta}_{\mathrm{BP}}^{(\mathrm{SC})}$. 

It depends on the initial condition $u(x,0)=u_{\mathrm{init}}(x)$ whether 
$\tilde{\beta}_{\mathrm{BP}}^{(\mathrm{SC})}$ is equal to the BP threshold. 
In other words, it depends on the initial condition whether the state converges 
to a stable non-uniform solution when the non-uniform solution exists. 
As a trivial example, let us consider the initial condition $u_{0}(x)=u_{\mathrm{r}}$. 
Even if there is a non-uniform stationary solution to 
(\ref{phenomenological_system}) the state $u(x,t)$ never converges to the 
non-uniform solution as $t\to\infty$, since the initial state 
is a stable stationary solution to (\ref{phenomenological_system}). 
When the initial state is smaller than the left stable solution 
$u_{\mathrm{l}}$ in a bulk region far from the boundaries, on the other hand, 
the state is expected to converge toward a stable non-uniform stationary 
solution if the non-uniform solution exists. 
Unfortunately, we could not prove the 
convergence toward a stable non-uniform stationary solution under the latter 
initial condition. If we could prove it, we would be able to present the 
complete answer to the question: When can the state arrive 
at the right stable solution $u_{\mathrm{r}}$ for all $x$ as $t\to\infty$?  

\begin{remark} \label{remark_review} 
The condition presented in Theorem~\ref{theorem_review}, i.e.\  
the global stability of the potential $V(u)$ at $u=u_{\mathrm{r}}$ may not be 
necessary for strictly positive $\gamma$, whereas it is sufficient for any 
$\gamma>0$: The uniform solution may be the unique stable stationary solution 
for strictly positive $\gamma$, when $\beta$ is slightly larger than 
$\tilde{\beta}_{\mathrm{BP}}^{(\mathrm{SC})}$. 
Let us consider the limit $\gamma\to\infty$ 
to present an intuitive understanding of this statement. The first 
term in the integrand of (\ref{energy_functional}) 
is negligible in the limit $\gamma\to\infty$. Thus, the energy 
functional~(\ref{energy_functional}) has the unique stable solution that 
satisfies $\partial u/\partial x=0$ and $u(\pm1)=u_{\mathrm{r}}$ or 
equivalently $u(x)=u_{\mathrm{r}}$. This observation implies that 
for sufficiently large $\gamma$ the state converges to the uniform solution 
as $t\to\infty$, regardless of $\beta$. The smoothing effect with $\gamma>0$, 
given by the second term in the integrand of (\ref{energy_functional}), helps 
the state converge to the uniform solution for $\beta$ slightly larger than 
$\tilde{\beta}_{\mathrm{BP}}^{(\mathrm{SC})}$. 
\end{remark}

\begin{corollary} \label{corollary} 
Suppose that the boundary is fixed to $\bar{u}>u_{\mathrm{r}}$. 
If $u_{\mathrm{r}}$ is the unique global stable solution of $V(u)$, 
a solution $u(x)$ that satisfies $u(x)\geq u_{\mathrm{r}}$ for all $x$ 
is the unique stationary solution to the 
system~(\ref{phenomenological_system}) for any $\gamma>0$. 
\end{corollary}
\begin{IEEEproof}
See Section~\ref{sec5_2}. 
\end{IEEEproof}
Theorem~\ref{theorem4} follows from Theorem~\ref{theorem5} and 
Corollary~\ref{corollary}. 

\subsection{Proof of Theorem~\ref{theorem_review}}
\label{sec5_2} 
Let $g(u)$ denote a monotonically increasing 
function that satisfies 
\begin{equation} \label{function_G} 
g'(u) = \sqrt{B(u)}. 
\end{equation} 
Letting $\tilde{u}=g(u)$ transforms the  
system~(\ref{phenomenological_system}) into 
\begin{equation} \label{simple_system}  
\frac{\partial\tilde{u}}{\partial t} = 
A(u)B(u)\left[
 \gamma^{2}\frac{\partial^{2}\tilde{u}}{\partial x^{2}} 
 - \frac{V'(u)}{\sqrt{B(u)}} 
\right], 
\end{equation} 
with $u=g^{-1}(\tilde{u})$. The newly introduced variable $\tilde{u}$ 
corresponds to the normal coordinate system in differential 
geometry~\cite{Kobayashi96}. 

Let us introduce an effective potential energy function $U(\tilde{u})$ 
that satisfies 
\begin{equation} \label{effective_potential} 
 U'(\tilde{u}) = \frac{V'(u)}{\sqrt{B(u)}},  
\end{equation}
with $u=g^{-1}(\tilde{u})$. It is straightforward to confirm 
$U(\tilde{u})=V(g^{-1}(\tilde{u}))+C$, with a constant $C$. 
Note that $\tilde{u}_{\mathrm{r}}=g(u_{\mathrm{r}})$ is the global stable 
solution of the effective potential $U(\tilde{u})$ if and only if 
$u_{\mathrm{r}}$ is the global stable solution of the original one $V(u)$. 

We first prove the sufficiency of Theorem~\ref{theorem_review}, 
i.e.\ the uniform solution is the unique stable stationary solution 
if $\tilde{u}_{\mathrm{r}}$ is the unique global stable solution of 
$U(\tilde{u})$. The following result is valid for any $\gamma>0$. 

\begin{theorem}[Takeuchi et al.\ 2012] \label{theorem1_review} 
Suppose that the boundary is fixed to the right 
stable solution $\tilde{u}_{\mathrm{r}}$ of the potential $U(\tilde{u})$.  
If $\tilde{u}_{\mathrm{r}}$ is the unique global stable solution of the 
effective potential $U(\tilde{u})$, the uniform solution 
$\tilde{u}(x)=\tilde{u}_{\mathrm{r}}$ is the unique stationary solution 
to (\ref{simple_system}). 
\end{theorem}
\begin{IEEEproof}[Proof of Theorem~\ref{theorem1_review}] 
We follow \cite{Takeuchi111} to prove Theorem~\ref{theorem1_review}. 
A stationary solution $\tilde{u}(x)$ to (\ref{simple_system}) satisfies  
\begin{equation} \label{stationary_system} 
\gamma^{2}\frac{d^{2}\tilde{u}}{d x^{2}}
= U'(\tilde{u}), 
\end{equation} 
with the boundary condition $\tilde{u}(\pm1)=\tilde{u}_{\mathrm{r}}$. 
Integrating (\ref{stationary_system}) after multiplying both sides by 
$d\tilde{u}/dx$, we obtain 
\begin{equation} \label{energy_conservation} 
\frac{\gamma^{2}}{2}\left(
 \frac{d\tilde{u}}{dx}
\right)^{2} = U(\tilde{u}) + C,  
\end{equation}
with a constant $C$. Since $\tilde{u}_{\mathrm{r}}$ is the global stable 
solution of $U(\tilde{u})$, the boundary condition 
$\tilde{u}(\pm1)=\tilde{u}_{\mathrm{r}}$ and the positivity of the 
left-hand side (LHS) on (\ref{energy_conservation}) imply 
$C \geq - U(\tilde{u}_{\mathrm{r}})$. 
Let us prove $C=-U(\tilde{u}_{\mathrm{r}})$. 
From the symmetry of the boundary-value problem~(\ref{stationary_system}) 
with $\tilde{u}(\pm1)=\tilde{u}_{\mathrm{r}}$, any solution $\tilde{u}(x)$ 
is symmetric about the axis~$x=0$, i.e.\ an even function 
$\tilde{u}(-x)=\tilde{u}(x)$. The point $x=0$ is the middle point of the 
interval $[-1,1]$. Furthermore, any stationary solution $\tilde{u}(x)$ must 
be continuously differentiable since it is a solution to the second-order 
differential equation~(\ref{stationary_system}). Thus, we find 
$d\tilde{u}/dx|_{x=0}=0$. 
Evaluating (\ref{energy_conservation}) at $x=0$ yields 
$U(\tilde{u}(0))=-C\leq U(\tilde{u}_{\mathrm{r}})$. 
Combining this result with the global stability of $\tilde{u}_{\mathrm{r}}$, 
i.e.\ $U(\tilde{u}(0))\geq U(\tilde{u}_{\mathrm{r}})$, we obtain  
$C=-U(\tilde{u}(0))=-U(\tilde{u}_{\mathrm{r}})$.  
Note that the uniqueness of the global stable solution implies 
$\tilde{u}(0)=\tilde{u}_{\mathrm{r}}$. 

We shall show that the uniform solution $\tilde{u}(x)=\tilde{u}_{\mathrm{r}}$ 
is the unique solution to (\ref{energy_conservation}) with 
$C=-U(\tilde{u}_{\mathrm{r}})$. 
We have decomposed the boundary-value problem on $[-1,1]$ into two 
equivalent subproblems on $[-1,0]$ and $[0,1]$. Repeating this argument 
infinitely, we find that $\tilde{u}(x)$ is equal to $\tilde{u}_{\mathrm{r}}$ at 
$x=k/2^{j}$ for all $\{k\in\mathbb{Z}:|k|\leq 2^{j}\}$ and all $j\geq0$. 
Since $\tilde{u}(x)$ is continuous, this 
observation implies $\tilde{u}(x)=\tilde{u}_{\mathrm{r}}$ for all $x$. 
Thus, the uniform solution $\tilde{u}(x)=\tilde{u}_{\mathrm{r}}$ is the unique 
solution to (\ref{energy_conservation}) or (\ref{stationary_system}). 
\end{IEEEproof}

\begin{remark}
Hassani et al.~\cite{Hassani12} presented an intuitive argument based on 
classical mechanics, and obtained results equivalent to 
Theorem~\ref{theorem1_review}. We shall review their intuitive argument.  
The differential equation~(\ref{stationary_system}) is regarded as 
the Newton equation of motion: The state $\tilde{u}(x)$ is regarded as the 
position of a particle with mass~$\gamma^{2}$ at time~$x$, moving 
subject to the potential energy $-U(\tilde{u})$. Note that $x$ corresponds to 
the temporal variable in this interpretation, although $x$ has been defined  
as the spatial variable in the original phenomenological 
system~(\ref{phenomenological_system}). 
Expression~(\ref{energy_conservation}) corresponds to the conservation of 
mechanical energy.  

The boundary condition $\tilde{u}(-1)=\tilde{u}_{\mathrm{r}}$ implies that 
the particle is on the right maximum of the inverted potential $-U(\tilde{u})$ 
at time~$x=-1$. The uniform solution corresponds to the situation under which 
the particle continues to stay on the right maximum. Can the other solutions 
exist? In other words, 
can the particle move from the initial position at time~$x=-1$ and return to the initial 
position at time~$x=1$? Since $\tilde{u}_{\mathrm{r}}$ 
is the global maximizer of the inverted potential $-U(\tilde{u})$, 
the conservation of mechanical energy implies that the velocity 
$\tilde{u}'(x)$ at time $x=0$ must be non-zero if the particle moves to 
some position $\tilde{u}(0)\neq \tilde{u}_{\mathrm{r}}$. 
From the symmetry of $\tilde{u}(x)$ about the axis $x=0$, however, the 
non-zero velocity $\tilde{u}'(x)\neq0$ at time $x=0$ indicates that 
$\tilde{u}(x)$ is non-differentiable at time $x=0$. This is a contradiction, 
since $\tilde{u}(x)$ is the solution to the second-order differential 
equation~(\ref{stationary_system}). Thus, it is impossible for the particle 
to move from the initial position and to return to the initial position at 
time~$x=1$. These observations imply that the uniform solution 
$\tilde{u}(x)=\tilde{u}_{\mathrm{r}}$ is the unique solution to 
(\ref{stationary_system}).  
\end{remark}

Let us prove Corollary~\ref{corollary}. The transformation of variables 
$\tilde{u}=g(u)$ maps $\bar{u}>u_{\mathrm{r}}$ to a point $g(\bar{u})$ greater 
than $\tilde{u}_{\mathrm{r}}$. Corollary~\ref{corollary} holds trivially from 
the physical intuition, since it is impossible for the particle to return to 
the initial position $g(\bar{u})>\tilde{u}_{\mathrm{r}}$ when the particle gets 
over the hill $\tilde{u}=\tilde{u}_{\mathrm{r}}$. The proof is formally given as 
follows: 
\begin{IEEEproof}[Proof of Corollary~\ref{corollary}]
The statement holds if $\tilde{u}(x)>\tilde{u}_{\mathrm{r}}$ for all $x$. Thus, 
we consider the case in which $\tilde{u}(x)\leq\tilde{u}_{\mathrm{r}}$ for some 
$x$. Since the stationary solution is continuous, 
$\tilde{u}(x)=\tilde{u}_{\mathrm{r}}$ holds at some points $x\in[0,1]$. 
Let $x_{0}\in[0,1]$ denote the maximum of such points. Thus, we find 
$\tilde{u}(x)>\tilde{u}_{\mathrm{r}}$ for all $x\in(x_{0},1]$. The symmetry of 
stationary solutions implies that $\tilde{u}(-x_{0})=\tilde{u}_{\mathrm{r}}$. 
This problem can be regarded as a boundary-value problem on 
$[-x_{0},x_{0}]\subset[-1,1]$ with the boundary condition 
$\tilde{u}(\pm x_{0})=\tilde{u}_{\mathrm{r}}$. Repeating the 
proof of Theorem~\ref{theorem1_review}, we find that 
$\tilde{u}(x)=\tilde{u}_{\mathrm{r}}$ for all $x\in[-x_{0},x_{0}]$. Thus, 
$\tilde{u}(x)\geq\tilde{u}_{\mathrm{r}}$ holds for all $x$. 
\end{IEEEproof}

Theorem~\ref{theorem1_review} implies that the sufficiency of 
Theorem~\ref{theorem_review} is correct. 
We next prove the necessity of Theorem~\ref{theorem_review}, i.e.\ 
there is a stable non-uniform stationary solution in the limit $\gamma\to0$, 
if $\tilde{u}_{\mathrm{r}}$ is the metastable solution of $U(\tilde{u})$. 
Let us focus on a non-uniform stationary solution $\tilde{u}(x)$ 
to (\ref{simple_system}) that satisfies $d\tilde{u}/dx<0$ 
(resp.\ $d\tilde{u}/dx>0$) for $x\in(-1,0)$ (resp.\ $x\in(0,1)$). 
The following theorem guarantees the existence of such a stable non-uniform 
stationary solution in the limit~$\gamma\to0$. 

\begin{figure}[t]
\begin{center}
\includegraphics[width=0.5\hsize]{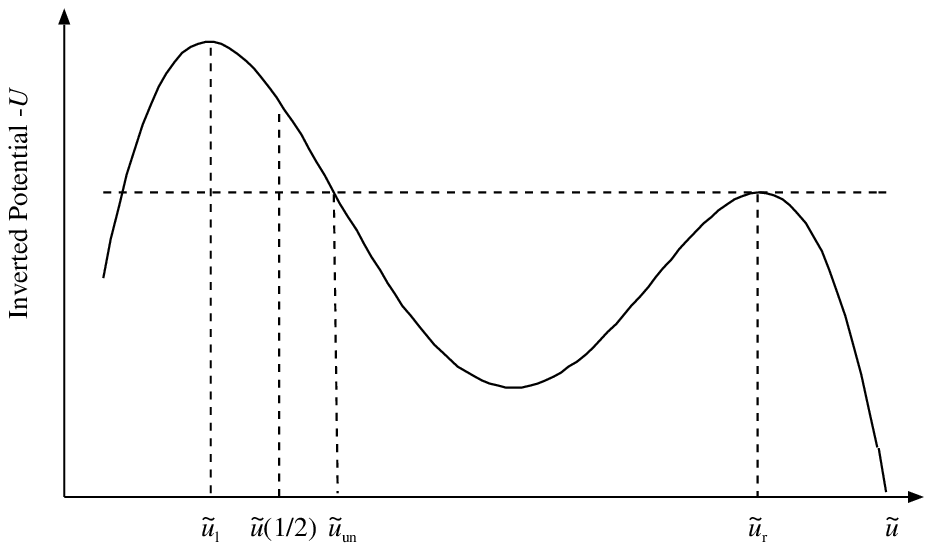}
\end{center}
\caption{
Inverted potential $-U(\tilde{u})$ for 
$\beta>\tilde{\beta}_{\mathrm{BP}}^{(\mathrm{SC})}$. 
}
\label{fig0} 
\end{figure}

\begin{theorem} \label{theorem2_review}
Suppose that the boundary is fixed to the right stable solution 
$\tilde{u}_{\mathrm{r}}$ of the potential $U(\tilde{u})$, and 
that $\tilde{u}_{\mathrm{r}}$ is the metastable solution of the effective 
potential $U(\tilde{u})$. 
Let us define $\tilde{u}_{\mathrm{l}}$ and $\tilde{u}_{\mathrm{un}}$ as 
$\tilde{u}_{\mathrm{l}}=g(u_{\mathrm{l}})$ and the point 
$\tilde{u}_{\mathrm{un}}\in(\tilde{u}_{\mathrm{l}},\tilde{u}_{\mathrm{r}})$ 
satisfying  $U(\tilde{u}_{\mathrm{un}})=U(\tilde{u}_{\mathrm{r}})$, respectively 
(See Fig.~\ref{fig0}). Then, for sufficiently small $\gamma>0$ there are two 
non-uniform stationary solutions $\tilde{u}_{\mathrm{s}}(x)$ and 
$\tilde{u}_{\mathrm{un}}(x)$ to (\ref{simple_system}). Furthermore, 
one stationary solution $\tilde{u}_{\mathrm{s}}(x)$ converges to 
$\tilde{u}_{\mathrm{l}}$ for $x\in(-1,1)$ in the limit $\gamma\to0$. The other 
stationary solution $\tilde{u}_{\mathrm{un}}(x)$ converges to 
$\tilde{u}_{\mathrm{r}}$ for $x\neq0$ in the limit $\gamma\to0$, whereas 
$\tilde{u}_{\mathrm{un}}(0)$ tends to $\tilde{u}_{\mathrm{un}}$ 
in the limit $\gamma\to0$. In particular, $\tilde{u}_{\mathrm{s}}(x)$ is 
stable in the limit $\gamma\to0$. 
\end{theorem}
\begin{IEEEproof}[Proof of Theorem~\ref{theorem2_review}] 
See Appendix~\ref{proof_theorem2_review}. 
\end{IEEEproof}

\begin{remark}
The two solutions $\tilde{u}_{\mathrm{s}}(x)$ and $\tilde{u}_{\mathrm{un}}(x)$ 
in the limit $\gamma\to0$ can be interpreted in terms of classical mechanics 
as follows: For $\tilde{u}_{\mathrm{s}}(x)$, a particle starts rolling down 
from the right maximum of the inverted potential $-U(\tilde{u})$ toward 
the left maximum at time $x=-1$. The velocity is infinitely large, when the 
mass is infinitely small. In a moment, the particle approaches the left 
maximum of the inverted potential with vanishing velocity. If the approaching 
velocity were finite, the particle would pass through the left maximum and 
roll down the left cliff. At time~$x=0$, the particle 
starts rolling down from the left maximum toward the right maximum with 
infinitely small velocity. Just before 
time $x=1$ the velocity becomes infinitely large, and returns to the right 
maximum of the inverted potential at time $x=1$. 

For the other solution $\tilde{u}_{\mathrm{un}}(x)$, a particle starts rolling 
down from the right maximum of the inverted potential to the left side with 
infinitely small velocity at time $x=-1$. Just before time $x=0$, the 
velocity of the particle becomes infinitely large, and stops at the point 
$\tilde{u}_{\mathrm{un}}$ at time $x=0$, because of the conservation of 
mechanical energy. If the initial velocity were finite, the particle could not 
stop at the point $\tilde{u}_{\mathrm{un}}$. 
The particle starts rolling down from the point $\tilde{u}_{\mathrm{un}}$ 
with infinitely large velocity, and approaches the right maximum with 
vanishing velocity at a moment. Then, the particle climbs the hill slowly, 
and returns to the right maximum of the inverted potential at time $x=1$. 
\end{remark}

Theorem~\ref{theorem2_review} is useful for plotting the stationary 
solutions to (\ref{phenomenological_system}). 
Intuitively, the non-uniform solution $\tilde{u}_{\mathrm{s}}(x)$ that 
converges to $\tilde{u}_{\mathrm{l}}$ for all $x\in(-1,1)$ in the limit 
$\gamma\to0$ represents the situation under which the state $\tilde{u}(x,t)$ 
is trapped around the left stable solution $\tilde{u}_{\mathrm{l}}$ as 
$t\to\infty$. We conjecture that $\tilde{u}_{\mathrm{un}}(x)$ is unstable, 
although we could not prove the instability. 


\section{Numerical Results} \label{sec6} 
\subsection{Density-Evolution Analysis} \label{sec6-1}
\begin{figure}[t]
\begin{center}
\includegraphics[width=0.5\hsize]{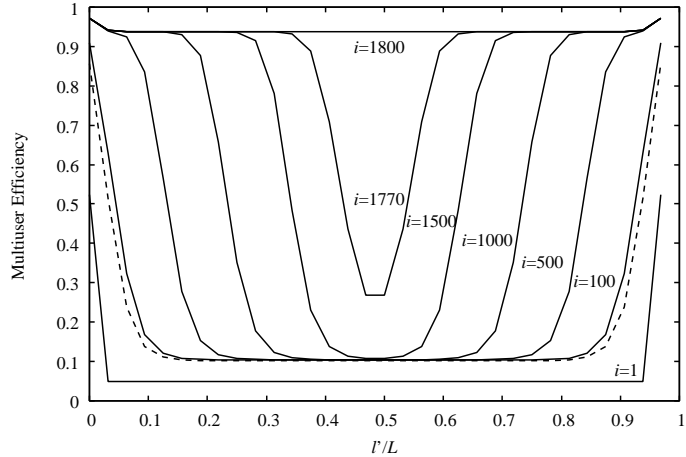}
\end{center}
\caption{
Multiuser efficiency versus $l'/L$ for $1/\sigma_{\mathrm{n}}^{2}=10$~dB, 
$L=32$, $W=1$, and $\beta_{\mathrm{init}}=0$. The solid lines denote the MEs for 
$\beta=1.97$. The dashed line shows the ME for $\beta=1.99$ and $i=10^{5}$. 
}
\label{fig4} 
\end{figure}

The DE equations~(\ref{SIR}) and (\ref{sigma_evolution}) are numerically 
solved to estimate the position of the BP threshold for the SC-SCDMA system. 
We focus on the ME 
$\eta_{l'}^{(i)}=\sigma_{\mathrm{n}}^{2}\mathrm{sir}_{l'}^{(i)}$ in 
iteration~$i$. Since the SIR $\mathrm{sir}_{l'}^{(i)}$ must be 
smaller than the SNR $1/\sigma_{\mathrm{n}}^{2}$, the ME takes a value between 
$0$ and $1$. 
Figure~\ref{fig4} shows the ME $\eta_{l'}^{(i)}$ for $\beta=1.97$ and 
$\beta=1.99$. The BP threshold $\beta_{\mathrm{BP}}^{(\mathrm{SC})}$ for the 
SC-SCDMA system based on Theorem~\ref{theorem4} and Remark~\ref{remark2} is 
given by $\beta_{\mathrm{BP}}^{(\mathrm{SC})}\approx1.982\,67$ for 
$1/\sigma_{\mathrm{n}}^{2}=10$~dB.   
For $\beta=1.97<\beta_{\mathrm{BP}}^{(\mathrm{SC})}$, the BP receiver first 
obtains reliable information about the data symbols at the boundaries $l'/L=0$ 
and $l'/L=1-1/L$, transmitted in the initialization phase. Then, 
the reliable information propagates toward the middle position $l'/L=1/2$ 
as $i$ increases. Eventually, the ME tends toward an almost uniform solution, 
which is close to $1$ for all positions $l'$. This result implies that the BP 
receiver can eliminate the MAI for $\beta=1.97$. 
For $\beta=1.99>\beta_{\mathrm{BP}}^{(\mathrm{SC})}$, on the 
other hand, the ME tends to a non-uniform solution after many iterations: The 
ME for $i=10^{5}$ is close to $0$ around the center $l'/L=1/2$, whereas it is 
close to $1$ near the boundaries $l'/L=0$ and $l'/L=1-1/L$. This observation 
implies that the system is interference-limited for $\beta=1.99$.  

\begin{figure}[t]
\begin{center}
\includegraphics[width=0.5\hsize]{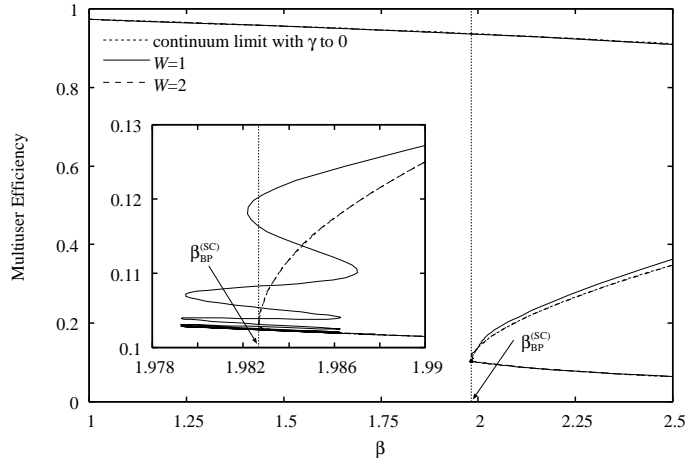}
\end{center}
\caption{
Multiuser efficiency at $l'/L=1/2$ versus $\beta$ for 
$1/\sigma_{\mathrm{n}}^{2}=10$~dB, $L=32$, and $\beta_{\mathrm{init}}=0$.  
}
\label{fig5} 
\end{figure}

In order to investigate the convergence speed of the continuum limit, 
we focus on the fixed-points to the DE 
equations~(\ref{SIR}) and (\ref{sigma_evolution}). 
See \cite{Kudekar111} for how to find the fixed-points to 
the DE equations~(\ref{SIR}) and (\ref{sigma_evolution}). 
Figure~\ref{fig5} shows the ME at the center $l'/L=1/2$, 
given via the fixed-points to (\ref{SIR}) and 
(\ref{sigma_evolution}). The MEs for $W=1$ and $W=2$ are 
represented by the solid and dashed lines, respectively. The dotted line 
shows the ME in the limit $L,W\to\infty$ with 
$\gamma=W/L\to0$, based on Theorem~\ref{theorem2_review}.  
The asymptotic\footnote{
Note that the term ``asymptotic'' in Section~\ref{sec6-1} implies the limit 
$\gamma\to0$ after taking the continuum limit, whereas we have so far used 
the term to mean the large-sparse-system limit.} 
ME is indistinguishable from that for $W=2$. 
In the limit $L, W\to\infty$ with $\gamma=W/L\to0$ 
the DE equations~(\ref{SIR}) and (\ref{sigma_evolution}) have the 
unique fixed-point when $\beta$ is smaller than 
$\beta_{\mathrm{BP}}^{(\mathrm{SC})}$, which is shown by the vertical line, 
whereas there are multiple fixed-points for 
$\beta>\beta_{\mathrm{BP}}^{(\mathrm{SC})}$. The ME for 
$W=1$ displays oscillating behavior around the BP threshold 
$\beta_{\mathrm{BP}}^{(\mathrm{SC})}$, as shown in the inset of Fig.~\ref{fig5}.  
The same phenomenon was observed in SC-LDPC codes~\cite{Kudekar111}. 
The wiggle decreases slightly the maximum of $\beta$ at which the DE 
equations~(\ref{SIR}) and (\ref{sigma_evolution}) for $W=1$ have a unique 
fixed-point. Since the amplitude of the wiggle 
decreases rapidly with the increase of $W$, the ME  
for $W=2$ is indistinguishable from the asymptotic one shown by 
the dotted line, except for a neighborhood of 
$\beta=\beta_{\mathrm{BP}}^{(\mathrm{SC})}$. This oscillating  
behavior around $\beta_{\mathrm{BP}}^{(\mathrm{SC})}$ seems 
to disappear in the limit $L,W\to\infty$ with $\gamma=W/L\to0$. 
These observations imply that the convergence to the asymptotic ME is so fast 
that the asymptotic result can provide good approximations for the SC-SCDMA 
systems with finite $L$ and $W$.  

Tables~\ref{table1} and~\ref{table2} list the BP thresholds for the 
SC-SCDMA system for SNRs $1/\sigma_{\mathrm{n}}^{2}=10$~dB and 
$1/\sigma_{\mathrm{n}}^{2}=12$~dB, respectively. The thresholds were estimated 
by solving the DE  equations~(\ref{SIR}) and (\ref{sigma_evolution}) 
numerically. Note that the listed values are not the average system load 
$\bar{\beta}$ but the system load $\beta$ in the communication phase. 
We find that the thresholds for $L=16$ are larger than the BP threshold 
$\beta_{\mathrm{BP}}^{(\mathrm{SC})}\approx 1.982\,67$ 
in the limit $L,W\to\infty$ with $\gamma=W/L\to0$, 
except for $W=1$. This observation is due to strictly positive $\gamma$: 
When $\beta$ is slightly larger than $\beta_{\mathrm{BP}}^{(\mathrm{SC})}$, 
as noted in Remark~\ref{remark_review}, the partial differential 
equation~(\ref{phenomenological_system}) for $\gamma>0$ may converge to an 
almost uniform solution, like the solution for $i=1800$ in Fig.~\ref{fig4}. 

\begin{table}[t]
\begin{center}
\caption{BP thresholds for the SC-SCDMA systems.   
$1/\sigma_{\mathrm{n}}^{2}=10$~dB and $\beta_{\mathrm{init}}=1$. 
$\beta_{\mathrm{BP}}$ and $\beta_{\mathrm{BP}}^{(\mathrm{SC})}$ are approximately 
given by $\beta_{\mathrm{BP}}\approx1.730\,78$ and 
$\beta_{\mathrm{BP}}^{(\mathrm{SC})}\approx1.982\,67$, respectively.}
\label{table1} 
\begin{tabular}{|c|c|c|c|c|c|}
\hline
\multicolumn{2}{|c|}{} & \multicolumn{4}{|c|}{$W$} \\ 
\cline{3-6} 
\multicolumn{2}{|c|}{} & 1 & 2 & 3 & 4 \\ 
\hline 
 & 16 & 1.97947 & 1.99150 & 2.04385 & 2.16470 \\ 
\cline{2-6} 
$L$ & 32 & 1.97925 & 1.98266 & 1.98321 & 1.98665 \\ 
\cline{2-6} 
 & 64 & 1.97925 & 1.98264 & 1.98267 & 1.98267 \\ 
\cline{2-6} 
 & 128 & 1.97925 & 1.98264 & 1.98267 & 1.98267 \\ 
\hline
\end{tabular}
\end{center}
\end{table}
\begin{table}[t]
\begin{center}
\caption{BP thresholds for the SC-SCDMA systems.   
$1/\sigma_{\mathrm{n}}^{2}=12$~dB and $\beta_{\mathrm{init}}=1$. 
$\beta_{\mathrm{BP}}$ and $\beta_{\mathrm{BP}}^{(\mathrm{SC})}$ are approximately 
given by $\beta_{\mathrm{BP}}\approx1.873\,44$ and 
$\beta_{\mathrm{BP}}^{(\mathrm{SC})}\approx2.507\,16$, respectively.}
\label{table2} 
\begin{tabular}{|c|c|c|c|c|c|}
\hline
\multicolumn{2}{|c|}{} & \multicolumn{4}{|c|}{$W$} \\ 
\cline{3-6} 
\multicolumn{2}{|c|}{} & 1 & 2 & 3 & 4 \\ 
\hline 
 & 16 & 2.38479 & 2.49386 & 2.53057 & 2.65917 \\ 
\cline{2-6} 
$L$ & 32 & 2.38479 & 2.49314 & 2.50589 & 2.50726 \\ 
\cline{2-6} 
 & 64 & 2.38479 & 2.49314 & 2.50588 & 2.50705 \\ 
\cline{2-6} 
 & 128 & 2.38479 & 2.49314 & 2.50588 & 2.50705 \\ 
\hline
\end{tabular}
\end{center}
\end{table}

\begin{figure}[t]
\begin{center}
\includegraphics[width=0.5\hsize]{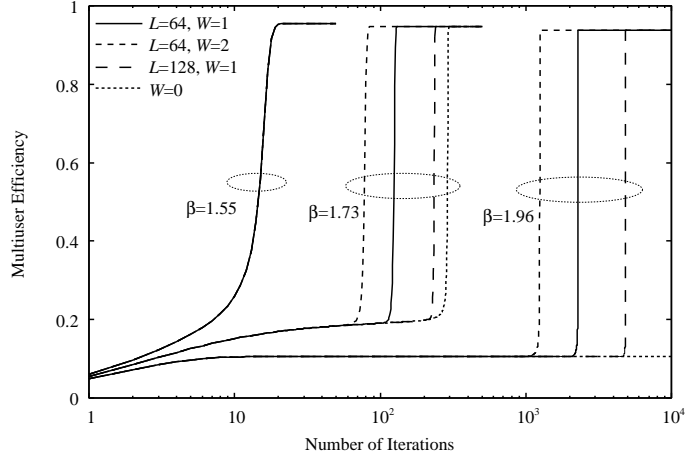}
\end{center}
\caption{
Multiuser efficiency at $l'/L=1/2$ versus the number of iterations for 
$1/\sigma_{\mathrm{n}}^{2}=10$~dB and $\beta_{\mathrm{init}}=0$. 
}
\label{fig6} 
\end{figure}

We have so far investigated the static properties of the DE 
equations~(\ref{SIR}) and (\ref{sigma_evolution}). We next consider 
the dynamic properties of the DE equations. Figure~\ref{fig6} shows 
the ME $\eta_{l'}^{(i)}$ at $l'/L=1/2$ as a 
function of the number of iterations $i$. 
The conventional BP threshold $\beta_{\mathrm{BP}}$ is approximately equal to 
$\beta_{\mathrm{BP}}\approx1.730\,78$ for SNR $1/\sigma_{\mathrm{n}}^{2}=10$~dB, while 
the BP threshold for the SC-SCDMA system is given by 
$\beta_{\mathrm{BP}}^{(\mathrm{SC})}\approx 1.982\,67$. 
The number of iterations required for convergence increases as $\beta$ grows. 
Interestingly, the SC-SCDMA systems converge to the stationary 
solutions more quickly than the uncoupled system $W=0$ for $\beta=1.73$, 
whereas all systems converge at the same speed 
for $\beta=1.55<\beta_{\mathrm{BP}}$. 
This observation is because the uncoupled system requires infinitely many  
iterations for convergence when $\beta$ tends to the BP threshold 
$\beta_{\mathrm{BP}}$ from below. The number of iterations required for 
$L=64$ and $W=2$ is roughly half the number of iterations for $L=64$ and 
$W=1$, while the one required for $L=128$ and $W=1$ is roughly twice. 
These results are consistent with the intuition that reliable information at 
the boundaries $l'=0$ and $l'=L-1$ should propagate toward the middle 
position $l'=L/2$ at a speed proportional to $\gamma=W/L$.  

\subsection{Numerical Simulations} 
\begin{figure}[t]
\begin{center}
\includegraphics[width=0.5\hsize]{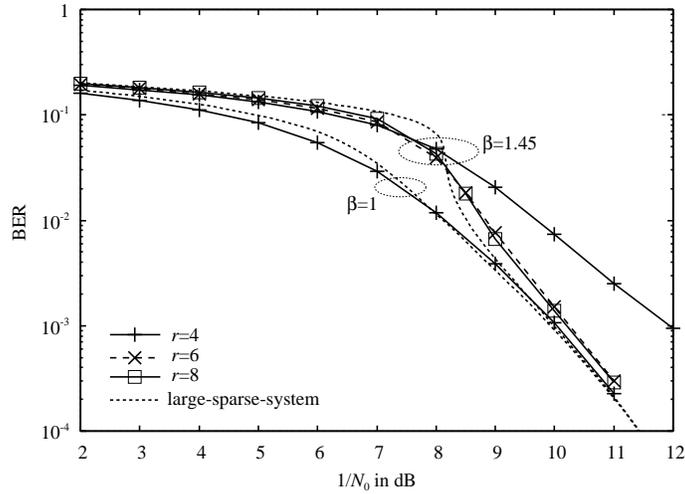}
\end{center}
\caption{
BER versus SNR for $K=1000$ and $W=0$. The number of iterations is equal to 
$40$. 
}
\label{fig7} 
\end{figure}

Numerical simulations of the BP receivers are presented. We first focus on  
uncoupled SCDMA systems, i.e.\ $W=0$. Figure~\ref{fig7} plots the BERs of the 
(exact) BP receiver. The results for $r=4,6,8$ are denoted by $\{+\}$, 
$\{\times\}$, and $\{\square\}$ connected with solid or dashed lines, 
respectively. The analytical results in 
the large-sparse-system limit are also shown by the dotted lines. When 
$\beta=1$, the large-sparse-system result can provide a good approximation of 
the BER for $r=4$. When $\beta=1.45$, however, there are gaps between the 
large-sparse-system result and the BERs for $r=4,6,8$ especially in the 
moderate-SNR regime. The large-sparse-system result provides a larger estimate 
than the actual BER in the low-SNR regime, whereas it predicts a smaller BER 
in the high-SNR regime. These observations imply that, as the system load 
grows, larger $r$ is required in order for the large-sparse-system 
result to provide good approximations for the actual BERs.  

\begin{figure}[t]
\begin{center}
\includegraphics[width=0.5\hsize]{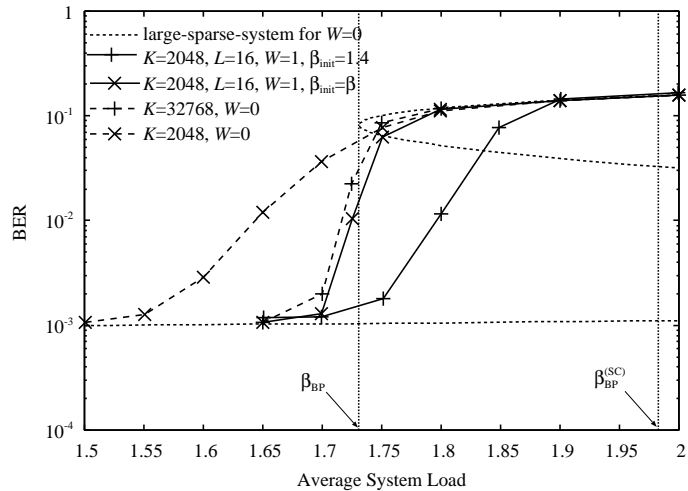}
\end{center}
\caption{
BER versus average system load for $1/\sigma_{\mathrm{n}}^{2}=10$~dB and 
$r=32$. The number of iterations is equal to $1000$. 
}
\label{fig9} 
\end{figure}

Figure~\ref{fig9} shows a comparison between the SCDMA systems with and 
without spatial coupling. We used the BP receiver with GA to reduce the 
computational complexity. The horizontal axis is the average system load 
given by (\ref{sum_rate}). The BER at the middle position $l'/L=1/2$ is 
plotted for the SC-SCDMA system. The BERs for 
the uncoupled SCDMA systems are denoted by $\{+\}$ or $\{\times\}$ connected 
with dashed lines. The BERs for the SC-SCDMA systems are represented by 
$\{+\}$ or $\{\times\}$ connected with solid lines. 
We find that the performance of the SC-SCDMA system with $K=2048$, $L=16$, 
$W=1$, and $\beta_{\mathrm{init}}=\beta$ is superior (resp.\ comparable) to that 
of the uncoupled SCDMA system with $2048$ (resp.\ $32768$) users. 
Note that the SC-SCDMA system performs the BP detection for every $L$ symbol 
periods, whereas the uncoupled SCDMA system does for every symbol period. 
However, this delay of detection does not necessarily result in the overall 
delay for coded systems. Since $L$ is commonly smaller than the code length, 
the overall delay is dominated by the decoding delay. Thus, the comparisons  
between the uncoupled SCDMA system with $2048$ users and the SC-SCDMA systems 
make sense for practical coded systems, although the detection delay for the 
SC-SCDMA systems is larger than for the uncoupled SCDMA system. 
These observations imply that the SCDMA system with 
one-dimensional coupling can accelerate the convergence speed toward the 
large-system limit, compared to the uncoupled SCDMA system with the same 
number of users.   
Furthermore, the SC-SCDMA system with $K=2048$, $L=16$, $W=1$, and 
$\beta_{\mathrm{init}}=1.4$ can provide a significant improvement in BER for 
high system loads, compared to the SC-SCDMA system with 
$\beta_{\mathrm{init}}=\beta$. The BERs for the SC-SCDMA system with 
$\beta_{\mathrm{init}}=1.4$ 
seem to be trapped in the top (bad) solution obtained from the 
large-sparse-system analysis, when the average system load 
is equal to $1.9$. This is due to finite $L$, $W$, $K$, and $N$: Substituting 
the BP threshold $\beta_{\mathrm{BP}}^{(\mathrm{SC})}\approx1.982\,67$ into 
$\beta$ in the average system load~(\ref{sum_rate}), we find that 
the corresponding average system load is approximately equal to $1.93$. The 
remaining gap $0.03$ seems to be due to finite $K$ and $N$. 

In order to achieve the improved BP threshold, $L$ must tend to infinity. 
Let us discuss the effect of increasing $L$ for finite $K$ and $N$.
If $K$ and $N$ were infinity, reliable information at the boundaries 
$l'=0$ and $l'=L-1$ could propagate to the adjacent positions successfully. 
For finite $K$ and $N$, however, it is probabilistic whether 
reliable information can propagate to the adjacent positions successfully. 
As $L$ increases, thus, it becomes difficult for reliable information to 
propagate to the middle position successfully. Increasing $K$ 
and $N$ results in a reduction of the probability with which the propagation 
of reliable information to the adjacent positions fails. 
These arguments imply that the system size required for achieving a BER 
close to the analytical one increases as the average system load gets 
closer to the improved BP threshold.  

\section{Conclusions} \label{sec7} 
The SC-SCDMA system has been proposed to improve the 
performance of iterative MUD based on BP. We have derived the two iterative 
receivers, one based on exact BP, and the other on approximate BP with GA. 
The two BP receivers can achieve the same performance in the 
large-sparse-system limit. The analysis of the DE equations 
for the two receivers implies that the BP threshold can be improved up 
to the IO threshold by spatial coupling. Numerical simulations imply that 
spatial coupling can provide a significant improvement in BER for a fixed 
finite-sized system especially in the region of high system loads, whereas 
a quite large system is required for approaching the IO threshold. 
  
We remark a capability of the phenomenological methodology for specifying the 
BP threshold for SC systems and a direction of future work to 
conclude this paper. The phenomenological result presented in 
Section~\ref{sec5} is applicable to characterizing the BP threshold for 
any SC system, if DE equations for the corresponding uncoupled 
system are described by one parameter and if DE equations for the SC system 
is included in the DE equations~(\ref{DE_v_dis}) and (\ref{DE_u_dis}). 
Of course, the presented method is not applicable to all SC systems. 
A further generalization of the phenomenological model is left as a future 
work.  


\appendices

\section{Proof of Theorem~\ref{theorem1}} 
\label{proof_theorem1} 
\subsection{Reparametrization}
In order to prove Theorem~\ref{theorem1}, we evaluate the evolution of the 
tentative marginal posterior probability~(\ref{BP}) in the large-sparse-system 
limit, following \cite{Guo08,Ikehara07}. The marginal posterior 
probability is a random variable on the space of probability distributions, 
because of the randomness of $\mathcal{Y}$ and $\boldsymbol{G}$.  
Since we have assumed BPSK, the marginal posterior probability~(\ref{BP}) can 
be represented with one parameter. Selecting the log likelihood ratio (LLR) as 
the parameter is suitable for proving Theorem~\ref{theorem1}. 
Evaluating the evolution of the tentative marginal posterior 
probability~(\ref{BP}) is equivalent to tracing the evolution of the pdf  
of the LLR.  

Let $L_{(n,l)\rightarrow(k,l')}^{(i)}$ denote the LLR for the 
message~(\ref{sum_step}) provided from the function node~$(n,l)$ to the 
variable node~$(k,l')$ in iteration~$i$, 
\begin{equation} \label{LLR} 
L_{(n,l)\rightarrow(k,l')}^{(i)} = \ln\frac{
 q_{n,l}^{(i)}(b_{k,l'}=1)
}
{
 q_{n,l}^{(i)}(b_{k,l'}=-1)
}. 
\end{equation} 
Furthermore, we write the LLR for the message~(\ref{product_step}) 
propagating along the same edge in the opposite direction as 
\begin{equation}
L_{(k,l')\rightarrow(n,l)}^{(i)} = \ln\frac{
 m_{n,l}^{(i)}(b_{k,l'}=1)
}
{
 m_{n,l}^{(i)}(b_{k,l'}=-1)
}. 
\end{equation}
The product step~(\ref{product_step}) can be represented as follows: 
\begin{equation} \label{target_LLR} 
L_{(k,l')\rightarrow(n,l)}^{(i)} 
= \sum_{(\tilde{n},\tilde{l})\in\partial(k,l')
\backslash(n,l)}L_{(\tilde{n},\tilde{l})\rightarrow(k,l')}^{(i)}.  
\end{equation}
The ACF property of the $(r,L,W)$-ensemble presented in Example~\ref{example3} 
guarantees that the incoming LLRs 
$\{L_{(\tilde{n},\tilde{l})\rightarrow(k,l')}^{(i)}\}$ are independent random 
variables in the large-system limit. Furthermore, the central limit theorem 
implies that the LLR~(\ref{target_LLR}) converges in law to a Gaussian 
random variable in the large-sparse-system limit, i.e.\ in the dense limit 
after taking the large-system limit. Thus, it is sufficient to evaluate 
the mean and variance of the LLR~(\ref{target_LLR}) conditioned 
on the data symbols $\{b_{k,l'}\}$ in the large-sparse-system limit. 
In the proof of Theorem~\ref{theorem1}, we always fix the data symbols and 
omit conditioning with respect to the data symbols. 

\subsection{Density Evolution} 
\subsubsection{Mean and Variance of (\ref{target_LLR})} 
In order to calculate the mean and variance of the LLR~(\ref{target_LLR}), 
we first define and calculate several quantities. 
Let us define $f_{j}(y)$ and $\tilde{f}_{j}^{(i)}(y)$ as 
\begin{equation} \label{f} 
f_{j}(y) = \left(
 \frac{y - I_{n,l,k,l'}}{\sigma_{\mathrm{n}}^{2}}
\right)^{j}g(y-I_{n,l,k,l'};\sigma_{\mathrm{n}}^{2}) 
\quad \hbox{for $j=0,1$,}  
\end{equation} 
\begin{IEEEeqnarray}{rl}
&\tilde{f}_{j}^{(i)}(y)
= \sum_{\{\tilde{b}_{\tilde{k},\tilde{l}'}^{(i)}\}}\left\{
 \left(
  \frac{y - \tilde{I}_{n,l,k,l'}^{(i)}}{\sigma_{\mathrm{n}}^{2}}
 \right)^{j} - \frac{\delta_{j,2}}{\sigma_{\mathrm{n}}^{2}}
\right\}\nonumber \\  
&\cdot g(y-\tilde{I}_{n,l,k,l'}^{(i)};\sigma_{\mathrm{n}}^{2}) 
\prod_{(\tilde{k},\tilde{l}')\in\partial(n,l)\backslash(k,l')}
m_{n,l}^{(i)}(\tilde{b}_{\tilde{k},\tilde{l}'}^{(i)}), \label{tilde_f}
\end{IEEEeqnarray}
for $j=0,1,2$. In (\ref{f}) and (\ref{tilde_f}), 
$\tilde{I}_{n,l,k,l'}^{(i)}$ and $I_{n,l,k,l'}$ are given by 
(\ref{postulated_interference}) and 
\begin{equation}
I_{n,l,k,l'} =
\sum_{(\tilde{k},\tilde{l}')\in\partial(n,l)\backslash(k,l')}
\frac{s_{n,l,\tilde{k},\tilde{l}'}b_{\tilde{k},\tilde{l}'}}
{\sqrt{(W+1)\bar{c}_{l,\tilde{k},\tilde{l}'}}},   
\end{equation}
respectively. Furthermore, $g(x;\sigma^{2})$ denotes the 
pdf~(\ref{Gauss_function}) for a zero-mean Gaussian random variable with 
variance $\sigma^{2}$. The two quantities~(\ref{f}) and~(\ref{tilde_f}) are 
used to Taylor-expand the RHS of (\ref{sum_step}) in the large-system limit. 

\begin{lemma} \label{lemma1_appen} 
Suppose that (\ref{large_spreading_matrix}) is picked up from the 
$(r,L,W)$-ensemble, presented in Example~\ref{example3}. Then, 
\begin{equation} \label{term1} 
\overline{\int_{-\infty}^{\infty}
\frac{\tilde{f}_{1}^{(i-1)}(y)}{\tilde{f}_{0}^{(i-1)}(y)}f_{1}(y)dy} 
\to \frac{1}{\tilde{\sigma}_{l}^{2}(i)},  
\end{equation}
\begin{equation} \label{term2}
\overline{\int_{-\infty}^{\infty}
\left(
 \frac{\tilde{f}_{1}^{(i-1)}(y)}{\tilde{f}_{0}^{(i-1)}(y)}
\right)^{2}f_{0}(y)dy} 
\to \frac{\sigma_{l}^{2}(i)}{\tilde{\sigma}_{l}^{4}(i)},  
\end{equation}
in the large-sparse-system limit, where the overlines in (\ref{term1}) and 
(\ref{term2}) represent the expectation with respect to 
(\ref{large_spreading_matrix}). 
In the RHSs of (\ref{term1}) and (\ref{term2}), $\sigma_{l}^{2}(i)$ and 
$\tilde{\sigma}_{l}^{2}(i)$ are respectively given by  
\begin{equation} \label{sigma} 
\sigma_{l}^{2}(i) = \sigma_{\mathrm{n}}^{2} 
+ \frac{\beta_{l}}{W+1}\sum_{l'=0}^{W}\xi_{(l-l')_{L}}^{(i-1)},
\end{equation}
\begin{equation} \label{tilde_sigma} 
\tilde{\sigma}_{l}^{2}(i) = \sigma_{\mathrm{n}}^{2} 
+ \frac{\beta_{l}}{W+1}\sum_{l'=0}^{W}\tilde{\xi}_{(l-l')_{L}}^{(i-1)}, 
\end{equation}
with    
\begin{equation} \label{xi} 
\xi_{l'}^{(i)} = \mathbb{E}\left[
 \left(
  b_{1,l'} - \mathbb{E}[\tilde{b}_{1,l'}^{(i)}]
 \right)^{2}
\right], 
\end{equation}
\begin{equation} \label{tilde_xi} 
\tilde{\xi}_{l'}^{(i)} = \mathbb{E}\left[ 
 \left(
  \tilde{b}_{1,l'}^{(i)}
  -\mathbb{E}[\tilde{b}_{1,l'}^{(i)}]
 \right)^{2}
\right]. 
\end{equation}
In (\ref{sigma}) and (\ref{tilde_sigma}), $\beta_{l}=K/N_{l}$ is equal to 
$\beta_{\mathrm{init}}$ for $l=0,\ldots,W-1$ and to $\beta$  
for $l=W,\ldots,L-1$, respectively. 
\end{lemma}
\begin{IEEEproof}[Proof of Lemma~\ref{lemma1_appen}] 
As we have noted in Section~\ref{sec_GA}, the central limit theorem implies 
that the postulated interference~(\ref{postulated_interference}) converges in 
law to a Gaussian random variable in the dense limit. The mean and 
variance of (\ref{postulated_interference}) are given by (\ref{mean}) and 
(\ref{variance}), respectively. 
Since (\ref{tilde_f}) depends on $\{\tilde{b}_{\tilde{k},\tilde{l}'}^{(i)}\}$ 
only through the postulated interference~(\ref{postulated_interference}), we 
calculate the marginalization in (\ref{tilde_f}) as the expectation with 
respect to the postulated interference to obtain  
\begin{equation}
\tilde{f}_{0}^{(i-1)}(y) = g(y-\tilde{\mu}_{n,l,k,l'}^{(i-1)};\sigma_{\mathrm{n}}^{2}
+\tilde{v}_{n,l,k,l'}^{(i-1)}), 
\end{equation}
\begin{equation}
\tilde{f}_{1}^{(i-1)}(y) 
= \frac{y-\tilde{\mu}_{n,l,k,l'}^{(i-1)}}{\sigma_{\mathrm{n}}^{2}+\tilde{v}_{n,l,k,l'}^{(i-1)}} 
g(y-\tilde{\mu}_{n,l,k,l'}^{(i-1)};\sigma_{\mathrm{n}}^{2}
+\tilde{v}_{n,l,k,l'}^{(i-1)}), 
\end{equation}
where $\tilde{\mu}_{n,l,k,l'}^{(i)}$ and $\tilde{v}_{n,l,k,l'}^{(i)}$ are 
given by (\ref{mean}) and (\ref{variance}), respectively.   
Substituting these expressions into (\ref{term1}) and (\ref{term2}) and  
performing the integrations with respect to $y$, we have 
\begin{equation} \label{term1_tmp1}
\int_{-\infty}^{\infty}
\frac{\tilde{f}_{1}^{(i-1)}(y)}{\tilde{f}_{0}^{(i-1)}(y)}f_{1}(y)dy 
\to \frac{1}{\sigma_{\mathrm{n}}^{2}+\tilde{v}_{n,l,k,l'}^{(i-1)}},  
\end{equation}
\begin{equation}
\int_{-\infty}^{\infty}
\left(
 \frac{\tilde{f}_{1}^{(i-1)}(y)}{\tilde{f}_{0}^{(i-1)}(y)}
\right)^{2}f_{0}(y)dy
\to \frac{\sigma_{\mathrm{n}}^{2}+(I_{n,l,k,l'}-\tilde{\mu}_{n,l,k,l'}^{(i-1)})^{2}}
{(\sigma_{\mathrm{n}}^{2}+\tilde{v}_{n,l,k,l'}^{(i-1)})^{2}},  
\end{equation}
in the large-sparse-system limit. 

In order to complete the proof, we take the expectation with respect to 
(\ref{large_spreading_matrix}), picked up from the 
$(r,L,W)$-ensemble. The weak law of large numbers implies that 
the variance~(\ref{variance}) converges in probability to the second term on 
the RHS of (\ref{tilde_sigma}) in the large-sparse-system limit. 
Since the RHS of (\ref{term1_tmp1}) is bounded, this implies 
that (\ref{term1}) holds. Next, the average of 
$(I_{n,l,k,l'}-\tilde{\mu}_{n,l,k,l'}^{(i-1)})^{2}$ over 
(\ref{large_spreading_matrix}) is equal to 
\begin{equation} \label{numerator}
\overline{
(I_{n,l,k,l'}-\tilde{\mu}_{n,l,k,l'}^{(i-1)})^{2}}
= \sum_{(\tilde{k},\tilde{l}')\in\partial(n,l)\backslash(k,l')}
\frac{
 (b_{\tilde{k},\tilde{l}'}- \mathbb{E}[\tilde{b}_{\tilde{k},\tilde{l}'}^{(i-1)}])^{2}
}{(W+1)\bar{c}_{l,\tilde{k},\tilde{l}'}},   
\end{equation}
which converges in probability to the second term on the RHS of 
(\ref{sigma}) in the large-sparse-system limit. 
This observation implies that (\ref{term2}) holds.    
\end{IEEEproof}

We shall evaluate the mean and variance of the 
LLR~(\ref{target_LLR}) in the large-sparse-system limit. For that purpose, 
we use Lemma~\ref{lemma1_appen} to calculate (\ref{LLR}) up to $O(r^{-1})$. 
Expanding the RHS of (\ref{sum_step}) with respect to 
$s_{n,l,k,l'}b_{k,l'}/\sqrt{(W+1)\bar{c}_{l,k,l'}}$ up to the second order yields  
\begin{IEEEeqnarray}{rl}
q_{n,l}^{(i)}(b_{k,l'}) =& \tilde{f}_{0}^{(i-1)}(y_{n,l}) 
+ \tilde{f}_{1}^{(i-1)}(y_{n,l})
\frac{s_{n,l,k,l'}b_{k,l'}}{\sqrt{(W+1)\bar{c}_{l,k,l'}}} \nonumber \\ 
+& \frac{\tilde{f}_{2}^{(i-1)}(y_{n,l})}{2} 
\frac{s_{n,l,k,l'}^{2}b_{k,l'}^{2}}{(W+1)\bar{c}_{l,k,l'}} + O(r^{-3/2}), 
\label{LLR_tmp} 
\end{IEEEeqnarray}
where $\tilde{f}_{j}^{(i)}(y)$ is given by (\ref{tilde_f}). Note that 
$\{b_{\tilde{k},\tilde{l}'}: (\tilde{k},\tilde{l}')\in\partial(n,l)
\backslash(k,l')\}$ in (\ref{sum_step}) are dummy variables, so that we have 
replaced them by $\{\tilde{b}_{\tilde{k},\tilde{l}'}^{(i)}\}$ to obtain 
(\ref{LLR_tmp}). Substituting (\ref{LLR_tmp}) into (\ref{LLR}) and expanding 
the obtained formula, we have 
\begin{equation} \label{LLR_tmp1} 
L_{(n,l)\rightarrow(k,l')}^{(i)} 
= \frac{\tilde{f}_{1}^{(i-1)}(y_{n,l})}{\tilde{f}_{0}^{(i-1)}(y_{n,l})} 
\frac{2s_{n,l,k,l'}}{\sqrt{(W+1)\bar{c}_{l,k,l'}}} + O(r^{-3/2}). 
\end{equation}
We note that no term proportional to $r^{-1}$ appears under the BPSK 
assumption, whereas the term does for general data 
symbols~\cite{Guo08,Ikehara07}. 

In order to calculate the mean and variance of (\ref{LLR_tmp1}), 
we use the following expansion: 
\begin{IEEEeqnarray}{rl}
&p\left(
 y_{n,l}\left|
  \frac{s_{n,l,k,l'}b_{k,l'}}{\sqrt{(W+1)\bar{c}_{l,k,l'}}}
  + I_{n,l,k,l'}
 \right.
\right) \nonumber \\ 
=& f_{0}(y_{n,l}) + f_{1}(y_{n,l})\frac{s_{n,l,k,l'}b_{k,l'}}
{\sqrt{(W+1)\bar{c}_{l,k,l'}}} + O(r^{-1}). 
\end{IEEEeqnarray}
Evaluating the mean of (\ref{LLR_tmp1}) with this expression, we obtain 
\begin{IEEEeqnarray}{rl}
&\mathbb{E}[L_{(n,l)\rightarrow(k,l')}^{(i)}] \nonumber \\ 
=& \overline{\int_{-\infty}^{\infty}
\frac{\tilde{f}_{1}^{(i-1)}(y_{n,l})}{\tilde{f}_{0}^{(i-1)}(y_{n,l})}
f_{1}(y_{n,l})dy_{n,l}}\frac{2b_{k,l'}}{(W+1)\bar{c}_{l,k,l'}} 
+ O(r^{-3/2}) \nonumber \\ 
=& \frac{2}{(W+1)\bar{c}_{l,k,l'}\tilde{\sigma}_{l}^{2}(i)}b_{k,l'} 
+ O(r^{-3/2}),  \label{mean_LLR}  
\end{IEEEeqnarray}
where we have used (\ref{term1}) in the derivation of the last 
equality. Similarly, we use (\ref{term2}) to calculate the variance of 
(\ref{LLR_tmp1}) as 
\begin{equation} \label{var_LLR}
\mathbb{V}[L_{(n,l)\rightarrow(k,l')}^{(i)}] 
= \frac{4\sigma_{l}^{2}(i)}{(W+1)\bar{c}_{l,k,l'}\tilde{\sigma}_{l}^{4}(i)}
+ O(r^{-3/2}). 
\end{equation}
The two expressions~(\ref{mean_LLR}) and (\ref{var_LLR}) imply that the 
mean and variance of the LLR~(\ref{target_LLR}) converge to 
\begin{equation} \label{target_mean_LLR} 
\mathbb{E}[L_{(k,l')\rightarrow(n,l)}^{(i)}]
\to 2\widetilde{\mathrm{sir}}_{l'}^{(i)}b_{k,l'},  
\end{equation}
\begin{equation} \label{target_variance_LLR} 
\mathbb{V}[L_{(k,l')\rightarrow(n,l)}^{(i)}] 
\to 4\mathrm{sir}_{l'}^{(i)}, 
\end{equation}
in the large-sparse-system limit, respectively, with 
\begin{equation} \label{sir} 
\mathrm{sir}_{l'}^{(i)} 
= \frac{1}{W+1}\sum_{\tilde{l}=0}^{W}
\frac{\sigma_{(\tilde{l}+l')_{L}}^{2}(i)}
{\tilde{\sigma}_{(\tilde{l}+l')_{L}}^{4}(i)}, 
\end{equation}
\begin{equation} \label{tilde_sir} 
\widetilde{\mathrm{sir}}_{l'}^{(i)} = \frac{1}{W+1}\sum_{\tilde{l}=0}^{W}
\frac{1}{\tilde{\sigma}_{(\tilde{l}+l')_{L}}^{2}(i)}, 
\end{equation}
where $\sigma_{l}^{2}(i)$ and $\tilde{\sigma}_{l}^{2}(i)$ are given by 
(\ref{sigma}) and (\ref{tilde_sigma}), respectively. In the derivation of 
(\ref{sir}) and (\ref{tilde_sir}), we have used the assumption that 
(\ref{large_spreading_matrix}) is picked up from the $(r,L,W)$-ensemble 
presented in Example~\ref{example3}. 

In summary, the LLR~(\ref{target_LLR}) converges in law to a Gaussian random 
variable with mean~(\ref{target_mean_LLR}) and 
variance~(\ref{target_variance_LLR}) in the large-sparse-system limit. 

\subsubsection{Decoupling} 
We shall present the asymptotic expression of the equivalent channel 
$p(b_{k,l'}^{(i)}|b_{k,l'})$ given by (\ref{equivalent_channel}). 
Let $z_{k,l'}^{(i)}$ denote a Gaussian 
random variable with mean $b_{k,l'}$ and variance 
$\mathrm{sir}_{l'}^{(i)}/(\widetilde{\mathrm{sir}}_{l'}^{(i)})^{2}$. 
The LLR of the tentative marginal posterior probability~(\ref{BP}) is 
statistically equivalent to the LLR~(\ref{target_LLR}) or 
$2\widetilde{\mathrm{sir}}_{l'}^{(i)}z_{k,l'}^{(i)}$ 
in the large-sparse-system limit.   
Since the pmf $p(x)$ for the BPSK random variable $x\in\{1,-1\}$ is 
proportional to $\exp(Lx/2)$, with $L=\ln\{p(x=1)/p(x=-1)\}$, we obtain 
\begin{IEEEeqnarray}{rl}
&p(b_{k,l'}^{(i)}|b_{k,l'}) \nonumber \\ 
\propto& \mathbb{E}_{z_{k,l'}^{(i)}}\left[
 \exp\left(
  \widetilde{\mathrm{sir}}_{l'}^{(i)}z_{k,l'}^{(i)}b_{k,l'}^{(i)} 
 \right)
\right] \nonumber \\ 
\propto& \int p(b_{k,l'}^{(i)} | z_{k,l'}^{(i)})
g\left(
 z_{k,l'}^{(i)}-b_{k,l'}; \frac{\mathrm{sir}_{l'}^{(i)}}
 {(\widetilde{\mathrm{sir}}_{l'}^{(i)})^{2}}
\right)dz_{k,l'}^{(i)}, \label{asymptotic_equivalent_channel_appen} 
\end{IEEEeqnarray}
in the large-sparse-system limit, where the posterior probability 
$p(b_{k,l'}^{(i)} | z_{k,l'}^{(i)})$ is given by 
\begin{equation} \label{posterior_probability} 
p(b_{k,l'}^{(i)} | z_{k,l'}^{(i)}) 
= \frac{
 g(z_{k,l'}^{(i)}-b_{k,l'}^{(i)}; 
 (\widetilde{\mathrm{sir}}_{l'}^{(i)})^{-1})
}
{
 p(z_{k,l'}^{(i)};(\widetilde{\mathrm{sir}}_{l'}^{(i)})^{-1}) 
}, 
\end{equation}
with 
\begin{equation}
p(z_{k,l'}^{(i)};(\widetilde{\mathrm{sir}}_{l'}^{(i)})^{-1}) 
= \sum_{b_{k,l'}^{(i)}=\pm1}
 g(z_{k,l'}^{(i)}-b_{k,l'}^{(i)}; 
 (\widetilde{\mathrm{sir}}_{l'}^{(i)})^{-1}). 
\end{equation}

In order to prove that the equivalent 
channel~(\ref{asymptotic_equivalent_channel_appen}) is equal to 
(\ref{asymptotic_equivalent_channel}), we show that  
$\mathrm{sir}_{l'}^{(i)}$ is given by (\ref{SIR}) and that 
$\widetilde{\mathrm{sir}}_{l'}^{(i)}$ is equal to $\mathrm{sir}_{l'}^{(i)}$. 
Since the LLR of the tentative marginal posterior probability~(\ref{BP}) is 
statistically equivalent to the LLR~(\ref{target_LLR}) in the 
large-sparse-system limit, the MSE~(\ref{xi}) and the posterior 
variance~(\ref{tilde_xi}) are equal to 
\begin{equation} \label{asymptotic_xi} 
\xi_{l'}^{(i)} = \mathbb{E}\left[
 \left(
  b_{1,l'} - \langle b_{1,l'}^{(i)} \rangle
 \right)^{2}
\right], 
\end{equation}
\begin{equation} \label{asymptotic_tilde_xi} 
\tilde{\xi}_{l'}^{(i)} = \mathbb{E}\left[ 
 \left(
  b_{1,l'}^{(i)}
  -\langle b_{1,l'}^{(i)} \rangle
 \right)^{2}
\right], 
\end{equation}
where the posterior mean $\langle b_{k,l'}^{(i)} \rangle$ is given by 
\begin{equation}
\langle b_{k,l'}^{(i)} \rangle 
= \sum_{b_{k,l'}^{(i)}=\pm 1}b_{k,l'}^{(i)}
p(b_{k,l'}^{(i)} | z_{k,l'}^{(i)}).  
\end{equation} 
Expressions~(\ref{sigma}), (\ref{tilde_sigma}), (\ref{sir}), 
(\ref{tilde_sir}), (\ref{asymptotic_xi}),  and (\ref{asymptotic_tilde_xi}) 
provide DE equations with respect to $\mathrm{sir}_{l'}^{(i)}$ and 
$\widetilde{\mathrm{sir}}_{l'}^{(i)}$. The initial condition 
$m_{n,l}^{(0)}(b_{k,l'})=1/2$ implies 
$\xi_{l'}^{(0)}=\tilde{\xi}_{l'}^{(0)}=1$, because of 
$\mathbb{E}[\tilde{b}_{k,l'}^{(0)}]=0$.

Let us prove that the DE equations reduce to those presented in 
Theorem~\ref{theorem1} by induction. When $i=1$, (\ref{sigma}) and 
(\ref{tilde_sigma}) implies $\sigma_{l}^{2}(1)=\tilde{\sigma}_{l}^{2}(1)$,  
given by (\ref{sigma_evolution}) with $i=1$. Furthermore, 
$\mathrm{sir}_{l'}^{(1)}=\widetilde{\mathrm{sir}}_{l'}^{(1)}$, given by 
(\ref{SIR}), holds from (\ref{sir}) and (\ref{tilde_sir}). Next, 
suppose that the statement holds for iteration~$i$. Since 
$\mathrm{sir}_{l'}^{(i)}=\widetilde{\mathrm{sir}}_{l'}^{(i)}$, the posterior 
probability~(\ref{posterior_probability}) is equal to that for the scalar 
AWGN channel~(\ref{AWGN}). Thus, the MSE~(\ref{asymptotic_xi}) and 
the posterior variance~(\ref{asymptotic_tilde_xi}) coincide with each other 
for iteration~$i$. This observation implies $\mathrm{sir}_{l'}^{(i+1)}
=\widetilde{\mathrm{sir}}_{l'}^{(i+1)}$, given by (\ref{SIR}) for 
iteration~$i+1$.  

\section{Proof of Lemma~\ref{lemma_stability_bulk}}
\label{deriv_gradient_system} 
Let $\hat{u}(\tilde{x},t)=u(\gamma \tilde{x},t)$ for 
$\tilde{x}\in[-\gamma^{-1},\gamma^{-1}]$. 
We first show that 
the partial differential equation~(\ref{single_system}) is represented as 
\begin{equation} \label{gradient_system_bulk} 
\frac{\partial \hat{u}}{\partial t} 
= - A(\hat{u}(\tilde{x},t))\frac{\delta H}{\delta\hat{u}}[\hat{u}(\cdot,t)]
(\tilde{x}), 
\end{equation}
with some energy functional $H$. In (\ref{gradient_system_bulk}), 
$\delta/\delta\hat{u}$ denotes the functional (Fr\'echet) derivative with 
respect to $\hat{u}$. Furthermore, the boundary condition 
$\hat{u}(\pm\gamma^{-1},t)=u_{\mathrm{max}}$ is imposed. 

Let us define the energy functional $H$ as 
\begin{equation} \label{energy_functional_bulk}
H[\hat{u}] 
= \int_{-\gamma^{-1}}^{\gamma^{-1}}\left[
 V(\hat{u}(\tilde{x})) + \frac{B(\hat{u}(\tilde{x}))}{2}\left(
  \frac{\partial\hat{u}}{\partial \tilde{x}}   
 \right)^{2}
\right]d\tilde{x}.
\end{equation}
For functions $\hat{u}(\tilde{x})$ and $w(\tilde{x})$, we expand the energy 
functional $H[\hat{u}+\epsilon w]$, given by (\ref{energy_functional_bulk}), 
around $\epsilon=0$ to obtain 
\begin{equation}
H[\hat{u}+\epsilon w]
= H[\hat{u}] 
+ \epsilon\frac{\partial H}{\partial\epsilon}[\hat{u}] + O(\epsilon^{2}), 
\end{equation} 
with 
\begin{IEEEeqnarray}{rl} 
\frac{\partial H}{\partial\epsilon}[\hat{u}] 
=& \int_{-\gamma^{-1}}^{\gamma^{-1}}\left[
 V'(\hat{u}(\tilde{x})) + \frac{B'(\hat{u}(\tilde{x}))}{2}\left(
  \frac{\partial \hat{u}}{\partial \tilde{x}}
 \right)^{2} 
\right]w(\tilde{x})d\tilde{x} \nonumber \\ 
&+ \int_{-\gamma^{-1}}^{\gamma^{-1}}B(\hat{u}(\tilde{x}))
\frac{\partial \hat{u}}{\partial \tilde{x}}
\frac{\partial w}{\partial \tilde{x}}d\tilde{x}. \label{variation} 
\end{IEEEeqnarray}
Since $\hat{u}(\tilde{x})+\epsilon w(\tilde{x})$ must satisfy the boundary 
conditions $\hat{u}(\pm\gamma^{-1})+\epsilon w(\pm\gamma^{-1})=u_{\mathrm{max}}$ 
around $\epsilon=0$, we impose the boundary conditions 
$\hat{u}(\pm\gamma^{-1})=u_{\mathrm{max}}$ and $w(\pm\gamma^{-1})=0$. 
Integrating by parts the last term in (\ref{variation}) 
with the boundary condition $w(\pm\gamma^{-1})=0$ yields 
\begin{equation} 
\frac{\partial H}{\partial\epsilon}[\hat{u}] 
= \int_{-\gamma^{-1}}^{\gamma^{-1}}\left\{
 V'(\hat{u}(\tilde{x})) - \mathfrak{L}[\hat{u}](\tilde{x})
\right\}w(\tilde{x})d\tilde{x}, 
\end{equation}
with $\mathfrak{L}$ defined in (\ref{operator}). 
This expression implies that the functional derivative of 
(\ref{energy_functional_bulk}) is given by 
\begin{equation} \label{functional_deriv} 
\frac{\delta H}{\delta \hat{u}}[\hat{u}](\tilde{x})  
= V'(\hat{u}(\tilde{x})) - \mathfrak{L}[\hat{u}](\tilde{x}). 
\end{equation}
Substituting (\ref{functional_deriv}) into (\ref{gradient_system_bulk}), 
we find that (\ref{gradient_system_bulk}) is equal to (\ref{single_system}) 
under the change of variables $x=\gamma\tilde{x}$.  

We next show that 
\begin{equation} \label{stability_bulk}
\frac{dH}{dt}[\hat{u}(\cdot,t)] 
= - \int_{-\gamma^{-1}}^{\gamma^{-1}}A(\hat{u}(\tilde{x},t))\left(
 \frac{\delta H}{\delta \hat{u}}[\hat{u}(\cdot,t)](\tilde{x})
\right)^{2}d\tilde{x} \leq 0, 
\end{equation}
where the equality holds if and only if $\delta H/\delta \hat{u}=0$, since 
$A(\cdot)$ given by (\ref{function_A}) is a positive function. 
This implies that the energy functional~(\ref{energy_functional_bulk}) is a 
Lyapunov functional~\cite{Michel08} on the space of twice continuously 
differentiable functions with the norm 
$\|u\|=\|u\|_{\infty}+\|u'\|_{\infty}+\|u''\|_{\infty}$. The space is 
known to be complete with respect to this norm. 
Lyapunov's direct method implies that Lemma~\ref{lemma_stability_bulk} 
follows from (\ref{stability_bulk}). The detailed proof is omitted since 
it is beyond the scope of this paper. 

Intuitively, the energy functional~(\ref{energy_functional_bulk}) 
monotonically decreases with the time-evolution of the state. Since 
(\ref{energy_functional_bulk}) is bounded below, $H[\hat{u}(\cdot,t)]$ 
converges to a finite value as $t\to\infty$, where $dH/dt=0$ holds and 
thus the functional derivative $\delta H/\delta \hat{u}$ vanishes. 
As $\gamma\to0$ the problem~(\ref{gradient_system_bulk}) for the interval 
$[-\gamma^{-1},\gamma^{-1}]$ reduces to that for 
the infinite interval $(-\infty,\infty)$. Nonetheless, the argument above 
should be valid as $\gamma\to0$, although a careful treatment for the region 
$|\tilde{x}|\gg1$ is required in considering the energy 
functional~(\ref{energy_functional_bulk}) and the functional derivative 
(\ref{stability_bulk}). These intuitive arguments imply that as $t\to\infty$ 
the state $\hat{u}(\tilde{x},t)$ should converge uniformly to a stationary 
state with respect to $\gamma>0$. 

Let us prove (\ref{stability_bulk}). Differentiating 
(\ref{energy_functional_bulk}) yields 
\begin{IEEEeqnarray}{rl}
&\frac{dH}{dt}[\hat{u}(\cdot,t)] 
\nonumber \\ 
=& \int_{-\gamma^{-1}}^{\gamma^{-1}}\left[
 V'(\hat{u}(\tilde{x},t)) + \frac{B'(\hat{u}(\tilde{x},t))}{2}
 \left(
  \frac{\partial \hat{u}}{\partial \tilde{x}}
 \right)^{2}
\right]\frac{\partial \hat{u}}{\partial t}d\tilde{x} \nonumber \\ 
&+ \int_{-\gamma^{-1}}^{\gamma^{-1}}B(\hat{u}(\tilde{x},t))
\frac{\partial \hat{u}}{\partial \tilde{x}}
\frac{\partial^{2}\hat{u}}{\partial t\partial \tilde{x}}d\tilde{x}. 
\end{IEEEeqnarray}
Integrating by parts the last term, we obtain 
\begin{equation} 
\frac{dH}{dt}[\hat{u}(\cdot,t)] = 
\int_{-\gamma^{-1}}^{\gamma^{-1}}
 \frac{\delta H}{\delta \hat{u}}[\hat{u}(\cdot,t)](\tilde{x})
 \frac{\partial \hat{u}}{\partial t}d\tilde{x}, 
\label{stability1} 
\end{equation}
with (\ref{functional_deriv}), 
where we have used $\partial u(\pm\gamma^{-1},t)/\partial t=0$, 
which is obtained from the boundary condition 
$\hat{u}(\pm\gamma^{-1},t)=u_{\mathrm{max}}$. 
Substituting (\ref{gradient_system_bulk}) into 
(\ref{stability1}) yields (\ref{stability_bulk}).

\section{Proof of Lemma~\ref{lemma_difference}}
\label{lemma_difference_proof} 
In order to explain the idea for proving Lemma~\ref{lemma_difference}, 
let us discretize the time derivative in (\ref{differential_system}) as  
$\partial\tilde{v}/\partial t\approx
(\tilde{v}(x,t+\delta) - \tilde{v}(x,t))/\delta$ for small $\delta>0$. 
The differential system~(\ref{differential_system}) is approximated by 
the discrete-time system $\tilde{v}_{n+1}(x)=\delta
\tilde{\mathfrak{G}}[\tilde{v}_{n}(\cdot)](x)$ for $n\in\mathbb{N}$, i.e.\ 
$\tilde{v}(x,n\delta)\approx\tilde{v}_{n}(x)$. 
Proposition~\ref{lemma_deriv} implies that, for sufficiently small $\gamma>0$, 
finite iterations of the system should result in a negligibly small change of 
the state $\tilde{v}_{n}(x)$ when the initial function 
$\tilde{v}_{0}(x)$ is smooth and very close to the solution $v_{\gamma}(x)$ of 
the fixed-point equation $v_{\gamma}(x)=\mathfrak{G}[v_{\gamma}(\cdot)](x)$. Since 
$\tilde{v}(x,t_{0})\approx\tilde{v}_{n_{0}}(x)$ for 
$n_{0}=t_{0}/\delta$, the state of the differential 
system~(\ref{differential_system}) at time $t=t_{0}$ should be very close to 
the initial state as long as $t_{0}$ is finite. The proof of 
Lemma~\ref{lemma_difference} is based on this intuition.  

Let $\tilde{v}_{n}(x)=\tilde{v}(x,\delta n)$ for $\delta>0$. 
We define two coupled sequences $\epsilon_{n}(x)$ and $\rho_{n}(x)$ of 
functions for $x\in\mathbb{R}$ by 
\begin{equation} \label{epsilon}
\epsilon_{n}(x) = \delta\left\{
 \rho_{n}(x)+(\kappa_{n}(x) + \epsilon_{1})\chi_{[-1,1]}(x) 
\right\}, 
\end{equation}
\begin{equation} 
\rho_{n+1}(x) = \rho_{n}(x) 
+ A\overline{\epsilon_{n}(x)} + \epsilon_{n}(x), \label{rho}
\end{equation}
for $A>0$ and $\epsilon_{1}>0$, with
\begin{equation}
\overline{\epsilon_{n}(x)} 
= \frac{1}{(2\gamma)^{2}}\int_{[-\gamma,\gamma]^{2}}\epsilon_{n}
(x+\omega_{1}+\omega_{2})d\omega_{1}d\omega_{2}.  
\end{equation}
In (\ref{epsilon}), $\chi_{[-1,1]}(x)$ denotes the indicator function of  
the interval $[-1,1]\subset\mathbb{R}$. 
and  the function $\kappa_{n}(x)$ is given by 
\begin{equation}
\kappa_{n}(x) 
= \left|
 \mathfrak{G}[\tilde{v}(\cdot,\delta n)](x)
 - \tilde{\mathfrak{G}}[\tilde{v}(\cdot,\delta n)](x)
\right|. 
\end{equation}

Since $A$ and $\epsilon_{1}$ are contained in (\ref{epsilon}) and (\ref{rho}), 
we note that $\epsilon_{n}(x)$ and $\rho_{n}(x)$ depend on $A$ and 
$\epsilon_{1}$. We first prove the following lemma. 
\begin{lemma} \label{lemma8} 
For any $x\in[-1,1]$, $\epsilon_{0}>0$, and any $\epsilon_{1}>0$, 
there exist some $A>0$ and $\delta>0$ such that 
\begin{equation} \label{former_bound} 
|\tilde{v}_{n}(x) - \mathfrak{G}[\tilde{v}_{n}](x)| < \rho_{n}(x),  
\end{equation}
\begin{equation} \label{latter_bound}
|\tilde{v}_{n+1}(x) - \tilde{v}_{n}(x)|<\epsilon_{n}(x), 
\end{equation}
with $\rho_{0}(x)=\epsilon_{0}\chi_{[-1,1]}(x)$. 
\end{lemma}  
\begin{IEEEproof}
Since we focus on $x\in[-1,1]$, $\chi_{[-1,1]}(x)=1$ holds. 
We first prove that the latter bound~(\ref{latter_bound}) is correct 
if the former bound~(\ref{former_bound}) holds. 
We use the mean-value theorem~\cite{Strang10} to find that for any 
$\epsilon_{1}>0$ there exists some $\delta>0$ such that 
\begin{equation}  \label{bound} 
\left|
 \tilde{v}(x,t+\delta) - \tilde{v}(x,t)
\right| <\delta\left|
 \frac{\partial}{\partial t}\tilde{v}(x,t)
\right| + \delta\epsilon_{1}.   
\end{equation}
From (\ref{differential_system}) and (\ref{bound}), we obtain 
\begin{IEEEeqnarray}{rl} 
&|\tilde{v}_{n+1}(x) - \tilde{v}_{n}(x)| 
\nonumber \\ 
<& \delta|\tilde{v}_{n}(x) - \tilde{\mathfrak{G}}[\tilde{v}_{n}](x)| 
+ \delta\epsilon_{1} \nonumber \\ 
<& \delta\left\{
 |\tilde{v}_{n}(x) - \mathfrak{G}[\tilde{v}_{n}](x)|
 + |\mathfrak{G}[\tilde{v}_{n}](x) 
 - \tilde{\mathfrak{G}}[\tilde{v}_{n}](x)| + \epsilon_{1}
\right\}, 
\nonumber \\ 
\label{recursion}
\end{IEEEeqnarray}
which implies the latter bound~(\ref{latter_bound}) holds if the former 
bound~(\ref{former_bound}) is correct.   

The proof is by induction. 
For $n=0$ (\ref{former_bound}) holds by selecting sufficiently small 
$\epsilon_{\mathrm{init}}>0$, because of 
$|v_{\mathrm{init}}(x)-v_{\gamma}(x,\infty)|<\epsilon_{\mathrm{init}}$. Thus, 
the latter bound~(\ref{latter_bound}) is also correct.  
Suppose that (\ref{former_bound}) holds for some $n$. 
Thus, the latter bound~(\ref{latter_bound}) is correct. 
Using the triangle inequality yields 
\begin{IEEEeqnarray}{rl}
&|\tilde{v}_{n+1}(x) - \mathfrak{G}[\tilde{v}_{n+1}](x)| 
\nonumber \\ 
<& |\tilde{v}_{n+1}(x)-\tilde{v}_{n}(x)|  
+ |\tilde{v}_{n}(x) - \mathfrak{G}[\tilde{v}_{n}](x)| 
\nonumber \\  
&+ \left|
 \mathfrak{G}[\tilde{v}_{n}](x) - \mathfrak{G}[\tilde{v}_{n+1}](x)
\right|.   \label{upper_bound_n} 
\end{IEEEeqnarray}
Since the functions $\varphi$ and $\psi$ are continuously differentiable, 
they are Lipschitz-continuous---there exist some positive constants 
$L_{\varphi}>0$ and $L_{\psi}>0$ such that 
$|\varphi(u_{1})-\varphi(u_{2})|<L_{\varphi}|u_{1}-u_{2}|$ and 
$|\psi(v_{1})-\psi(v_{2})|<L_{\psi}|v_{1}-v_{2}|$ for all 
$u_{1}, u_{2}\in\tilde{\mathcal{D}}$ and $v_{1}, v_{2}\in\mathcal{D}$. 
From (\ref{latter_bound}) and the definition of $\mathfrak{G}$ given via 
(\ref{FP_v}) and (\ref{FP_u}), it is possible to prove that 
\begin{IEEEeqnarray}{rl}
&\left|
 \mathfrak{G}[\tilde{v}_{n+1}](x)-\mathfrak{G}[\tilde{v}_{n}](x)
\right| 
\nonumber \\ 
<& \frac{A}{(2\gamma)^{2}}\int_{[-\gamma,\gamma]^{2}}
\epsilon_{n}(x+\omega_{1}+\omega_{2})d\omega_{1}d\omega_{2}, 
\end{IEEEeqnarray} 
with $A=\beta L_{\varphi}L_{\psi}$. This implies that the RHS of 
(\ref{upper_bound_n}) is bounded from above by $\rho_{n+1}(x)$.  
By induction, (\ref{former_bound}) and (\ref{latter_bound}) hold for any $n$.  
\end{IEEEproof}

We next prove that 
\begin{equation} \label{epsilon_bound} 
\lim_{\gamma\to0}\frac{1}{\delta}\int_{-1}^{1}\sup_{n\leq n_{0}}\epsilon_{n}(x)dx=0,  
\end{equation}
with $n_{0}=t_{0}/\delta\in\mathbb{N}$. 
Lemma~\ref{lemma_difference} follows immediately from (\ref{epsilon_bound}). 
Using the triangle inequality and Lemma~\ref{lemma8} yield  
\begin{IEEEeqnarray}{rl}
&\int_{-1}^{1}|\tilde{v}(x,t_{0}) - \tilde{v}(x,0)|dx 
\nonumber \\ 
<& \sum_{n=0}^{n_{0}-1}\int_{-1}^{1}|\tilde{v}_{n+1}(x) - \tilde{v}_{n}(x)|dx 
\nonumber \\  
<&\frac{t_{0}}{\delta}\int_{-1}^{1}\sup_{n<n_{0}}\epsilon_{n}(x)dx, 
\label{difference_bound}
\end{IEEEeqnarray}
where we have used (\ref{latter_bound}). Applying (\ref{epsilon_bound}) to 
(\ref{difference_bound}) implies that Lemma~\ref{lemma_difference} holds.  

In order to prove (\ref{epsilon_bound}), we let 
\begin{equation} \label{rho_transform} 
\hat{\rho}_{n}(\omega) 
= \int_{-\infty}^{\infty}\rho_{n}(x)e^{-\mathrm{i}\omega x}dx, 
\end{equation}
\begin{equation}
\hat{\kappa}_{n}(\omega) 
= \int_{-1}^{1}\kappa_{n}(x)e^{-\mathrm{i}\omega x}dx. 
\end{equation}
Applying the Fourier transform to both sides of (\ref{rho}), 
and then substituting (\ref{epsilon}) into the obtained expression, we obtain 
\begin{equation}
\hat{\rho}_{n+1}(\omega) 
= \hat{\rho}_{n}(\omega) 
+ \delta\hat{C}_{1}(\omega)   
\left(
 \hat{\rho}_{n}(\omega) + \hat{\kappa}_{n}(\omega) 
 + 2\epsilon_{1}\frac{\sin\omega}{\omega}
\right), \label{single_rho} 
\end{equation} 
with 
\begin{equation}
\hat{C}_{1}(\omega) 
= A\left(
 \frac{\sin\gamma\omega}{\gamma\omega}
\right)^{2}+1. 
\end{equation}
Proposition~\ref{lemma_deriv} implies that for any $\epsilon>0$ there exists 
some $\gamma_{0}>0$ such that $|\hat{\kappa}_{n}(\omega)|<\epsilon$ for 
all $\gamma\in(0,\gamma_{0})$, $n\leq n_{0}$, and all $\omega\in\mathbb{R}$. 
Solving (\ref{single_rho}) yields 
\begin{equation}
|\hat{\rho}_{n}(\omega)| 
\leq \epsilon_{0}\left|
 \frac{\sin\omega}{\omega}
\right| + \hat{C}_{2}(\omega)\left\{
 \left(
  1 + \delta|\hat{C}_{1}(\omega)| 
 \right)^{n} - 1
\right\}, 
\end{equation}
with 
\begin{equation}
\hat{C}_{2}(\omega)
= \epsilon + (\epsilon_{0} + 2\epsilon_{1})\left|
 \frac{\sin\omega}{\omega} 
\right|. 
\end{equation}
Using $n\leq n_{0}=t_{0}/\delta$ and the well-known fact 
that the function $f(x)=(1+1/x)^{x}$ for $x>0$ monotonically 
increases toward $e$, we obtain 
\begin{equation}
|\hat{\rho}_{n}(\omega)| 
< \epsilon_{0}\left|
 \frac{\sin\omega}{\omega}
\right| 
+ \hat{C}_{2}(\omega)\left(
 e^{t_{0}|\hat{C}_{1}(\omega)|} - 1
\right). 
\end{equation}
Taking $\omega\to0$ yields 
\begin{equation}
\lim_{\omega\to0}|\hat{\rho}_{n}(\omega)|  
< \epsilon_{0} + (\epsilon+\epsilon_{0}+2\epsilon_{1})\left(
 e^{t_{0}(A+1)} - 1
\right),  
\end{equation} 
which implies $\lim_{\gamma\to0}\sup_{n\leq n_{0}}
\lim_{\omega\to0}|\hat{\rho}_{n}(\omega)|=0$. Since 
$\rho_{n}(x)\geq0$, from (\ref{rho_transform}) we arrive at 
\begin{equation}
\lim_{\gamma\to0}\int_{-1}^{1}\sup_{n\leq n_{0}}\rho_{n}(x)dx = 0. 
\end{equation}
From (\ref{epsilon}), this implies (\ref{epsilon_bound}). 

\section{Proof of Theorem~\ref{theorem2_review}}
\label{proof_theorem2_review} 
We start with proving the following lemma. 
\begin{lemma} \label{lemma_review}  
Suppose that $\tilde{u}_{\mathrm{r}}$ is the 
metastable solution of the effective potential $U(\tilde{u})$. 
If there is a solution 
$\tilde{u}(0)\in(\tilde{u}_{\mathrm{l}},\tilde{u}_{\mathrm{un}})$ to the 
fixed-point equation $F(\tilde{u}(0))=1/\gamma$, with 
\begin{equation} \label{function_F} 
F(x) = \int_{x}^{\tilde{u}_{\mathrm{r}}}
\frac{dy}
{\sqrt{2\{U(y)-U(\tilde{u}(0))\}}}, 
\end{equation}
then there is a non-uniform stationary solution $\tilde{u}(x)$ to 
(\ref{simple_system}). Furthermore, the solution 
$\tilde{u}(x)$ satisfies 
\begin{equation} \label{non-uniform_solution} 
F(\tilde{u}(x)) = \frac{1-|x|}{\gamma}. 
\end{equation}
\end{lemma}
\begin{IEEEproof}[Proof of Lemma~\ref{lemma_review}] 
We confirm that $\tilde{u}(x)$ satisfying (\ref{non-uniform_solution}) is 
a stationary solution to (\ref{simple_system}). 
It is obvious that the solution~(\ref{non-uniform_solution}) satisfies 
the boundary condition $\tilde{u}(\pm1)=\tilde{u}_{\mathrm{r}}$. 
Let us show that the solution~(\ref{non-uniform_solution}) is a solution to 
(\ref{energy_conservation}).   
Differentiating (\ref{non-uniform_solution}) with respect to $x$ yields 
\begin{equation} \label{non_uniform_solution_deriv} 
\frac{\gamma}{\sqrt{2}}\frac{d\tilde{u}}{dx} = \left\{ 
\begin{array}{cl}  
\sqrt{U(\tilde{u}) - U(\tilde{u}(0))} & \hbox{for $x>0$} \\ 
-\sqrt{U(\tilde{u}) - U(\tilde{u}(0))} & \hbox{for $x<0$,} 
\end{array}
\right.
\end{equation}
which is equivalent to (\ref{energy_conservation}) with the constant 
$C=-U(\tilde{u}(0))$. Furthermore, it is straightforward to show that 
the LHS of (\ref{non_uniform_solution_deriv}) is differentiable 
with respect to $x$. These observations imply that the 
solution~(\ref{non-uniform_solution}) is indeed a non-uniform 
stationary solution to (\ref{simple_system}).   
\end{IEEEproof} 

\begin{remark} 
The conservation of mechanical energy explains why $\tilde{u}(0)$ must 
be between $\tilde{u}_{\mathrm{l}}$ and $\tilde{u}_{\mathrm{un}}$, as defined 
in Lemma~\ref{lemma_review}. Let us assume 
that there is a solution $\tilde{u}(x)$ to the Newton 
equation~(\ref{stationary_system}) such that the solution is on the right 
maximum of the inverted potential $-U(\tilde{u})$ at $x=-1$, and arrives at  
some position $\tilde{u}(0)>\tilde{u}_{\mathrm{un}}$ at $x=0$. 
The velocity $u'(0)$ at $x=0$ must be zero, because of 
the differentiability of $\tilde{u}(x)$ at $x=0$.  
However, the definition of $\tilde{u}_{\mathrm{un}}$, i.e.\ 
$U(\tilde{u}_{\mathrm{un}})=U(\tilde{u}_{\mathrm{r}})$ implies 
$-U(\tilde{u}(0))<-U(\tilde{u}_{\mathrm{r}})$, which breaks the conservation 
of mechanical energy: 
\begin{equation}
\frac{\gamma^{2}}{2}\tilde{u}'(0)^{2} - U(\tilde{u}(0)) < 
\frac{\gamma^{2}}{2}\tilde{u}'(-1)^{2} - U(\tilde{u}_{\mathrm{r}}). 
\end{equation} 
Thus, $\tilde{u}(0)$ must be smaller than $\tilde{u}_{\mathrm{un}}$. 

The same argument explains that $\tilde{u}(0)$ must be larger than 
$\tilde{u}_{\mathrm{l}}$. Let us assume $\tilde{u}(0)<\tilde{u}_{\mathrm{l}}$.  
Then, the conservation of mechanical energy implies that the velocity 
$\tilde{u}'(0)$ at $x=0$ must be non-zero, since $\tilde{u}_{\mathrm{l}}$ 
is the global maximizer of the inverted potential $-U(\tilde{u})$. 
However, the non-zero velocity $\tilde{u}'(0)$ contradicts the 
differentiability of $\tilde{u}(x)$ at $x=0$. 
Thus, $\tilde{u}(0)$ must be larger than $\tilde{u}_{\mathrm{l}}$. 
From the arguments above, $\tilde{u}(0)$ must be between 
$\tilde{u}_{\mathrm{l}}$ and $\tilde{u}_{\mathrm{un}}$. 
\end{remark} 

In order to prove the first part of Theorem~\ref{theorem2_review}, 
we show that $F(\tilde{u}(0))$ is bounded for all 
$\tilde{u}(0)\in(\tilde{u}_{\mathrm{l}},\tilde{u}_{\mathrm{un}})$, and that 
$F(\tilde{u}(0))$ tends to infinity 
as $\tilde{u}(0)\to\tilde{u}_{\mathrm{l}}$ or 
as $\tilde{u}(0)\to\tilde{u}_{\mathrm{un}}$. Lemma~\ref{lemma_review} and   
these properties of $F(\tilde{u}(0))$ imply that there are two 
non-uniform stationary solutions to (\ref{simple_system}) for sufficiently 
small $\gamma>0$. 

We shall show the former property of $F(\tilde{u}(0))$. 
Let $\tilde{u}_{0}$ denote a value between $\tilde{u}(0)$ and 
the unstable solution of the effective potential $U(\tilde{u})$. 
Splitting the interval of integration 
$(\tilde{u}(0),\tilde{u}_{\mathrm{r}})$ into the two intervals 
$(\tilde{u}(0),\tilde{u}_{0})$ and $(\tilde{u}_{0},\tilde{u}_{\mathrm{r}})$ 
yields 
\begin{IEEEeqnarray}{rl} 
F(\tilde{u}(0)) =& \int_{\tilde{u}(0)}^{\tilde{u}_{0}}
\frac{dy}
{\sqrt{2\{U(y)-U(\tilde{u}(0))\}}} 
\nonumber \\ 
&+ \int_{\tilde{u}_{0}}^{\tilde{u}_{\mathrm{r}}}\frac{
dy}{\sqrt{2\{U(y)-U(\tilde{u}(0))\}}}. 
\label{function_F_tmp} 
\end{IEEEeqnarray}
The condition $\tilde{u}(0)<\tilde{u}_{\mathrm{un}}$ implies that the second 
term is bounded, because of $U(y)>U(\tilde{u}(0))$ for all 
$y\in[\tilde{u}_{0},\tilde{u}_{\mathrm{r}}]$. Thus, we focus on the first term. 
Let $\bar{u}$ denote an appropriately chosen value between $\tilde{u}(0)$ 
and $\tilde{u}_{0}$. From the mean-value theorem~\cite{Strang10}, we obtain 
\begin{IEEEeqnarray}{rl}
& \int_{\tilde{u}(0)}^{\tilde{u}_{0}}\frac{dy}
{\sqrt{2\{U(y)-U(\tilde{u}(0))\}}} \nonumber \\ 
=& \int_{\tilde{u}(0)}^{\tilde{u}_{0}}\frac{dy}
{\sqrt{2U'(\bar{u})(y-\tilde{u}(0))}} \nonumber \\ 
<& \sup_{\tilde{u}\in(\tilde{u}(0),\tilde{u}_{0})}\frac{1}
{\sqrt{U'(\tilde{u})}}\int_{\tilde{u}(0)}^{\tilde{u}_{0}}\frac{dy}
{\sqrt{2(y-\tilde{u}(0))}} \nonumber \\ 
=& \sqrt{2(\tilde{u}_{0}-\tilde{u}(0))}
\sup_{\tilde{u}\in(\tilde{u}(0),\tilde{u}_{0})}
\frac{1}{\sqrt{U'(\tilde{u})}}, \label{first_term_bound} 
\end{IEEEeqnarray}
which is bounded, because of $\tilde{u}(0)>\tilde{u}_{\mathrm{l}}$. Thus, 
we find that $F(\tilde{u}(0))$ is bounded for all  
$\tilde{u}(0)\in(\tilde{u}_{\mathrm{l}},\tilde{u}_{\mathrm{un}})$. 

We next show the latter property of $F(\tilde{u}(0))$. 
The upper bound~(\ref{first_term_bound}) on the first term of 
(\ref{function_F_tmp}) diverges as $\tilde{u}(0)\to 
\tilde{u}_{\mathrm{l}}$, because of $U'(\tilde{u}_{\mathrm{l}})=0$.  
It is straightforward to show that the first term of (\ref{function_F_tmp}) 
tends to infinity as $\tilde{u}(0)\to \tilde{u}_{\mathrm{l}}$. 
On the other hand, the second term of (\ref{function_F_tmp}) diverges 
as $\tilde{u}(0)\to \tilde{u}_{\mathrm{un}}$, owing to 
$U(\tilde{u}_{\mathrm{r}})=U(\tilde{u}_{\mathrm{un}})$. These observations 
imply that $\tilde{u}(0)=\tilde{u}_{\mathrm{l}}$ and 
$\tilde{u}(0)=\tilde{u}_{\mathrm{un}}$ are solutions to the fixed-point 
equation $F(\tilde{u}(0))=1/\gamma$ in the limit $\gamma\to0$. 

We have shown that there are two non-uniform stationary solutions 
$\tilde{u}_{\mathrm{s}}(x)$ and $\tilde{u}_{\mathrm{un}}(x)$ to 
(\ref{simple_system}) for sufficiently small $\gamma>0$, and that 
$\tilde{u}_{\mathrm{s}}(0)$ and $\tilde{u}_{\mathrm{un}}(0)$ tend to 
$\tilde{u}_{\mathrm{l}}$ and $\tilde{u}_{\mathrm{un}}$ in the limit $\gamma\to0$, 
respectively. We next prove that $\tilde{u}_{\mathrm{s}}(x)$ converges to 
$\tilde{u}_{\mathrm{l}}$ for $x\in(-1,1)$ in the limit $\gamma\to0$, and that 
$\tilde{u}_{\mathrm{un}}(x)$ tends to $\tilde{u}_{\mathrm{r}}$ for $x\neq0$ in 
the limit $\gamma\to0$. 
Since the stationary solutions are even functions, 
without loss of generality, we focus on the interval $(0,1]$.  
Differentiating (\ref{non-uniform_solution}) for $x\in(0,1]$ with respect 
to $x$ yields  
\begin{equation} \label{derivative} 
\frac{\tilde{u}'(x)}{\sqrt{2\{U(\tilde{u}(x)) - U(\tilde{u}(0))\}}} 
= \frac{1}{\gamma}, 
\end{equation}
where we have used (\ref{function_F}). For the stationary solution 
$\tilde{u}(x)=\tilde{u}_{\mathrm{s}}(x)$, the denominator on the LHS of 
(\ref{derivative}) should tend to zero as $\gamma\to0$. 
Since $\tilde{u}_{\mathrm{l}}$ is the global stable solution of the 
potential $U(\tilde{u})$, the stationary solution $\tilde{u}_{\mathrm{s}}(x)$ 
tends to $\tilde{u}(0)=\tilde{u}_{\mathrm{l}}$ for $x\in(0,1]$ in the 
limit $\gamma\to0$. Similarly, we find that $\tilde{u}_{\mathrm{un}}(x)$ 
converges to $\tilde{u}_{\mathrm{r}}$ or $\tilde{u}_{\mathrm{un}}$ for 
$x\in(0,1]$ in the limit $\gamma\to0$, because of 
$U(\tilde{u}_{\mathrm{un}})=U(\tilde{u}_{\mathrm{r}})$. 

We shall prove the convergence of $\tilde{u}_{\mathrm{un}}(x)$ toward 
$\tilde{u}_{\mathrm{r}}$ for $x\in(0,1]$ in the limit $\gamma\to0$. 
For that purpose, we show $\tilde{u}_{\mathrm{un}}(x)>\tilde{u}_{\mathrm{un}}$ 
for $x\in(0,\epsilon)$, with sufficiently small $\epsilon>0$. 
Combining this property and $\tilde{u}_{\mathrm{un}}'(x)\geq0$ for 
$x\in(0,1]$, obtained from (\ref{derivative}), we find 
$\tilde{u}_{\mathrm{un}}(x)>\tilde{u}_{\mathrm{un}}$ for $x\in(0,1]$. 
This implies that $\tilde{u}_{\mathrm{un}}(x)$ tends to $\tilde{u}_{\mathrm{r}}$ 
for $x\in(0,1]$ in the limit $\gamma\to0$. 

Let us prove $\tilde{u}_{\mathrm{un}}(x)>\tilde{u}_{\mathrm{un}}$ 
for $x\in(0,\epsilon)$, for sufficiently small $\epsilon>0$. 
The mean-value theorem implies 
\begin{equation} \label{expansion} 
\tilde{u}_{\mathrm{un}}'(x) 
= \tilde{u}_{\mathrm{un}}'(0) 
+ \tilde{u}_{\mathrm{un}}''(\bar{x})x, 
\end{equation}
for $x\in(0,\epsilon)$ and some $\bar{x}\in(0,x)$. 
Substituting $\tilde{u}_{\mathrm{un}}'(0)=0$ and 
(\ref{stationary_system}) into (\ref{expansion}), we obtain 
\begin{equation}
\tilde{u}_{\mathrm{un}}'(x) 
= \frac{U'(\tilde{u}_{\mathrm{un}}(\bar{x}))}{\gamma^{2}}x. 
\end{equation}
This expression implies that $\tilde{u}_{\mathrm{un}}'(x)$ is strictly positive   
for $x\in(0,\epsilon)$, because of 
$U'(\tilde{u}_{\mathrm{un}}(\bar{x}))>0$. Thus, we find 
$\tilde{u}_{\mathrm{un}}(x)>\tilde{u}_{\mathrm{un}}(0)
=\tilde{u}_{\mathrm{un}}$ in the neighborhood $(0,\epsilon)$. 
This observation corresponds to the physical fact that a free particle cannot 
continue to stay at a point on a smooth slope. 

We complete the proof of Theorem~\ref{theorem2_review}, by 
analyzing the stability of the non-uniform stationary solution 
$\tilde{u}_{\mathrm{s}}(x)$ in the limit $\gamma\to0$. 
Repeating the derivation of (\ref{gradient_system}), we find 
that (\ref{simple_system}) can be represented as 
\begin{equation}
\frac{\partial\tilde{u}}{\partial t} 
= -A(g^{-1}(\tilde{u}))B(g^{-1}(\tilde{u}))
\frac{\delta\tilde{H}}{\delta\tilde{u}}[\tilde{u}(\cdot,t)](x), 
\end{equation}
with 
\begin{equation} \label{another_energy_functional} 
\tilde{H}[\tilde{u}] 
= \int_{-1}^{1}\left[
 U(\tilde{u}(x)) + \frac{\gamma^{2}}{2}\left(
  \frac{\partial\tilde{u}}{\partial x} 
 \right)^{2}
\right]dx. 
\end{equation}
Since $A(u)$ and $B(u)$ are positive, $\tilde{u}_{\mathrm{s}}(x)$ is stable if 
$\tilde{u}_{\mathrm{s}}(x)$ is a (local) minimizer of the energy 
functional~(\ref{another_energy_functional}). We shall prove 
\begin{equation} \label{global_stability} 
\lim_{\gamma\to0}\tilde{H}[\tilde{u}_{\mathrm{s}}] 
= 2U(\tilde{u}_{\mathrm{l}}), 
\end{equation} 
which implies that $\tilde{u}_{\mathrm{s}}(x)$ attains the global minimum  
of (\ref{another_energy_functional}) in the limit $\gamma\to0$, 
since $\tilde{u}_{\mathrm{l}}$ is the global stable solution of the 
potential $U(\tilde{u})$. Thus, the solution $\tilde{u}_{\mathrm{s}}(x)$ is 
stable in the limit $\gamma\to0$. 

Let us prove (\ref{global_stability}). Substituting 
(\ref{non_uniform_solution_deriv}) into (\ref{another_energy_functional}) 
yields 
\begin{equation}
\tilde{H}[\tilde{u}_{\mathrm{s}}]
= 2\int_{-1}^{1}U(\tilde{u}_{\mathrm{s}}(x))dx 
- 2U(\tilde{u}_{\mathrm{s}}(0)).  
\end{equation} 
Since the integrand $U(\tilde{u}_{\mathrm{s}}(x))$ is obviously bounded 
for all $x\in[-1,1]$, the dominated convergence theorem implies 
\begin{equation}
\lim_{\gamma\to0}\tilde{H}[\tilde{u}_{\mathrm{s}}]
= 2\int_{-1}^{1}\lim_{\gamma\to0}U(\tilde{u}_{\mathrm{s}}(x))dx 
- 2U(\tilde{u}_{\mathrm{l}}) 
= 2U(\tilde{u}_{\mathrm{l}}),  
\end{equation} 
where we have used the fact that $\tilde{u}_{\mathrm{s}}(x)$ converges to 
$\tilde{u}_{\mathrm{l}}$ for $x\in(-1,1)$ in the limit $\gamma\to0$.  

\section*{Acknowledgment}
K.\ Takeuchi would like to thank Hiroshi Nagaoka for useful comments.  

\ifCLASSOPTIONcaptionsoff
  \newpage
\fi



\bibliographystyle{IEEEtran}
\bibliography{IEEEabrv,kt-it2011}

\begin{thebibliography}{10}
\providecommand{\url}[1]{#1}
\csname url@samestyle\endcsname
\providecommand{\newblock}{\relax}
\providecommand{\bibinfo}[2]{#2}
\providecommand{\BIBentrySTDinterwordspacing}{\spaceskip=0pt\relax}
\providecommand{\BIBentryALTinterwordstretchfactor}{4}
\providecommand{\BIBentryALTinterwordspacing}{\spaceskip=\fontdimen2\font plus
\BIBentryALTinterwordstretchfactor\fontdimen3\font minus
  \fontdimen4\font\relax}
\providecommand{\BIBforeignlanguage}[2]{{%
\expandafter\ifx\csname l@#1\endcsname\relax
\typeout{** WARNING: IEEEtran.bst: No hyphenation pattern has been}%
\typeout{** loaded for the language `#1'. Using the pattern for}%
\typeout{** the default language instead.}%
\else
\language=\csname l@#1\endcsname
\fi
#2}}
\providecommand{\BIBdecl}{\relax}
\BIBdecl

\bibitem{Adachi98}
F.~Adachi, M.~Sawahashi, and H.~Suda, ``Wideband {DS-CDMA} for next-generation
  mobile communications systems,'' \emph{{IEEE} Commun. Mag.}, vol.~36, no.~9,
  pp. 56--69, Sep. 1998.

\bibitem{Dahlman98}
E.~Dahlman, B.~Gudmundson, M.~Nilsson, and J.~Sk\"old, ``{UMTS/IMT}-2000 based
  on wideband {CDMA},'' \emph{{IEEE} Commun. Mag.}, vol.~36, no.~9, pp. 70--80,
  Sep. 1998.

\bibitem{Ojanperae98}
T.~Ojanper\"a and R.~Prasad, ``An overview of air interface multiple access for
  {IMT-2000/UMTS},'' \emph{{IEEE} Commun. Mag.}, vol.~36, no.~9, pp. 82--95,
  Sep. 1998.

\bibitem{Verdu98}
S.~Verd\'u, \emph{Multiuser Detection}.\hskip 1em plus 0.5em minus 0.4em\relax
  New York: Cambridge University Press, 1998.

\bibitem{Lupas89}
R.~Lupas and S.~Verd\'u, ``Linear multiuser detectors for synchronous
  code-division multiple-access channels,'' \emph{{IEEE} Trans. Inf. Theory},
  vol.~35, no.~1, pp. 123--136, Jan. 1989.

\bibitem{Xie90}
Z.~Xie, R.~T. Short, and C.~K. Rushforth, ``A family of suboptimum detectors
  for coherent multiuser communications,'' \emph{{IEEE} J. Sel. Areas Commun.},
  vol.~8, no.~4, pp. 683--690, May 1990.

\bibitem{Madhow94}
U.~Madhow and M.~L. Honig, ``{MMSE} interference suppression for
  direct-sequence spread-spectrum {CDMA},'' \emph{{IEEE} Trans. Commun.},
  vol.~42, no.~12, pp. 3178--3188, Dec. 1994.

\bibitem{Tse99}
D.~N.~C. Tse and S.~V. Hanly, ``Linear multiuser receivers: effective
  interference, effective bandwidth and user capacity,'' \emph{{IEEE} Trans.
  Inf. Theory}, vol.~45, no.~2, pp. 641--657, Mar. 1999.

\bibitem{Verdu99}
S.~Verd\'u and S.~{Shamai (Shitz)}, ``Spectral efficiency of {CDMA} with random
  spreading,'' \emph{{IEEE} Trans. Inf. Theory}, vol.~45, no.~2, pp. 622--640,
  Mar. 1999.

\bibitem{Tanaka02}
T.~Tanaka, ``A statistical-mechanics approach to large-system analysis of
  {CDMA} multiuser detectors,'' \emph{{IEEE} Trans. Inf. Theory}, vol.~48,
  no.~11, pp. 2888--2910, Nov. 2002.

\bibitem{Mueller04}
R.~R. M\"{u}ller and W.~H. Gerstacker, ``On the capacity loss due to separation
  of detection and decoding,'' \emph{{IEEE} Trans. Inf. Theory}, vol.~50,
  no.~8, pp. 1769--1778, Aug. 2004.

\bibitem{Guo05}
D.~Guo and S.~Verd\'u, ``Randomly spread {CDMA}: Asymptotics via statistical
  physics,'' \emph{{IEEE} Trans. Inf. Theory}, vol.~51, no.~6, pp. 1983--2010,
  Jun. 2005.

\bibitem{Pearl88}
J.~Pearl, \emph{Probabilistic Reasoning in Intelligent Systems: Networks of
  Plausible Inference}.\hskip 1em plus 0.5em minus 0.4em\relax San Francisco,
  CA: Morgan Kaufmann, 1988.

\bibitem{Richardson08}
T.~Richardson and R.~Urbanke, \emph{Modern Coding Theory}.\hskip 1em plus 0.5em
  minus 0.4em\relax New York: Cambridge University Press, 2008.

\bibitem{Wang99}
X.~Wang and H.~V. Poor, ``Iterative (turbo) soft interference cancellation and
  decoding for coded {CDMA},'' \emph{{IEEE} Trans. Commun.}, vol.~47, no.~7,
  pp. 1046--1061, Jul. 1999.

\bibitem{Boutros02}
J.~Boutros and G.~Caire, ``Iterative multiuser joint decoding: Unified
  framework and asymptotic analysis,'' \emph{{IEEE} Trans. Inf. Theory},
  vol.~48, no.~7, pp. 1772--1793, Jul. 2002.

\bibitem{Caire04}
G.~Caire, R.~R. M\"uller, and T.~Tanaka, ``Iterative multiuser joint decoding:
  Optimal power allocation and low-complexity implementation,'' \emph{{IEEE}
  Trans. Inf. Theory}, vol.~50, no.~9, pp. 1950--1973, Sep. 2004.

\bibitem{Kabashima03}
Y.~Kabashima, ``A {CDMA} multiuser detection algorithm on the basis of belief
  propagation,'' \emph{J. Phys. A: Math. Gen.}, vol.~36, no.~43, pp.
  11\,111--11\,121, Oct. 2003.

\bibitem{Takeuchi13}
K.~Takeuchi and S.~Horio, ``Iterative multiuser detection and decoding with
  spatially coupled interleaving,'' \emph{IEEE Wireless Commun. Lett.}, vol.~2,
  no.~6, pp. 619--622, Dec. 2013.

\bibitem{Montanari06}
A.~Montanari and D.~N.~C. Tse, ``Analysis of belief propagation for non-linear
  problems: The example of {CDMA} (or: How to prove {Tanaka's} formula),'' in
  \emph{Proc. 2006 IEEE Inf. Theory Workshop}, Punta del Este, Uruguay, Mar.
  2006, pp. 160--164.

\bibitem{Yoshida06}
M.~Yoshida and T.~Tanaka, ``Analysis of sparsely-spread {CDMA} via statistical
  mechanics,'' in \emph{Proc. 2006 IEEE Int. Symp. Inf. Theory}, Seattle, USA,
  Jul. 2006, pp. 2378--2382.

\bibitem{Raymond07}
J.~Raymond and D.~Saad, ``Sparsely spread {CDMA}---a statistical
  mechanics-based analysis,'' \emph{J. Phys. A: Math. Theor.}, vol.~40, no.~41,
  pp. 12\,315--12\,333, Oct. 2007.

\bibitem{Guo08}
D.~Guo and C.-C. Wang, ``Multiuser detection of sparsely spread {CDMA},''
  \emph{{IEEE} J. Sel. Areas Commun.}, vol.~26, no.~3, pp. 421--431, Apr. 2008.

\bibitem{Kudekar111}
S.~Kudekar, T.~Richardson, and R.~Urbanke, ``Threshold saturation via spatial
  coupling: Why convolutional {LDPC} ensembles perform so well over the
  {BEC},'' \emph{{IEEE} Trans. Inf. Theory}, vol.~57, no.~2, pp. 803--834, Feb.
  2011.

\bibitem{Felstrom99}
A.~J. Felstr\"om and K.~S. Zigangirov, ``Time-varying periodic convolutional
  codes with low-density parity-check matrix,'' \emph{{IEEE} Trans. Inf.
  Theory}, vol.~45, no.~6, pp. 2181--2191, Sep. 1999.

\bibitem{Lentmaier10}
M.~Lentmaier and G.~P. Fettweis, ``On the thresholds of generalized {LDPC}
  convolutional codes based on protographs,'' in \emph{Proc. 2010 IEEE Int.
  Symp. Inf. Theory}, Austin, TX, USA, Jun. 2010, pp. 709--713.

\bibitem{Hassani10}
S.~H. Hassani, N.~Macris, and R.~Urbanke, ``Coupled graphical models and their
  thresholds,'' in \emph{Proc. 2010 IEEE Inf. Theory Workshop}, Dublin,
  Ireland, Aug.--Sep. 2010.

\bibitem{Hassani12}
------, ``Chains of mean field models,'' \emph{J. Stat. Mech.}, no.~2, p.
  P02011, Feb. 2012.

\bibitem{Takeuchi111}
K.~Takeuchi, T.~Tanaka, and T.~Kawabata, ``A phenomenological study on
  threshold improvement via spatial coupling,'' \emph{IEICE Trans.
  Fundamentals}, vol. E95-A, no.~5, pp. 974--977, May 2012.

\bibitem{Yedla12}
A.~Yedla, Y.-Y. Jian, P.~S. Nguyen, and H.~D. Pfister, ``A simple proof of
  threshold saturation for coupled scalar recursions,'' in \emph{Proc. 7th Int.
  Symp. Turbo Codes \& Iter. Inf. Process.}, Gothenburg, Sweden, Aug. 2012.

\bibitem{Kudekar12}
S.~Kudekar, T.~Richardson, and R.~Urbanke, ``Wave-like solutions of general
  one-dimensional spatially coupled systems,'' \emph{{\rm submitted to} IEEE
  Trans. Inf. Theory}, 2012, [Online]. Available:
  http://arxiv.org/abs/1208.5273.

\bibitem{Schlegel132}
C.~Schlegel and M.~Burnashev, ``Thresholds of spatially coupled systems via
  {Lyapunov's} method,'' in \emph{Proc. 2013 IEEE Inf. Theory Workshop},
  Seville, Spain, Sep. 2013.

\bibitem{Kudekar102}
S.~Kudekar and H.~D. Pfister, ``The effect of spatial coupling on compressive
  sensing,'' in \emph{Proc. 48th Annual Allerton Conf. Commun. Control \&
  Computing}, Los Alamos, USA, Sep.--Oct. 2010, pp. 347--353.

\bibitem{Krzakala12}
F.~Krzakala, M.~M\'ezard, F.~Sausset, Y.~F. Sun, and L.~Zdeborov\'a,
  ``Statistical-physics-based reconstruction in compressed sensing,''
  \emph{Phys. Rev. X}, vol.~2, pp. 021\,005--1--18, May 2012.

\bibitem{Donoho13}
D.~L. Donoho, A.~Javanmard, and A.~Montanari, ``Information-theoretically
  optimal compressed sensing via spatial coupling and approximate message
  passing,'' \emph{IEEE Trans. Inf. Theory}, vol.~59, no.~11, pp. 7434--7464,
  Nov. 2013.

\bibitem{Hagiwara11}
M.~Hagiwara, K.~Kasai, H.~Imai, and K.~Sakaniwa, ``Spatially coupled
  quasi-cyclic quantum {LDPC} codes,'' in \emph{Proc. 2011 IEEE Int. Symp. Inf.
  Theory}, Saint Petersburg, Russia, Aug. 2011, pp. 638--642.

\bibitem{Kasai11}
K.~Kasai and K.~Sakaniwa, ``Spatially-coupled {MacKay-Neal} codes and
  {Hsu-Anastasopoulos} codes,'' \emph{IEICE Trans. Fundamentals}, vol. E94-A,
  no.~11, pp. 2161--2168, Nov. 2011.

\bibitem{Kudekar112}
S.~Kudekar and K.~Kasai, ``Threshold saturation on channels with memory via
  spatial coupling,'' in \emph{Proc. 2011 IEEE Int. Symp. Inf. Theory}, Saint
  Petersburg, Russia, Aug. 2011, pp. 2562--2566.

\bibitem{Kudekar113}
------, ``Spatially coupled codes over the multiple access channel,'' in
  \emph{Proc. 2011 IEEE Int. Symp. Inf. Theory}, Saint Petersburg, Russia, Aug.
  2011, pp. 2816--2820.

\bibitem{Rathi11}
V.~Rathi, R.~Urbanke, M.~Andersson, and M.~Skoglund, ``Rate-equivocation
  optimal spatially coupled {LDPC} codes for the {BEC} wiretap channel,'' in
  \emph{Proc. 2011 IEEE Int. Symp. Inf. Theory}, Saint Petersburg, Russia, Aug.
  2011, pp. 2393--2397.

\bibitem{Schlegel11}
C.~Schlegel and D.~Truhachev, ``Multiple access demodulation in the lifted
  signal graph with spatial coupling,'' in \emph{Proc. 2011 IEEE Int. Symp.
  Inf. Theory}, Saint Petersburg, Russia, Aug. 2011, pp. 2989--2993.

\bibitem{Schlegel131}
------, ``Multiple access demodulation in the lifted signal graph with spatial
  coupling,'' \emph{IEEE Trans. Inf. Theory}, vol.~59, no.~4, pp. 2459--2470,
  Apr. 2013.

\bibitem{Takeuchi112}
K.~Takeuchi, T.~Tanaka, and T.~Kawabata, ``Improvement of {BP}-based {CDMA}
  multiuser detection by spatial coupling,'' in \emph{Proc. 2011 IEEE Int.
  Symp. Inf. Theory}, Saint Petersburg, Russia, Aug. 2011, pp. 1489--1493.

\bibitem{Uchikawa11}
H.~Uchikawa, K.~Kasai, and K.~Sakaniwa, ``Spatially coupled {LDPC} codes for
  decode-and-forward in erasure relay channel,'' in \emph{Proc. 2011 IEEE Int.
  Symp. Inf. Theory}, Saint Petersburg, Russia, Aug. 2011, pp. 1474--1478.

\bibitem{Yedla11}
A.~Yedla, H.~D. Pfister, and K.~R. Narayanan, ``Universality for the noisy
  {Slepian-Wolf} problem via spatial coupling,'' in \emph{Proc. 2011 IEEE Int.
  Symp. Inf. Theory}, Saint Petersburg, Russia, Aug. 2011, pp. 2567--2571.

\bibitem{Aref11}
V.~Aref and R.~Urbanke, ``Universal rateless codes from coupled {LT} codes,''
  in \emph{Proc. 2011 IEEE Inf. Theory Workshop}, Paraty, Brazil, Oct. 2011,
  pp. 277--281.

\bibitem{Truhachev12}
D.~Truhachev, ``Achieving {AWGN} multiple access channel capacity with spatial
  graph coupling,'' \emph{{IEEE} Commun. Lett.}, vol.~16, no.~5, pp. 585--588,
  May 2012.

\bibitem{Guo052}
D.~Guo, S.~{Shamai (Shitz)}, and S.~Verd\'u, ``Mutual information and minimum
  mean-square error in {Gaussian} channels,'' \emph{{IEEE} Trans. Inf. Theory},
  vol.~51, no.~4, pp. 1261--1282, Apr. 2005.

\bibitem{Korada11}
S.~B. Korada and A.~Montanari, ``Applications of the {Lindeberg} principle in
  communications and statistical learning,'' \emph{{IEEE} Trans. Inf. Theory},
  vol.~57, no.~4, pp. 2440--2450, Apr. 2011.

\bibitem{Ash99}
R.~B. Ash and C.~A. Dol\'eans-Dade, \emph{Probability \& Measure Theory},
  2nd~ed.\hskip 1em plus 0.5em minus 0.4em\relax San Diego: Academic Press,
  1999.

\bibitem{Strang10}
G.~Strang, \emph{Calculus}, 2nd~ed.\hskip 1em plus 0.5em minus 0.4em\relax
  Wellesley: Wellesley-Cambridge Press, 2010.

\bibitem{Amari00}
S.~Amari and H.~Nagaoka, \emph{Methods of Information Geometry}.\hskip 1em plus
  0.5em minus 0.4em\relax Providence, RI, USA: American Mathematical Society,
  2000.

\bibitem{Kobayashi96}
S.~Kobayashi and K.~Nomizu, \emph{Foundations of Differential Geometry},
  {Wiley} classics library~ed.\hskip 1em plus 0.5em minus 0.4em\relax New York:
  Wiley, 1996.

\bibitem{Ikehara07}
T.~Ikehara and T.~Tanaka, ``Decoupling principle in belief-propagation-based
  {CDMA} multiuser detection algorithm,'' in \emph{Proc. 2007 IEEE Int. Symp.
  Inf. Theory}, Nice, France, Jun. 2007, pp. 2081--2085.

\bibitem{Michel08}
A.~N. Michel, L.~Hou, and D.~Liu, \emph{Stability of Dynamical Systems:
  Continuous, Discontinuous, and Discrete Systems}.\hskip 1em plus 0.5em minus
  0.4em\relax Boston, MA, USA: Birkh\"auser, 2008.

\end{thebibliography}
\end{document}